\documentclass[11pt,a4paper]{article}

\usepackage{amsmath,amssymb,amsthm,mathtools}
\usepackage[T1]{fontenc}
\usepackage{microtype}
\usepackage{geometry}
\usepackage{xcolor}
\usepackage{hyperref}
\usepackage{cleveref}
\usepackage{tcolorbox}
\usepackage{booktabs}
\usepackage{array}
\usepackage{colortbl}
\usepackage{enumitem}
\usepackage{tikz}
\usetikzlibrary{arrows.meta,positioning,calc,decorations.pathreplacing,matrix}
\usepackage{titlesec}
\usepackage{fancyhdr}
\usepackage{mdframed}
\usepackage{xspace}
\usepackage{todonotes}
\usepackage{thm-restate}
\usetikzlibrary{arrows.meta,calc,positioning}
\definecolor{mdfBlue}{HTML}{176B91}
\definecolor{mdfPurple}{HTML}{8A3F78}
\definecolor{mdfOrange}{HTML}{C77718}
\definecolor{mdfInk}{HTML}{24333D}
\definecolor{mdfGray}{HTML}{6B747B}
\definecolor{mdfLight}{HTML}{E0E5E8}
\tikzset{
  mdf vertex/.style={circle,draw=mdfBlue,fill=mdfBlue,inner sep=0pt,minimum size=4.6pt},
  mdf witness/.style={circle,draw=mdfOrange,fill=white,line width=1pt,inner sep=0pt,minimum size=5.3pt},
  mdf route/.style={draw=mdfBlue,line width=1.1pt,line cap=round},
  mdf pair/.style={draw=mdfPurple,densely dashed,line width=1pt,line cap=round},
  mdf aux/.style={draw=mdfGray,line width=.65pt},
  mdf arrow/.style={-{Stealth[length=2mm]},draw=mdfGray,line width=.75pt},
  mdf panel/.style={font=\fontsize{9}{11}\selectfont\bfseries,text=mdfInk,anchor=west},
  mdf note/.style={font=\fontsize{8}{9.5}\selectfont,text=mdfInk,align=center,inner sep=2pt},
  mdf small/.style={font=\fontsize{7}{8.5}\selectfont,text=mdfGray,align=center,inner sep=2pt},
  mdf box/.style={draw=mdfLight,rounded corners=2pt,fill=mdfBlue!3,line width=.6pt},
}

\newcommand{\MethodsPreservationDrawing}{%
\begin{tikzpicture}[x=1cm,y=1cm,font=\fontsize{9}{11}\selectfont,text=mdfInk]
  \path[use as bounding box] (0,-1.5) rectangle (15.5,2.55);
  \node[mdf panel] at (0,2.25) {(a) A witness gives a tree path};
  \node[mdf panel] at (5.35,2.25) {(b) The preservation graph};
  \node[mdf panel] at (10.65,2.25) {(c) Longer shortest paths};
  \draw[mdfLight] (5.05,-.85)--(5.05,2.02);
  \draw[mdfLight] (10.35,-.85)--(10.35,2.02);
  \begin{scope}[shift={(.55,.65)}]
    \coordinate (x) at (0,0); \coordinate (z) at (1.75,0); \coordinate (y) at (3.55,0);
    \draw[mdf route] (x)--node[above=3pt] {$a$}(z)--node[above=3pt] {$b$}(y);
    \draw[mdf pair] (x) to[bend left=38] node[above=3pt] {$d(x,y)=a+b$}(y);
    \node[mdf vertex,label=below:$x$] at (x) {};
    \node[mdf witness,label=below:$z$] at (z) {};
    \node[mdf vertex,label=below:$y$] at (y) {};
    \node[mdf note] at (1.78,-1.03) {$xz,zy\in E(T),\quad xy\notin E(T)$};
  \end{scope}
  \begin{scope}[shift={(5.9,.65)}]
    \coordinate (x) at (0,0); \coordinate (z) at (1.75,0); \coordinate (y) at (3.55,0);
    \draw[mdf pair] (x) to[bend left=38] node[above=3pt] {$xy\in E(H_X)$}(y);
    \node[mdf vertex,label=below:$x$] at (x) {};
    \node[mdf witness,label=below:$z$] at (z) {};
    \node[mdf vertex,label=below:$y$] at (y) {};
    \node[mdf note] at (1.78,-1.03) {$U=\{x,y\}$; the witness $z$ is isolated};
  \end{scope}
  \begin{scope}[shift={(11.05,.65)}]
    \coordinate (x) at (0,0); \coordinate (a) at (1.1,0);
    \coordinate (b) at (2.2,0); \coordinate (y) at (3.3,0);
    \draw[mdf route] (x)--(a)--(b)--(y);
    \draw[mdf pair] (x) to[bend left=35] (b);
    \draw[mdf pair] (a) to[bend right=35] (y);
    \foreach \p in {x,a,b,y}\node[mdf vertex] at (\p) {};
    \node[mdf note] at (-.07,-.34) {$x$};
    \node[mdf note] at (1.1,.65) {$v_1$};
    \node[mdf note] at (2.2,-.65) {$v_2$};
    \node[mdf note] at (3.48,.25) {$y$};
    \node[mdf note] at (1.65,-1.03) {$xv_2,v_1y\in E(H_X)$; all four lie in $U$};
  \end{scope}
  \draw[mdf route] (.4,-1.2)--(.95,-1.2);
  \node[anchor=west,font=\fontsize{7}{8.5}\selectfont] at (1.05,-1.2) {edges of a shortest path};
  \draw[mdf pair] (5.5,-1.2)--(6.05,-1.2);
  \node[anchor=west,font=\fontsize{7}{8.5}\selectfont] at (6.15,-1.2) {pairs in $E(H_X)$};
  \node[mdf vertex] at (10.35,-1.2) {};
  \node[anchor=west,font=\fontsize{7}{8.5}\selectfont] at (10.5,-1.2) {in $U$};
  \node[mdf witness] at (12,-1.2) {};
  \node[anchor=west,font=\fontsize{7}{8.5}\selectfont] at (12.15,-1.2) {outside $U$};
\end{tikzpicture}%
}
\newcommand{\MethodsPreservationFigure}{%
\begin{figure}[t]
  \centering\resizebox{\linewidth}{!}{\MethodsPreservationDrawing}
  \caption{A preservation edge records a pair, rather than a route.
  In (a)--(b), $X=\{x,z,y\}$, $a,b>0$, and $z\in X\setminus\{x,y\}$ witnesses
  $\{x,y\}$. The tree path $xzy$ strongly preserves this pair, while $H_X$
  contains only the edge $xy$: a useful witness can be isolated in $H_X$.
  In (c), $xv_1v_2y$ is a shortest path. Its subpaths imply
  $xv_2,v_1y\in E(H_X)$, so all four vertices belong to the non-isolated
  set $U$. Only the two preservation edges needed for this conclusion are shown.}
  \label{fig:methods-preservation}
\end{figure}%
}

\newcommand{\MethodsWinDrawing}{%
\begin{tikzpicture}[x=1cm,y=1cm,font=\fontsize{9}{11}\selectfont,text=mdfInk]
  \path[use as bounding box] (0,-4.55) rectangle (15.5,3.05);
  \node[mdf note] at (7.75,2.8) {Assume $n>6k$ and $1\leq\ell<\lceil n/2\rceil$};
  \node[mdf panel] at (.25,2.25) {(a) A large pattern in $H_X$};
  \begin{scope}[shift={(1.1,.9)}]
    \foreach \i in {0,1,3}{
      \draw[mdf pair] (\i*.72,0)--(\i*.72,.65);
      \node[mdf vertex] at (\i*.72,0) {};
      \node[mdf vertex] at (\i*.72,.65) {};
    }
    \node at (1.44,.33) {$\cdots$};
    \node[mdf note] at (1.1,-.7) {matching of size $2k$};
  \end{scope}
  \node[mdf small] at (4.25,1.15) {or};
  \begin{scope}[shift={(5.7,1.22)}]
    \foreach \ang in {15,75,150,215,290}{
      \draw[mdf pair] (0,0)--(\ang:.7);
      \node[mdf vertex] at (\ang:.7) {};
    }
    \node[mdf vertex] at (0,0) {};
    \node[mdf note] at (0,-1.02) {degree $\geq k$};
  \end{scope}
  \draw[mdf arrow] (7.05,1.25)--(7.85,1.25);
  \node[mdf panel] at (8.05,2.25) {(b) A small preserving forest};
  \begin{scope}[shift={(8.45,1.2)}]
    \foreach \s in {0,1}{
      \begin{scope}[xshift=\s*3.1cm]
        \draw[mdf route] (0,0)--(.95,0)--(1.9,0);
        \draw[mdf pair] (0,0) to[bend left=35] (1.9,0);
        \node[mdf vertex] at (0,0) {};
        \node[circle,draw=mdfGray,fill=white,line width=.9pt,inner sep=0pt,minimum size=5.3pt] at (.95,0) {};
        \node[mdf vertex] at (1.9,0) {};
      \end{scope}
    }
    \node[mdf small] at (2.5,0) {$\cdots$};
    \node[mdf note] at (2.5,-.85) {$|E(F)|\leq2k$; $F$ strongly preserves $k$ pairs in $R$};
  \end{scope}
  \draw[mdf arrow] (11.1,-.12)--(11.1,-.55);
  \node[mdf box,minimum width=15cm,minimum height=1.1cm] at (7.75,-1.15) {};
  \node[mdf panel] at (.5,-.88) {(c) Reserve the extra pairs before packing trees};
  \node[mdf note] at (7.75,-1.38)
    {$L=R\cup E(F),\qquad |L|\leq 3k<n/2\qquad\Longrightarrow\qquad K_X-L\text{ has }\ell\text{ edge-disjoint spanning trees}$};
  \draw[mdf arrow] (7.75,-1.75)--(7.75,-2.15);
  \node[mdf note] at (7.75,-2.45)
    {Insert $F$ into the first tree using only $F$ and that tree's edges};
  \node[mdf note] at (7.75,-3.02)
    {$F\subseteq T_1,\qquad E(T_i)\cap E(T_j)=\varnothing\ (i\ne j),\qquad E(T_i)\cap R=\varnothing\ \text{for every }i$};
  \node[mdf box,fill=mdfPurple!4,minimum width=13.1cm,minimum height=1.02cm] at (7.75,-3.95) {};
  \node at (7.75,-3.92)
    {$\displaystyle \#\text{preserved pairs}\ \geq\
      \underbrace{\ell(n-1)}_{\text{distinct tree edges}}\ +\
      \underbrace{k}_{\text{pairs in }R}$};
\end{tikzpicture}%
}
\newcommand{\MethodsWinFigure}{%
\begin{figure}[t]
  \centering\resizebox{\linewidth}{!}{\MethodsWinDrawing}
  \caption{From a large preservation graph to a yes-instance.
  The matching or degree condition yields a forest $F$ with at most $2k$
  edges strongly preserving a set $R$ of $k$ pairs; the displayed forest is
  schematic. Identifying pairs with edges of $K_X$, reserve both $R$ and
  $E(F)$ before applying the tree-packing lemma. Extending the first packed
  tree to contain $F$ preserves edge-disjointness, and no resulting tree
  uses an edge of $R$. Thus the $k$ extra pairs are distinct from all
  $\ell(n-1)$ tree-edge pairs.}
  \label{fig:methods-win}
\end{figure}%
}

\newcommand{\MethodsWitnessTypesDrawing}{%
\begin{tikzpicture}[x=1cm,y=1cm,font=\fontsize{9}{11}\selectfont,text=mdfInk]
  \path[use as bounding box] (0,-2.35) rectangle (15.5,3.15);
  \node[mdf panel] at (.2,2.85) {(a) A shared outside witness};
  \node[mdf panel] at (7.3,2.85) {(b) Witness signatures and equivalence};
  \node[mdf box,minimum width=2.9cm,minimum height=3cm] at (1.9,.7) {};
  \node[mdf note] at (1.9,1.93) {$U$};
  \coordinate (x) at (.95,1.45); \coordinate (y) at (.95,-.1);
  \coordinate (u) at (2.8,1.45); \coordinate (v) at (2.8,-.1);
  \coordinate (w) at (5.3,.675);
  \foreach \a/\b in {x/y,u/v} \draw[mdf pair] (\a)--(\b);
  \foreach \a in {x,y,u,v} \draw[mdf route] (\a)--(w);
  \node[mdf vertex,label=above:$x$] at (x) {};
  \node[mdf vertex,label=below:$y$] at (y) {};
  \node[mdf vertex,label=above:$u$] at (u) {};
  \node[mdf vertex,label=below:$v$] at (v) {};
  \node[mdf witness,label=right:{$w\notin U$}] at (w) {};
  \node[mdf note] at (3.2,-1.08) {$\sigma(w)=\{xy,uv\}$};
  \draw[mdfLight] (6.85,-1.35)--(6.85,2.55);
  \node[mdf small] at (11.35,2.25) {pair columns $R=E(H_X)$ (schematic)};
  \foreach \c/\lab in {0/xy,1/uv,2/st}
    \node[mdf note] at (9.4+\c*1.15,1.76) {$\lab$};
  \node[mdf note] at (13.7,1.76) {signature};
  \fill[mdfOrange!9,rounded corners=2pt] (7.6,.24) rectangle (15.3,1.4);
  \foreach \r/\lab in {0/w_1,1/w_2,2/w_3,3/w_4}{
    \node[mdf note] at (8.2,1.11-\r*.55) {$\lab$};
    \foreach \c in {0,1,2}
      \draw[mdfLight,fill=white] (9.4+\c*1.15,1.11-\r*.55) circle (2.1pt);
  }
  \foreach \r/\c in {0/0,0/1,1/0,1/1,2/1,2/2,3/0}
    \fill[mdfPurple] (9.4+\c*1.15,1.11-\r*.55) circle (2.1pt);
  \foreach \r/\sig in {0/{\{xy,uv\}},1/{\{xy,uv\}},2/{\{uv,st\}},3/{\{xy\}}}
    \node[mdf note] at (13.7,1.11-\r*.55) {$\sig$};
  \node[mdf small] at (11.35,-1.08) {$w_1,w_2$ are equivalent: $\sigma(w_1)=\sigma(w_2)$};
  \node[mdf box,minimum width=15cm,minimum height=.82cm] at (7.75,-1.88) {};
  \node[mdf note] at (7.75,-1.88)
    {$|R|\leq2k^2\quad\Longrightarrow\quad
      \underbrace{2^{|R|}}_{\text{at most this many types}}\ \times\
      \underbrace{2k^2}_{\text{representatives per type}}\quad\Longrightarrow\quad
      |A^*|\leq 2k^2\,2^{2k^2}$};
\end{tikzpicture}%
}
\newcommand{\MethodsWitnessTypesFigure}{%
\begin{figure}[t]
  \centering\resizebox{\linewidth}{!}{\MethodsWitnessTypesDrawing}
  \caption{The witness bottleneck in general metrics.
  In (a), take $d(x,y)=d(u,v)=2$ and all other distances between distinct
  points to be $1$. Then $E(H_X)=\{xy,uv\}$, and the isolated vertex $w$
  witnesses both pairs. Solid edges illustrate the two preserving paths.
  In (b), a filled entry means that a witness serves the column's pair;
  this signature table is schematic. Witnesses with identical signatures
  form an equivalence class. Since a minimal preservation forest family
  needs at most $2k^2$ distinct witnesses outside $U$, keeping at most
  $2k^2$ representatives of each class suffices for the search.}
  \label{fig:methods-witness-types}
\end{figure}%
}

\newcommand{\MethodsCompressionDrawing}{%
\begin{tikzpicture}[x=1cm,y=1cm,font=\fontsize{9}{11}\selectfont,text=mdfInk]
  \path[use as bounding box] (0,-6.85) rectangle (15.5,3.05);
  \node[mdf panel] at (.2,2.75) {(a) General metric: two outside witnesses};
  \node[mdf panel] at (8.15,2.75) {(b) Strict convexity forces both into $U$};
  \draw[mdfLight] (7.75,-1.4)--(7.75,2.45);
  \begin{scope}[shift={(.9,.75)}]
    \coordinate (x) at (0,0); \coordinate (y) at (5.4,0);
    \coordinate (a) at (2.7,1); \coordinate (b) at (2.7,-1);
    \draw[mdf route] (x)--node[above left] {$1$}(a)--node[above right] {$1$}(y);
    \draw[mdf route] (x)--node[below left] {$1$}(b)--node[below right] {$1$}(y);
    \draw[mdf aux] (a)--node[right=3pt,pos=.3,fill=white,inner sep=1pt] {$1$}(b);
    \draw[mdf pair] (x)--node[above=2pt,pos=.25] {$2$}(y);
    \node[mdf vertex,label=left:$x$] at (x) {};
    \node[mdf vertex,label=right:$y$] at (y) {};
    \node[mdf witness,label=above:{$w_1$}] at (a) {};
    \node[mdf witness,label=below:{$w_2$}] at (b) {};
    \node[mdf note] at (2.7,-1.83) {$E(H_X)=\{xy\},\qquad w_1,w_2\notin U$};
  \end{scope}
  \begin{scope}[shift={(8.7,.75)}]
    \coordinate (x) at (0,0); \coordinate (a) at (1.65,0);
    \coordinate (b) at (3.3,0); \coordinate (y) at (4.95,0);
    \draw[mdf route] (x)--(a)--(b)--(y);
    \draw[mdf pair] (x) to[bend left=37] node[above=3pt] {$xw_2\in E(H_X)$}(b);
    \draw[mdf pair] (a) to[bend right=37] node[below=3pt] {$w_1y\in E(H_X)$}(y);
    \node[mdf vertex,label=below left:{$x$}] at (x) {};
    \node[mdf vertex,label=above:{$w_1$}] at (a) {};
    \node[mdf vertex,label=below:{$w_2$}] at (b) {};
    \node[mdf vertex,label=above right:{$y$}] at (y) {};
    \node[mdf note] at (2.48,-1.83) {A common shortest path orders $x,w_1,w_2,y$};
  \end{scope}
  \draw[mdfLight] (.2,-1.5)--(15.3,-1.5);
  \node[mdf panel] at (.2,-1.88) {(c) Add the witness edges};
  \node[mdf panel] at (6.05,-1.88) {(d) Retain a sparse support for every forest};
  \begin{scope}[shift={(.7,-3.12)}]
    \coordinate (x) at (0,0); \coordinate (z) at (2.05,0); \coordinate (y) at (4.1,0);
    \draw[mdf route] (x)--(z)--(y);
    \draw[mdf pair] (x) to[bend left=24] node[above=1pt] {$xy\in E(H_X)$} (y);
    \node[mdf vertex,label=below:$x$] at (x) {};
    \node[mdf witness,label=below:$z$] at (z) {};
    \node[mdf vertex,label=below:$y$] at (y) {};
    \node[mdf small] at (2.05,-.72) {$z$ is the unique witness: add $xz,zy$ to $H_X^*$};
  \end{scope}
  \draw[mdf arrow] (5.2,-2.95)--(5.8,-2.95);
  \node[mdf note] at (10.65,-2.45)
    {$|U\cup W|<4k^2,\qquad |E(H_X^*)|<6k^2$};
  \node[mdf note] at (10.65,-3.02)
    {Every minimal preservation forest $F_i\subseteq H_X^*$};
  \node[mdf note] at (10.65,-3.62)
    {$L=S\cup\bigcup_i E(F_i)\subseteq E(H_X^*)$};
  \node[mdf box,minimum width=15cm,minimum height=1.02cm] at (7.75,-4.62) {};
  \node[mdf note] at (7.75,-4.39)
    {$U\cup W\subseteq X',\quad |X'|=12k^2,\quad \ell'=\min\{\ell,|E(H_X)|\}$};
  \node[mdf note] at (7.75,-4.87)
    {$|L|<6k^2=|X'|/2\quad\Longrightarrow\quad\text{the forests extend to spanning trees}$};
  \node[mdf panel] at (.2,-5.58) {(e) Preserve betweenness and compress the distances};
  \node[mdf note] at (7.75,-6.03)
    {$d(x,y)+d(y,z)=d(x,z)\quad\Longleftrightarrow\quad d'(x,y)+d'(y,z)=d'(x,z)$};
  \node[mdf note] at (7.75,-6.55)
    {$d'(x,y)\in\mathbb{Z}_{>0}\ (x\ne y),\qquad d'(x,y)\leq 2^{O(k^4\log k)},\qquad O(k^8\log k)\text{ bits in total}$};
\end{tikzpicture}%
}
\newcommand{\MethodsCompressionFigure}{%
\begin{figure}[t]
  \centering\resizebox{\linewidth}{!}{\MethodsCompressionDrawing}
  \caption{The two compression stages in \Cref{thm:kernel-euclid}.
  In (a), $d(x,y)=2$ and all other distances are $1$. Both $w_1,w_2$ witness
  $\{x,y\}$, but both are isolated in $H_X$; the drawing specifies an
  abstract metric, not Euclidean distances. In (b), strict convexity puts
  any two distinct witnesses on a common shortest $x$--$y$ path. After
  exchanging their names if necessary, its order is $x,w_1,w_2,y$.
  The subpaths give $xw_2,w_1y\in E(H_X)$, hence $w_1,w_2\in U$.
  In (c), the unique witness $z$ contributes its two path edges to $H_X^*$;
  $W$ denotes the set of all special witnesses, including those in $U$.
  In (d), after excluding the easy cases, every minimal preservation
  forest lies in this sparse graph. Retaining $U\cup W$ and padding to
  $12k^2$ points leaves room to extend the forests while reserving the
  strongly preserved pairs $S$. In (e), an integer metric $d'$ preserves
  exactly the same betweenness relations, hence also the instance's answer
  and strict convexity. The distance bound follows from a linear system
  with $\binom{|X'|}{2}$ variables and Hadamard's inequality; for rational
  input distances, the result is a polynomial kernel.}
  \label{fig:methods-compression}
\end{figure}%
}

\newcommand{\MethodsKernelLowerBoundDrawing}{%
\begin{tikzpicture}[x=1cm,y=1cm,font=\fontsize{9}{11}\selectfont,text=mdfInk]
  \path[use as bounding box] (0,-5.35) rectangle (15.5,3.05);
  \node[mdf panel] at (.2,2.75) {(a) One set gives a preserving star};
  \node[mdf panel] at (7.5,2.75) {(b) An exact cover gives a tree};
  \draw[mdfLight] (7.15,-1.62)--(7.15,2.4);
  \begin{scope}[shift={(.7,.65)}]
    \coordinate (r) at (0,1.05);
    \coordinate (s) at (2.25,.7);
    \coordinate (p) at (0,-.8); \coordinate (q) at (2.25,-.8);
    \coordinate (u) at (4.5,-.8); \coordinate (h) at (5.3,1.05);
    \draw[mdf pair] (r) to[bend right=32] (p);
    \draw[mdf pair] (r) to[bend right=12] (q);
    \draw[mdf pair] (r) .. controls (1.7,1.85) and (4.6,.95) .. (u);
    \draw[mdf pair] (p) to[bend right=15] (q);
    \draw[mdf pair] (q) to[bend right=15] (u);
    \draw[mdf pair] (p) to[bend right=24]
      node[below=2pt] {$4=2+2$} (u);
    \foreach \v in {r,p,q,u} \draw[mdf route] (s)--(\v);
    \node[mdf small,fill=white] at (1.13,-.05) {$2$};
    \node[mdf small,fill=white] at (2.25,-.05) {$2$};
    \draw[mdf aux] (s)--node[above=3pt] {$3$} (h);
    \node[mdf witness,label=above:{$s_i$}] at (s) {};
    \node[mdf vertex,label=above:{$r$}] at (r) {};
    \node[mdf vertex,label=below:{$u_p$}] at (p) {};
    \node[mdf vertex,label=below:{$u_q$}] at (q) {};
    \node[mdf vertex,label=below:{$u_s$}] at (u) {};
    \node[mdf vertex,label=above:{$u_h\notin S_i$}] at (h) {};
    \node[mdf small] at (2.7,-2.0)
      {$S_i=\{u_p,u_q,u_s\}$; all solid blue edges have length $2$};
  \end{scope}
  \begin{scope}[shift={(7.75,.65)}]
    \coordinate (r) at (3.1,1.25);
    \coordinate (a) at (1.2,.2); \coordinate (b) at (4.75,.2);
    \coordinate (c) at (6.65,1.25);
    \fill[mdfBlue!5,rounded corners=2pt] (.05,-1.3) rectangle (2.35,-.42);
    \fill[mdfBlue!5,rounded corners=2pt] (3.6,-1.3) rectangle (5.9,-.42);
    \foreach \i in {0,1,2}{
      \draw[mdf pair,line width=.7pt] (r)
        to[bend right=14] (.35+\i*.85,-.7);
      \draw[mdf pair,line width=.7pt] (r)
        to[bend left=14] (3.9+\i*.85,-.7);
    }
    \foreach \base in {.35,3.9}{
      \draw[mdf pair,line width=.7pt] (\base,-.7) to[bend right=18] (\base+.85,-.7);
      \draw[mdf pair,line width=.7pt] (\base+.85,-.7) to[bend right=18] (\base+1.7,-.7);
      \draw[mdf pair,line width=.7pt] (\base,-.7) to[bend right=33] (\base+1.7,-.7);
    }
    \draw[mdf route] (a)--(r)--(b);
    \draw[mdf aux] (r)--(c);
    \foreach \i in {0,1,2}{
      \draw[mdf route] (a)--(.35+\i*.85,-.7);
      \draw[mdf route] (b)--(3.9+\i*.85,-.7);
      \node[mdf vertex] at (.35+\i*.85,-.7) {};
      \node[mdf vertex] at (3.9+\i*.85,-.7) {};
    }
    \node[mdf vertex,label=above:{$r$}] at (r) {};
    \node[mdf witness,label=left:{$s_{i_1}$}] at (a) {};
    \node[mdf witness,label=right:{$s_{i_2}$}] at (b) {};
    \node[mdf witness,label=right:{$s_j$}] at (c) {};
    \node[mdf small] at (4.85,1.7) {attach unused centers};
    \node[mdf note] at (1.2,-1.5) {$S_{i_1}$};
    \node[mdf note] at (4.75,-1.5) {$S_{i_2}$};
    \node[mdf small] at (3.4,-2.0)
      {Selected stars share only $r$; their leaf sets partition $U$};
  \end{scope}
  \node[mdf box,fill=mdfPurple!4,minimum width=15cm,minimum height=.9cm] at (7.75,-2.2) {};
  \node[mdf note] at (7.75,-2.2)
    {$|U|=kt,\qquad \ell=1,\qquad k'=k\binom{t+1}{2},\qquad
      \operatorname{pres}(T)\geq \underbrace{|X|-1}_{\text{tree edges}}+
      \underbrace{k'}_{\text{extra pairs}}$};
  \node[mdf panel] at (.2,-3.02) {(c) Reaching the threshold forces an exact cover};
  \node[mdf note] at (7.75,-3.5)
    {In a minimal preservation forest, $d_j=\deg_F(s_{i_j})$ satisfies
      $2\leq d_j\leq t+1$ and $\sum_{j=1}^p(d_j-1)\leq kt$};
  \node[mdf note] at (7.75,-4.13)
    {$k\binom{t+1}{2}\leq |R|\leq
      \sum_{j=1}^p\binom{d_j}{2}\leq
      \frac{t+1}{2}\sum_{j=1}^p(d_j-1)\leq k\binom{t+1}{2}$};
  \node[mdf note] at (7.75,-4.86)
    {Equality forces $p=k$ and every $d_j=t+1$; acyclicity forces the selected sets to be disjoint};
\end{tikzpicture}%
}
\newcommand{\MethodsKernelLowerBoundFigure}{%
\begin{figure}[t]
  \centering\resizebox{\linewidth}{!}{\MethodsKernelLowerBoundDrawing}
  \caption{The polynomial parameter transformation in \Cref{thm:lb-kernel}.
  In (a), $s_i$ witnesses every pair in $\binom{S_i\cup\{r\}}{2}$:
  the star has edges of length $2$, and its leaves are at mutual distance
  $4$. All six strongly preserved pairs are drawn dashed.
  An edge from $s_i$ to an element outside $S_i$ has length $3$.
  In the complete metric, all distinct set centers have mutual distance
  $2$, and every center has distance $2$ from $r$; all distinct points of
  $U\cup\{r\}$ have mutual distance $4$.
  Panel (b) illustrates two selected sets of size three, with all twelve
  strongly preserved pairs drawn dashed; in general,
  an exact cover by $k$ sets yields $k\binom{t+1}{2}$ extra pairs, and
  unused centers can be attached to $r$ as leaves.
  Conversely, in (c), $R$ is the set of strongly preserved pairs and
  $F$ is a minimal forest preserving $R$. Equality in the counting bound
  forces exactly $k$ full stars. They all contain $r$, so overlapping
  selected sets would create a cycle. Thus their leaf sets form an exact
  cover of $U$. The drawing represents an abstract metric.}
  \label{fig:methods-kernel-lower-bound}
  \label{fig:kernel_lb}
\end{figure}%
}

\newcommand{\MethodsNPHardnessDrawing}{%
\begin{tikzpicture}[x=1cm,y=1cm,font=\fontsize{9}{11}\selectfont,text=mdfInk]
  \path[use as bounding box] (0,-6.5) rectangle (15.5,4.55);
  \node[mdf panel] at (.2,4.2) {(a) Encode sets by vertical four-point paths};
  \node[mdf panel] at (8.65,4.2) {(b) A full column gains three pairs};
  \draw[mdfLight] (8.35,-.6)--(8.35,3.9);
  \begin{scope}[shift={(.8,.2)}]
    \foreach \x in {.35,1.7,3.05}
      \draw[mdf aux,densely dotted] (\x,-.05)--(\x,3.25);
    \foreach \i/\lab in {0/{L},1/{u_{i_1}},2/{u_{i_2}},3/{u_{i_3}}}{
      \draw[mdf aux] (.15,\i)--(4.4+\i,\i);
      \node[mdf note,anchor=east] at (.05,\i) {$\lab$};
      \fill[mdfBlue!5,rounded corners=2pt]
        (3.5+\i,\i-.18) rectangle (4.5+\i,\i+.18);
      \foreach \dx in {0,.16,.8}
        \node[mdf vertex] at (3.6+\i+\dx,\i) {};
      \node[mdf small] at (4.07+\i,\i) {$\cdots$};
    }
    \draw[draw=mdfOrange,line width=1.2pt] (1.7,0)--(1.7,3);
    \foreach \x/\y in {.35/0,1.7/0,3.05/0,.35/1,.35/3,3.05/2,3.05/3}
      \node[mdf vertex] at (\x,\y) {};
    \foreach \y in {1,2,3}
      \node[mdf vertex] at (1.7,\y) {};
    \node[mdf small] at (4,.32) {$R_0$};
    \node[mdf small] at (5,1.32) {$R_{i_1}$};
    \node[mdf small] at (6,2.32) {$R_{i_2}$};
    \node[mdf small] at (7,3.32) {$R_{i_3}$};
    \foreach \x/\lab in {.35/{\widehat S_1},1.7/{\widehat S_j},3.05/{\widehat S_m}}
      \node[mdf note] at (\x,-.4) {$\lab$};
    \foreach \x in {1.02,2.37}\node[mdf small] at (\x,-.4) {$\cdots$};
    \node[mdf small] at (5.6,-.4) {$|R_i|=3m$};
  \end{scope}
  \begin{scope}[shift={(9.05,2.1)}]
    \coordinate (v0) at (0,0); \coordinate (v1) at (1.75,0);
    \coordinate (v2) at (3.5,0); \coordinate (v3) at (5.25,0);
    \draw[mdf pair] (v0) to[bend left=28] (v2);
    \draw[mdf pair] (v1) to[bend left=28] (v3);
    \draw[mdf pair] (v0) to[bend right=24] (v3);
    \draw[draw=mdfOrange,line width=1.2pt] (v0)--(v1)--(v2)--(v3);
    \foreach \v/\lab in {v0/{p_{j,0}},v1/{p_{j,i_1}},v2/{p_{j,i_2}},v3/{p_{j,i_3}}}
      \node[mdf vertex,label=below:{$\lab$}] at (\v) {};
    \node[mdf note] at (2.625,1.22) {$\widehat S_j$: drawn horizontally here};
    \node[mdf note] at (2.625,-1.2) {$\binom{4}{2}-3=3$ strongly preserved pairs};
    \node[mdf small] at (2.625,-1.85) {All three non-edge pairs are drawn dashed};
  \end{scope}
  \draw[mdfLight] (.2,-.85)--(15.3,-.85);
  \node[mdf panel] at (.2,-1.23) {(c) Contract the connected horizontal rows};
  \node[mdf panel] at (8.1,-1.23) {(d) The threshold forces full columns};
  \draw[mdfLight] (7.75,-4.65)--(7.75,-1.55);
  \begin{scope}[shift={(3.9,-2.05)}]
    \coordinate (root) at (0,0);
    \coordinate (a1) at (-1.9,-.6); \coordinate (a2) at (-2.6,-1.35);
    \coordinate (a3) at (-2.6,-2.15);
    \coordinate (b1) at (1.9,-.6); \coordinate (b2) at (2.6,-1.35);
    \coordinate (b3) at (2.6,-2.15);
    \draw[draw=mdfOrange,line width=1.2pt] (root)--(a1)--(a2)--(a3);
    \draw[draw=mdfOrange,line width=1.2pt] (root)--(b1)--(b2)--(b3);
    \node[mdf vertex,label=above:{$P_0$}] at (root) {};
    \foreach \v/\lab in {a1/P_1,a2/P_2,a3/P_3}
      \node[mdf vertex,label=left:{$\lab$}] at (\v) {};
    \foreach \v/\lab in {b1/P_4,b2/P_5,b3/P_6}
      \node[mdf vertex,label=right:{$\lab$}] at (\v) {};
    \node[mdf small] at (-.95,-1.12) {selected\\$\widehat S_{j_1}$};
    \node[mdf small] at (.95,-1.12) {selected\\$\widehat S_{j_2}$};
    \node[mdf small] at (0,-2.5) {$S_{j_1}=\{u_1,u_2,u_3\}$, $S_{j_2}=\{u_4,u_5,u_6\}$};
  \end{scope}
  \node[mdf note] at (11.65,-1.9) {Forcing sets make every horizontal row connected};
  \node[mdf note] at (11.65,-2.45) {After contraction: at most $n=3k$ vertical edges};
  \node[mdf note] at (11.65,-3.15)
    {$\begin{array}{c|cc}
       \text{vertices in a column component}&3&4\\
       \hline
       \text{edges used}&2&3\\
       \text{extra pairs (at most)}&1&3
     \end{array}$};
  \node[mdf note] at (11.65,-4.1) {Gaining $3k$ pairs requires $k$ full four-point paths};
  \node[mdf small] at (11.65,-4.6) {Two selected sets sharing an element would create a cycle};
  \node[mdf box,fill=mdfPurple!4,minimum width=15cm,minimum height=1.2cm] at (7.75,-5.65) {};
  \node[mdf note] at (7.75,-5.38)
    {$q(s)=\binom{s}{2}-s+1,\qquad B=q(4m)+n\,q(3m+3),\qquad k'=B+3k$};
  \node[mdf note] at (7.75,-5.94)
    {$\text{exact cover by }k\text{ sets}\quad\Longleftrightarrow\quad
      \operatorname{pres}(T)\geq |X|-1+k'\qquad(\ell=1)$};
\end{tikzpicture}%
}
\newcommand{\MethodsNPHardnessFigure}{%
\begin{figure}[!t]
  \centering\resizebox{\linewidth}{!}{\MethodsNPHardnessDrawing}
  \caption{The planar NP-hardness reduction in \Cref{thm:nphard}.
  In (a), an element $u_i$ is represented by a horizontal row, and
  $S_j=\{u_{i_1},u_{i_2},u_{i_3}\}$ by the four points of
  $\widehat S_j$, including its point on the spine $L$.
  The forcing sets $R_i$ occupy disjoint ranges of columns; only a few
  points, rows, and set columns are shown. Spacings are schematic:
  the actual construction uses polynomially bounded integer coordinates
  in a general-position grid, with no non-axis-parallel collinear triples.
  In (b), the straight path through a selected column strongly preserves
  its three non-edge pairs. In (c), contracting the horizontal paths
  illustrates an exact cover with $n=6$ and $k=2$: the two selected
  columns connect all seven row vertices without a cycle.
  Conversely, the forcing sets ensure that every row is connected in a
  minimal preservation forest. In (d), after contracting these rows,
  at most $n=3k$ vertical edges remain. A three-vertex column component
  uses two edges but gains at most one pair; a four-vertex component
  uses three edges and gains at most three pairs. Reaching $3k$ pairs
  therefore requires exactly $k$ full straight column paths.
  Acyclicity makes their sets disjoint, hence an exact cover.
  Here $B$ is the maximum horizontal contribution, since the spine has
  $4m$ points and each element row has $3m+3$ points.}
  \label{fig:methods-nphardness}
  \label{fig:npc}
\end{figure}%
}

\allowdisplaybreaks

\definecolor{cblue}{HTML}{1E3A6E}
\definecolor{cteal}{HTML}{0F6E56}
\definecolor{cred}{HTML}{8B2000}
\definecolor{clight}{HTML}{EEF2F7}
\definecolor{cteallight}{HTML}{E6F4F0}
\definecolor{credlight}{HTML}{FAF0EE}
\definecolor{cborder}{HTML}{BBCCE0}
\definecolor{ctborder}{HTML}{9ECFC0}
\definecolor{crborder}{HTML}{DEB8A8}
\definecolor{cgray}{HTML}{555555}
\definecolor{clightgray}{HTML}{F7F7F5}
\definecolor{crulegray}{HTML}{CCCCCC}

\definecolor{accent}{RGB}{33,73,116}
\definecolor{accentline}{RGB}{110,146,180}
\definecolor{accentbg}{RGB}{244,248,252}
\definecolor{softbg}{RGB}{249,250,252}

\hypersetup{
  colorlinks=true,
  linkcolor=accent,
  urlcolor=accent,
  citecolor=accent
}

\newtheoremstyle{bluestyle}{9pt}{6pt}{\itshape}{}{\bfseries\color{cblue}}{.}{.5em}{}
\newtheoremstyle{tealstyle}{9pt}{6pt}{\normalfont}{}{\bfseries\color{cteal}}{.}{.5em}{}
\newtheoremstyle{remarkstyle}{6pt}{4pt}{\normalfont}{}{\itshape\color{cgray}}{.}{.5em}{}

\theoremstyle{bluestyle}
\newtheorem{theorem}{Theorem}[section]
\newtheorem{lemma}[theorem]{Lemma}

\theoremstyle{tealstyle}
\newtheorem{definition}[theorem]{Definition}
\newtheorem{observation}[theorem]{Observation}

\newtheorem{claim}[theorem]{Claim}

\theoremstyle{remarkstyle}

\newcommand{\DCT}{\textsc{Distance Coverage by Trees}\xspace}

\tcbuselibrary{skins,breakable}
\tcbset{
  keybox/.style={enhanced,breakable,
    colback=clight,colframe=cborder,boxrule=0.5pt,arc=4pt,
    left=10pt,right=10pt,top=7pt,bottom=7pt,
    borderline west={3pt}{0pt}{cblue}},
  tealbox/.style={enhanced,breakable,
    colback=cteallight,colframe=ctborder,boxrule=0.5pt,arc=4pt,
    left=10pt,right=10pt,top=7pt,bottom=7pt,
    borderline west={3pt}{0pt}{cteal}},
  redbox/.style={enhanced,breakable,
    colback=credlight,colframe=crborder,boxrule=0.5pt,arc=4pt,
    left=10pt,right=10pt,top=7pt,bottom=7pt,
    borderline west={3pt}{0pt}{cred}},
  graybox/.style={enhanced,breakable,
    colback=clightgray,colframe=crulegray,boxrule=0.5pt,arc=4pt,
    left=10pt,right=10pt,top=7pt,bottom=7pt},
}

\titleformat{\section}
  {\large\bfseries\color{cblue}}{\thesection}{1em}{}
  [\vspace{1pt}{\color{crulegray}\rule{\linewidth}{0.5pt}}]
\titleformat{\subsection}
  {\normalsize\bfseries\color{cgray}}{\thesubsection}{1em}{}
\titleformat{\subsubsection}
  {\normalsize\itshape\color{cgray}}{\thesubsubsection}{1em}{}

\hypersetup{colorlinks,linkcolor=cblue,citecolor=cteal,urlcolor=cblue,
  pdftitle={Distance Coverage by Spanning Trees}}

\newcommand{\R}{\mathbb{R}}
\newcommand{\pres}{\operatorname{pres}}

\newcommand{\Oh}{\mathcal{O}}

\newcommand{\F}{\mathcal{F}}
\newcommand{\T}{\mathcal{T}}
\newcommand{\Pcal}{\mathcal{P}}
\newcommand{\Scal}{\mathcal{S}}

\newcommand{\FPT}{\textup{FPT}\xspace}

\newenvironment{problembox}{\par\medskip\noindent\begin{minipage}{\linewidth}\hrule\smallskip}{\smallskip\hrule\end{minipage}\par\medskip}

\title{\LARGE\bfseries\color{cblue}{Distance Coverage by Spanning Trees}}

\usepackage{authblk}

\author[1]{Akanksha Agrawal}
\author[2]{Fedor V. Fomin\thanks{Supported by the Research Council of Norway under the BWCA project, reference 314528,  and European Research Council (ERC) via grant NewPC, reference 101199930.}}
\author[2]{Petr A. Golovach\thanks{Supported by the Research Council of Norway under the BWCA  (grant no.~314528) and Extreme-Algorithms (grant no~355137) projects.}}
\author[3]{Saket Saurabh}
\author[2]{Kirill Simonov}

\affil[1]{Indian Institute of Technology Madras, Chennai, India}
\affil[2]{University of Bergen, Bergen, Norway}
\affil[3]{Institute of Mathematical Sciences, Chennai, India}

\date{}

\begin{document}

\maketitle

\begin{abstract}
\begin{tcolorbox}[graybox]
Spanning trees provide sparse representations of finite metrics, supporting
efficient routing and distance queries. Classical tree embeddings and tree
covers seek to approximate pairwise distances. We study a complementary
objective: how many pairwise distances can be preserved \emph{exactly} by
a small family of spanning trees? 

For an $n$-point metric and
$\ell<\lceil n/2\rceil$, the Nash-Williams--Tutte theorem guarantees
$\ell$ edge-disjoint spanning trees, which collectively preserve at least
$\ell(n-1)$ distinct pairs. This motivates \DCT: given a metric
$(X,d)$ and integers $\ell<\lceil n/2\rceil$ and $k$, decide whether $\ell$ spanning trees
can collectively preserve at least
$
  \ell(n-1)+k
$
distinct pairs. The parameter $k$ measures the improvement over this
universal combinatorial guarantee.

We prove that \DCT is fixed-parameter tractable parameterized by $k$.
For strictly convex metrics, a class containing all finite Euclidean
metrics and, more generally, metrics induced by point sets in strictly
convex normed spaces, we give a polynomial-time reduction to an equivalent
instance on $O(k^2)$ input points. For rational-valued distances, this
yields a polynomial kernel within the class of strictly convex metrics.
In contrast, on general metrics no polynomial kernel exists unless
$\mathrm{coNP}\subseteq\mathrm{NP/poly}$, even for a single tree and
integer distances bounded by~$4$. Finally, the problem is already
\textup{NP}-hard for a single tree on point sets in the Euclidean plane
with integer coordinates.
\end{tcolorbox}
\end{abstract}

\newpage 
\tableofcontents
\bigskip
{\color{crulegray}\rule{\linewidth}{0.5pt}}

\newpage

\section{Introduction}
\label{sec:intro}
 
 Representing a metric by a small collection of trees is a well-studied
approach to simplifying its distance structure. Each pair of points may
use a different tree: the distance represented by the collection is the
minimum of the distances provided by its individual trees. This viewpoint
underlies the study of tree covers and collective tree spanners, with
applications to routing and distance queries
\cite{GuptaKR05,DraganYL06,DraganY10,BartalFN22}.

More formally, let $(X,d)$ be a finite metric space with $|X|=n$, where $X$ is a set of points and $d\colon X\times X\rightarrow\R_{\geq 0}$ is a distance function (metric).
We view $(X,d)$ as the complete weighted graph $G=K_X$ on the vertex set $X$, where
each edge $xy$ is assigned weight $d(x,y)$.
A spanning tree $T$ of $G$ inherits its edge lengths from $d$ and induces the
tree metric $d_T$. Since
$
d(x,y)\le d_T(x,y)
$
for all $x,y\in X$, a pair $\{x,y\}$ is said to be \emph{preserved} (or covered) by $T$ if
$
d_T(x,y)=d(x,y).
$
Equivalently, the unique $x$--$y$ path in $T$ realizes the original metric
distance exactly.

A family $\mathcal{T}=\{T_1,\ldots,T_\ell\}$ is an
\emph{exact spanning tree cover} if for every $x,y\in X$ the distance $d(x,y)$ is preserved by some $T_i$.
Thus, every pair has a shortest path in at least one tree of the family.
Exact tree covers have been studied for several graph classes~\cite{GuptaKR05,ChangCLMST23}.
A complementary line of work allows distortion, either additive, as in
collective tree spanners~\cite{DraganYL06,DraganY10}, or multiplicative,
and investigates how many trees suffice to approximate every distance
within a prescribed factor~\cite{BartalFN22,ChangCLMST26}.

We study the \emph{maximum-coverage counterpart of exact tree
covers}. When the number of available trees is prescribed, an exact cover
of all pairs may be impossible. This leads to the following question:
\begin{quote}
\emph{Given a budget of $\ell$ spanning trees, how many pairwise distances
can be preserved exactly?}
\end{quote}
Each pair is counted once, regardless of how many trees preserve it.

This objective has a direct routing interpretation. Spanning trees are
natural network backbones: they connect $n$ sites using only $n-1$
edges and support compact routing schemes~\cite{FraigniaudG01,ThorupZ01}.
Suppose that $X$ is a set of participating sites, $d(x,y)$ is the
communication cost between sites, and communication uses a small
collection of tree overlays, where each overlay edge $xy$ has cost
$d(x,y)$. A preserved pair can communicate along the unique path in
one overlay at cost exactly $d(x,y)$. Exact tree covers provide this
guarantee for every pair; under a fixed tree budget, our objective
maximizes the number of pairs receiving this guarantee.

Every spanning tree preserves at least its own $n-1$ edges and therefore
guarantees $n-1$ preserved pairs. In general, however, a tree may preserve
many additional pairs. This naturally leads to the following optimization
questions: \emph{How can one find a spanning tree that preserves the maximum
number of pairs?} More generally, \emph{how can one find $\ell$ spanning trees that
collectively preserve the maximum number of pairs}?

By the classical Nash-Williams theorem~\cite{Nash-Williams64}, the edge set of the complete $n$-vertex graph can be partitioned into at most $\lceil\frac{n}{2}\rceil$ forests. Therefore, the problem is trivial for $\ell\geq\lceil\frac{n}{2}\rceil$. On the other hand, for 
$\ell<\lceil\frac{n}{2}\rceil$, the complete $n$-vertex graph contains $\ell$ edge-disjoint spanning trees by the results of Nash-Williams~\cite{NashWilliams61} and Tutte~\cite{Tutte61}. As each spanning tree preserves at least $n-1$ pairs, we have that there are $\ell$ spanning trees preserving at least $\ell(n-1)$ pairs.  Thus, $\ell(n-1)$ constitutes a natural benchmark motivating the following above-guarantee parameterization.
\begin{problembox}
\textsc{\DCT}

\smallskip

\emph{Input:} A finite metric space $(X,d)$ and integers $\ell$ and $k\geq 0$ with $1\leq \ell<\lceil\frac{n}{2}\rceil$.

\smallskip

\emph{Question:} Do there exist $\ell$ spanning trees which collectively preserve at
least $\ell(n-1)+k$ distinct pairs?
\end{problembox}

The parameter $k$ measures the improvement over coverage guaranteed
solely by selecting distinct tree edges. Our results investigate the
algorithmic complexity of obtaining this improvement and the role of
geometric structure in reducing the instance size.

\subparagraph{Our results.}
We begin by establishing the computational hardness of \DCT. The problem
remains \textup{NP}-hard even in the highly restricted setting of a single
tree ($\ell=1$) and when the input metric is induced by a finite set of
points in the Euclidean plane.
\begin{restatable}{theorem}{NPhard}
\label{thm:nphard}
\DCT is \textup{NP}-hard even for $\ell=1$ when the input metric is the
Euclidean metric induced by a finite point set in $\R^2$ with integer coordinates.
\end{restatable}

We remark that this result holds for points with integer coordinates whose absolute values are upper-bounded by a polynomial in $n$.
\textup{NP}-hardness immediately raises the question of fixed-parameter
tractability. Our main algorithmic result establishes that the problem is \textup{FPT} parameterized by $k$ only.

\begin{restatable}{theorem}{FPTalg}
\label{thm:fpt}
\DCT can be solved in
$2^{k^{O(1)}}\cdot n^{O(1)}$ time.
\end{restatable}

As is common in the computational geometry area, we assume the real RAM computational model in this theorem as our algorithm performs additions and comparisons of the values of $d(x,y)$ for the input metric space. More precisely, the algorithm checks the betweenness for triples $x,y,z\in X$, that is, we check  whether $d(x,y)+d(y,z)=d(x,z)$.

Following the standard workflow of parameterized complexity, the next natural
question after establishing \textup{FPT} is whether the problem admits a
polynomial kernel in~$k$. 

\begin{restatable}{theorem}{NoKernel}
 \label{thm:lb-kernel}
\DCT on general metrics admits no polynomial kernel,
even when restricted to $\ell=1$ and integer-valued metric spaces with the distances at most $4$,
unless $\mathrm{coNP}\subseteq\mathrm{NP}/\mathrm{poly}$. 
\end{restatable}

\begin{figure}[t]
  \centering
  \resizebox{\linewidth}{!}{%
\begin{tikzpicture}[x=1cm,y=1cm,font=\fontsize{9}{11}\selectfont,text=mdfInk]
  \node[mdf panel] at (0,2.55) {(a) Euclidean: strictly convex};
  \node[mdf panel] at (6.6,2.55) {(b) $\ell_1$: not strictly convex};
  \draw[mdfLight] (6.3,-1.3)--(6.3,2.3);
  \begin{scope}[shift={(.4,.35)}]
    \coordinate (u) at (0,0);   \coordinate (x) at (1.6,0);
    \coordinate (y) at (3.2,0); \coordinate (v) at (5.2,0);
    \draw[mdf route] (u)--(x)--(y)--(v);
    \draw[mdf pair] (u) to[bend left=32] node[above=2pt] {$d(u,v)=d(u,x)+d(x,y)+d(y,v)$} (v);
    \node[mdf vertex,label=below:$u$] at (u) {};
    \node[mdf witness,label=below:$x$] at (x) {};
    \node[mdf witness,label=below:$y$] at (y) {};
    \node[mdf vertex,label=below:$v$] at (v) {};
    \node[mdf note] at (2.6,-1.05) {$uxyv$ is a shortest $u$--$v$-path in $K_X$};
  \end{scope}
  \begin{scope}[shift={(8.3,-.65)},scale=1.25]
    \draw[mdfLight,line width=.4pt] (-.4,-.15) grid[step=.5] (2.4,2.15);
    \coordinate (u) at (0,0); \coordinate (v) at (2,1.6);
    \coordinate (x) at (2,0); \coordinate (y) at (0,1.6);
    \draw[mdf route] (u)--(x)--(v);
    \draw[mdf route] (u)--(y)--(v);
    \draw[mdf pair] (x)--(y);
    \node[mdf vertex,label=below left:$u$] at (u) {};
    \node[mdf vertex,label=above right:$v$] at (v) {};
    \node[mdf witness,label=below right:$x$] at (x) {};
    \node[mdf witness,label=above left:$y$] at (y) {};
  \end{scope}
  \node[mdf note,anchor=west] at (11.4,1.35) {$d(u,v)=d(u,x)+d(x,v)$};
  \node[mdf note,anchor=west] at (11.4,.95) {$\phantom{d(u,v)}=d(u,y)+d(y,v)$};
  \node[mdf note,anchor=west] at (11.4,.35) {but $d(x,y)=d(u,v)$,};
  \node[mdf note,anchor=west] at (11.4,-.05) {so neither $uxyv$ nor $uyxv$};
  \node[mdf note,anchor=west] at (11.4,-.45) {is a shortest $u$--$v$-path};
\end{tikzpicture}}
  \caption{Strict convexity}
  \label{fig:strict-convexity}
\end{figure}
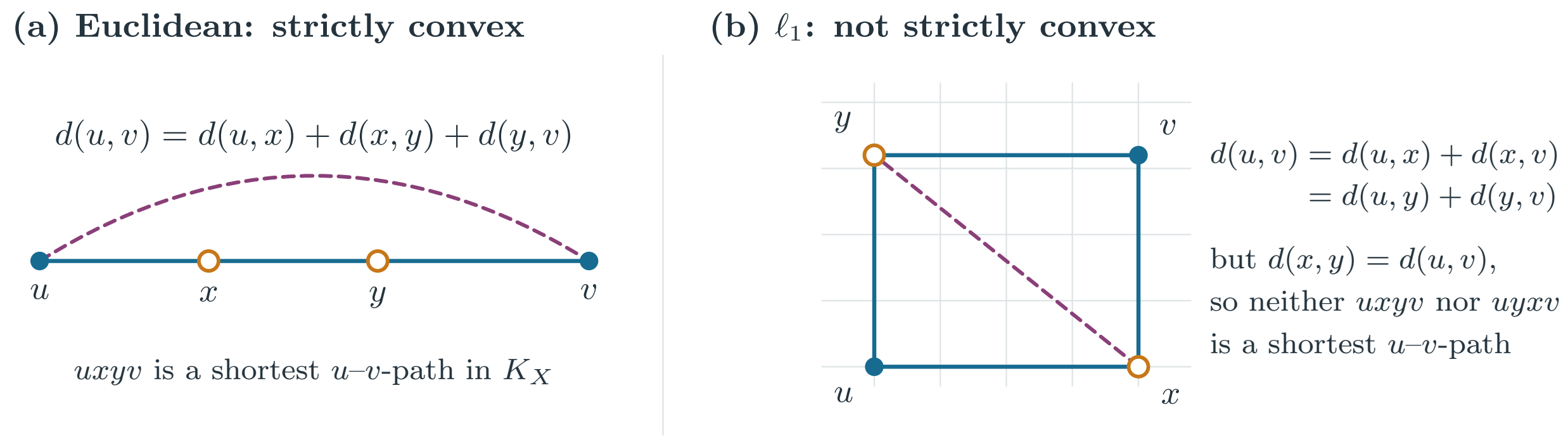

Euclidean metrics are of particular importance in this context. Not only are
they the most natural and well-studied class of metric spaces, but they also
carry the geometric rigidity that makes polynomial kernelization possible. In fact, we show a more general result. A finite metric space $(X,d)$ is said to be \emph{strictly convex} if for any four pairwise distinct points $u,v,x,y\in X$ such that $d(u,v)=d(u,x)+d(x,v)=d(u,y)+d(y,v)$, either $uxyv$ or $uyxv$ is a shortest $u$--$v$-path in $G=K_X$.  See \Cref{fig:strict-convexity}. Note that this definition is a bit non-standard as we deal with finite metric spaces. However, if $X$ is a finite set of points in a strictly convex normed vector space, we have the required property. In particular, if $X\subseteq \R^{p}$ and $d(x,y)=\|x-y\|_2$ for $x,y\in X$, then $(X,d)$ is strictly convex.

\begin{restatable}{theorem}{Compression}
\label{thm:kernel-euclid}
There is a polynomial-time algorithm that, given an instance
\((X,d,\ell,k)\) of \DCT where $(X,d)$ is a finite strictly convex metric space, 
either 
solves the problem, or 
 outputs an equivalent
instance \((X',d',\ell',k)\), where $X'\subseteq X$ is of size  
$\Oh(k^2)$ and $d'$ is an integer-valued strictly convex metric such that $d'(x,y)=2^{\Oh(k^4\log k)}$ for $x,y\in X'$. Furthermore, if the values of $d(x,y)$ are restricted to be rational for $x,y\in X$,  
\DCT admits a polynomial kernel of bit-size $\Oh(k^8\log k)$.
\end{restatable}

We remark that while the algorithm always outputs an instance in a strictly convex metric space, in the Euclidean case it may output an instance in a non-Euclidean metric space. Furthermore, note that we claim the existence of a polynomial kernel only for a rational-valued metric $d$. This is due to the fact that the algorithm performs operations (additions and comparisons) over the values of $d(x,y)$ which may be real in the general case.
\subsection{Related work}

\subparagraph{Representing metrics by trees.}

Classical tree spanners seek a spanning tree that approximates all
pairwise distances~\cite{CaiCorneil95}. A single tree may incur linear
distortion in the worst case~\cite{RabinowichR98}, motivating
representations by distributions over trees. Bartal obtained
probabilistic embeddings of finite metrics into tree metrics with
distortion $O(\log^2 n)$~\cite{Bartal96}, and Fakcharoenphol, Rao,
and Talwar achieved the optimal $O(\log n)$
bound~\cite{FakcharoenpolRT03}.

A related classical question asks whether a finite metric can be
represented exactly by a weighted tree. In this model, the input
points are represented by vertices of a tree that may contain
additional, unlabeled Steiner vertices, and the edge lengths are
chosen as part of the representation. Buneman's four-point
characterization gives a polynomial-time recognition criterion
for such tree metrics~\cite{Buneman74}. This model differs from
ours, where the vertex set is exactly $X$ and every tree edge
$xy$ has the prescribed length $d(x,y)$. Gupta showed that Steiner
vertices can be removed from a weighted tree representation at
the cost of a constant-factor increase in
distortion~\cite{Gupta01}; this guarantee does not preserve
distances exactly.

Low-stretch spanning trees provide another approach. Elkin, Emek,
Spielman, and Teng obtained spanning trees with average edge
stretch $O(\log^2 n\log\log n)$~\cite{ElkinEST05}, and Abraham,
Bartal, and Neiman subsequently obtained nearly optimal
bounds~\cite{AbrahamBN08}. Tree representations also underlie
network-design algorithms, including R\"acke's hierarchical
decompositions and logarithmic congestion
guarantees~\cite{Racke02,Racke08}.

\subparagraph{Exact tree covers and partial distance coverage.}

Exact preservation by a collection of trees has been studied
through \emph{exact tree covers}: every pair must have its
distance preserved in at least one tree. Gupta, Kumar, and
Rastogi introduced tree-cover constructions for routing and
obtained, in particular, exact spanning-tree covers of planar
graphs using $O(\sqrt n)$ trees~\cite{GuptaKR05}.
For unweighted $K_r$-minor-free graphs of diameter~$\Delta$, Chang et
al.\ obtained exact spanning tree covers with $2^{O_r(\Delta)}$
trees~\cite{ChangCLMST23}.
Collective tree spanners study the more general requirement that
every pair be approximated within a specified multiplicative or
additive error in one of a small family of spanning
trees~\cite{DraganYL06,DraganYC06,DraganY10,DraganK14}.
Our question concerns the complementary tradeoff: with the
number of trees fixed, how many pairs can be preserved exactly?

The closest predecessor to our single-tree objective is
\emph{Single Tree Effectiveness}, introduced by Yallouz et
al.~\cite{YallouzTKBGO17}. Given a positively weighted graph
$G$ and a set of demand pairs $D$, they seek a spanning tree
$T\subseteq G$ maximizing the number of pairs
$\{u,v\}\in D$ satisfying $d_T(u,v)=d_G(u,v)$.
They   prove
NP-hardness for this problem. 
Taking $G=K_X$ with edge lengths given by $d$, and taking
$D=\binom{X}{2}$, yields our single-tree optimization objective.
Their hardness result for graph subtrees does not, by itself,
establish hardness for complete metric graphs or Euclidean
point sets.  

\subparagraph{Approximate tree covers in geometric and graph metrics.}

For fixed dimension $d$, the Dumbbell Theorem of Arya, Das,
Mount, Salowe, and Smid gives a $(1+\varepsilon)$-stretch tree
cover of any finite point set in $\mathbb{R}^d$ using
$O_d(\varepsilon^{-d}\log(1/\varepsilon))$
trees~\cite{AryaDMSS95}. Their construction permits geometric
Steiner points. Bartal, Fandina, and Neiman extended
near-isometric tree covers to doubling
metrics~\cite{BartalFN22}. Chang et al.~improved the Euclidean
bound to
$O_d(\varepsilon^{-(d-1)}\log(1/\varepsilon))$ trees without
Steiner points, which is optimal up to a logarithmic
factor~\cite{ChangCLMST26}. Bikeev, Kupavskii, and
Turevskii~\cite{BikeevKT26} and, independently, Le,
Milenkovi\'c, Solomon, and Zhang~\cite{LeMSZ26} showed that
two trees suffice for constant stretch in the Euclidean plane.

For graph metrics, Chang et al.~constructed constant-size
$(1+\varepsilon)$-stretch tree covers of planar metrics for
every fixed $\varepsilon>0$~\cite{ChangCLMST23}.
Chang et al.~\cite{ChangCLST25} subsequently obtained constant-size
$(1+\varepsilon)$-stretch covers using actual spanning trees
of doubling graphs, together with light-tree-cover and routing
applications. La et al.~obtained
$(1+\varepsilon)$-stretch covers of doubling metrics with
nearly optimal size and constant maximum degree in each
tree~\cite{LaEtAl26}. 
A recent preprint of Le, Pham, Than, and Tran shows that every
fixed-minor-free graph family admits a constant-stretch tree cover with
three trees~\cite{LePTT26}. This matches a lower bound of Chen, Tan, and
Xu, who showed that toroidal grids require three trees for constant
stretch; they also obtained polynomial lower bounds on the distortion of
covers with any fixed number of trees~\cite{ChenTX25}, subsequently
strengthened in a preprint of Huang~\cite{Huang26}. The trees in these
covers are dominating edge-weighted trees rather than spanning trees
of~$K_X$ with inherited edge lengths, and the results bound the stretch
of every pair.  

\subparagraph{Tree-metric fitting with minimum disagreements.}

Another closely related problem is $\ell_0$ tree-metric
fitting, where the objective is to minimize the number of
pairwise distances on which an output tree metric disagrees
with the input. Kipouridis gave a constant-factor
approximation and proved APX-hardness for this
problem~\cite{Kipouridis23}. The fitting model allows
additional vertices and freely chosen edge lengths, and the
formulation in~\cite{Kipouridis23} also permits zero distances
between distinct input labels. These freedoms distinguish it
from our spanning-tree model with inherited edge lengths.

A recent preprint of Bandyapadhyay et al.~gives a
$q^{\Oh(q)}n^{\Oh(1)}$-time algorithm for \emph{Tree Metric
Violation Distance}, parameterized by the number $q$ of
distances that may be changed to obtain a   tree
metric~\cite{BandyapadhyayEtAl26}.
Besides the different feasible representations, this
parameter measures closeness to preserving all pairs.
 
\subparagraph{Distance preservers and spanners.}

Exact preservation has also been studied through sparse
subgraphs. Early work of Bollob\'as, Coppersmith, and Elkin
considers distance preservers for pairs at sufficiently large
distance~\cite{BollobasCoppersmithElkin1998}. For a prescribed
pair set $P$ with $p=|P|$, Coppersmith and Elkin proved that
an undirected weighted graph admits a pairwise distance
preserver with $O(n+p\sqrt n)$
edges~\cite{CoppersmithE06}. Subsequent work studies the
tradeoffs between the number of prescribed pairs and the
size of a preserver~\cite{Bodwin2016,BodwinW16}, including
fault-tolerant variants~\cite{BoldwingGMW17}.

For unweighted graphs, additive spanners instead allow $d_H(u,v) \le d_G(u,v) + \beta$ for every
pair. Known constructions achieve additive error $+2$ with $O(n^{3/2})$
edges~\cite{AingworthCIM99}, $+4$ with $\tilde{O}(n^{7/5})$
edges~\cite{Chechik13}, and $+6$ with $O(n^{4/3})$
edges~\cite{BaswanaKMP10}; Woodruff~\cite{Woodruff10} gave a substantially
faster, nearly quadratic-time construction of $+6$ spanners.

\subparagraph{Metric betweenness.}
Our preservation graph is closely related to the classical betweenness
structure of a finite metric. A point~$z$ lies between~$x$ and~$y$ when
$d(x,y) = d(x,z) + d(z,y)$. Betweenness underlies the notion of lines in
finite metric spaces introduced by Chen and
Chv{\'a}tal~\cite{ChenChvatal08} in their proposed metric generalization
of the De Bruijn--Erd{\H{o}}s theorem. The underlying graph of the
betweenness structure joins two distinct points precisely when no third
point lies between them; see Szab{\'o}~\cite{Szabo23} and the references
therein. Thus our preservation graph~$H_X$ is the complement of this
underlying graph.

\subparagraph{Parameterized complexity and kernelization.}

Parameterization above guaranteed values was introduced by
Mahajan and Raman~\cite{MahajanR99}.
In our setting, the Nash-Williams--Tutte
theorem~\cite{NashWilliams61,Tutte61} supplies the guarantee
$\ell(n-1)$, and $k$ measures the additional exact coverage.
Our kernelization lower bounds use polynomial-parameter
transformations and known lower bounds for packing
problems~\cite{DellM12}, within the general framework of
kernelization lower
bounds~\cite{DellvM14,HermelinW14,BodlaenderJK14}.

\subsection{Technical overview}\label{sec:overview}

Our algorithms combine tools of parameterized complexity with new
combinatorial and geometric properties of specific objects. What ties these
together, and carries the actual content, is the {preservation graph} $H_X$. The combinatorial structure of $H_X$ drives the \FPT algorithms. For
kernelization on Euclidean, and more generally strictly convex, metrics, geometry
comes into the play.

\subparagraph{Preservation graph and witnesses: \Cref{fig:methods-preservation}.}
\MethodsPreservationFigure
The technical core of the paper is a single structural object, the
\emph{preservation graph} $H_X$.
To preserve more than $\ell(n-1)$ distances, we have to preserve pairs $\{x,y\}$ such that $xy$ is a non-edge in any tree of a solution to the considered instance of \DCT. In this case, we say that $\{x,y\}$ is \emph{strongly preserved}. Notice that $\{x,y\}$ may be strongly preserved if and only if the weighted graph $G=K_X$ has a shortest $x$--$y$-path with at least two edges. Equivalently, given a metric space $(X,d)$, $\{x,y\}$ may be strongly preserved if and only if there is  $z\in X$ distinct from $x$ and $y$, called a \emph{witness}, such that $d(x,y)=d(x,z)+d(z,y)$. We define the \emph{preservation} graph $H_X$ with the vertex set $X$ such that two distinct vertices $x$ and $y$ are adjacent if and only if there is a witness $z\neq x,y$ with $d(x,y)=d(x,z)+d(z,y)$. In words, the edges of $H_X$ are pairs that can be potentially strongly preserved in a solution. Our algorithmic results in~\Cref{thm:fpt} and \Cref{thm:kernel-euclid} are based on structural properties of the preservation graph. 

\subparagraph{The win condition:  \Cref{fig:methods-win}.
} \MethodsWinFigure

In our main structural~\Cref{lem:bigH}, we show that if $|X|> 6k$ and $H_X$ has either a matching of size at least $2k$ or a vertex of degree at least $k$, then $G=K_X$ has a forest $F$ with at most $2k$ edges that strongly preserves at least $k$ pairs. Then we apply
\Cref{lem:kn-minus-s}, which is a corollary of the classical Nash-Williams–-Tutte theorem~\cite{NashWilliams61,Tutte61} (see~\Cref{thm:NWT}), to argue that $G$ contains $\ell$ edge-disjoint trees $T_1,\ldots,T_\ell$ such that $F$ is a subforest of $T_1$, and the trees strongly preserve the $k$ pairs strongly preserved by $F$.  As the trees are edge-disjoint, they preserve $\ell(n-1)$ pairs corresponding to their edges. Thus, the trees $T_1,\ldots,T_\ell$ preserve at least $k+\ell(n-1)$ pairs. 
Consequently, once $H_X$ has more than
$(2k-1)k\leq 2k^2$ non-isolated vertices or more than $(2k-1)k\leq 2k^2$ edges, the instance is a yes-instance of \DCT.

\subparagraph{The FPT algorithm: \Cref{fig:methods-witness-types}.
}   \MethodsWitnessTypesFigure
We exploit~\Cref{lem:bigH} in our \FPT algorithm for \DCT parameterized by $k$ in \Cref{thm:fpt}. If the input metric space $(X,d)$ has bounded size, namely if $|X|\leq 14k^4$, then we solve the problem by brute force. Thus, we may assume that $|X|> 14k^4$. Then we construct $H_X$ and immediately decide whether we have a yes-instance because of~\Cref{lem:bigH}. If this is not the case, then $H_X$ is of bounded size. Then we can guess all strongly preserved pairs in a hypothetical solution and all edges of the trees of the solution with both endpoints in the set of non-isolated vertices of $H_X$. Next, we have to deal with the vertices and edges of the trees in the solution outside $G[U]$. The crucial observation that allows us to select them is that we need a vertex $w\notin U$ to strongly preserve a pair $\{x,y\}$ only if $w$ is a witness for this pair and, moreover, $w$ is a middle vertex of the path $xwy$ in one of the solution trees. To see this, note that if $\{x,y\}$ is strongly preserved in one of the trees containing a shortest $x$--$y$-path $P=xv_1\dots v_sy$ with at least two internal vertices, then the internal vertices are in $U$. 
So, in the following step of the algorithm, we guess the witnesses outside $U$ that are used to ensure the strong preservation of certain pairs. The difficult part here is that a vertex can serve as a witness for several pairs, and trees in a solution may have common edges. Nevertheless, we can bound the possible choice of these witnesses. First, the total number of the witnesses is upper bounded by the number of strongly preserved pairs, which is at most $|E(H_X)|\leq 2k^2$. Second, if we have a set of equivalent potential witnesses, that is, points of $X\setminus U$ that are witnesses for exactly the same pairs,
of size exceeding $2k^2$, then we can restrict our choice to an arbitrary subset of size $2k^2$. Thus, we can guess inclusion-minimal forests $F_1,\ldots,F_p$ for some
$p\leq \min\{|E(H_X)|,\ell\}$ that ensure the strong preservation of all pairs, which we guess to be strongly preserved in the solution. Furthermore, $|\bigcup_{i=1}^pE(F_i)|<\frac{n}{2}$. This allows us to apply  \Cref{lem:kn-minus-s}, our corollary of the Nash-Williams--Tutte theorem, to extend these forests to trees in such a way that the intersections of the edge sets are constrained to the forests. This way we obtain a solution in the case when a correct guess of $F_1,\ldots,F_p$ exists. Otherwise, if we are not able to find such forests, we conclude that we have a no-instance of \DCT.

\subparagraph{Compression: \Cref{fig:methods-compression}.} \MethodsCompressionFigure 
The bottleneck of our \FPT algorithm is the choice of witnesses for the strongly preserved pairs in a potential solution outside the set $U$ of non-isolated vertices of $H_X$. 
This is the place where geometry enters in the case of strictly convex metric spaces.
If $(X,d)$ is a strictly convex metric space and $w_1$ and $w_2$ are distinct witnesses for a pair $\{x,y\}$, then $w_1$ and $w_2$ are middle internal vertices of a shortest $x$--$y$-path $P$ in the weighted graph $G=K_X$. Assuming that $P=xw_1w_2y$, we obtain that $xw_2,w_1y\in E(H_X)$, that is, $w_1,w_2\in U$. This implies that for each pair $\{x,y\}$ corresponding to $xy\in E(H_X)$, we have at most one potential witness outside $U$. This observation gives a compression with $\Oh(k^4)$ points. However, more accurate arguments allow us to improve the result to $\Oh(k^2)$. 

For this, we introduce the \emph{enhanced preservation graph} $H_X^*$ with the set of vertices $X$. We say that $z\in X\setminus\{x,y\}$ is a \emph{special witness} for a pair $\{x,y\}\in \binom{X}{2}$ if $z$ is a \emph{unique} witness for the pair. Then the edge set of $H_X^*$ is
$E(H_X)\cup\bigcup_{z\in X}\{xz,zy\mid z\text{ is a special witness for }\{x,y\}\in\binom{X}{2}\}$.

The crucial property proved in~\Cref{lem:enhanced} is that, in the case of a strictly convex metric, the choice of minimal forests strongly preserving pairs in a potential solution is restricted to $H_X^*$. 

Thus, in our compression algorithm in~\Cref{thm:kernel-euclid}, we argue as follows. Given an instance $(X,d,\ell,k)$ of \DCT, we may assume that $n>12k^2$; otherwise, we already have the required bound for the size of $X$. Further, if $H_X$ has either at least $2k^2$ non-isolated vertices or at least $2k^2$ edges, then we report a yes-instance. Thus, we can assume that neither is the case. Then we consider the enhanced preservation graph $H_X^*$ which has at most $|U|+|E(H_X)|<4k^2$ non-isolated vertices and $3|E(H_X)|<6k^2$ edges. By complementing the set of non-isolated vertices of $H_X^*$ by additional points to ensure the extendability of forests strongly preserving pairs in a potential solution to spanning trees, we obtain a compressed instance $(X',d,\ell',k)$ of \DCT with at most $12k^2$ points.

In the next step, we replace the original distances by integer distances of bounded size. For this, we use the fact that replacing $d$ by a metric $d'$ maintaining betweenness results in an equivalent instance of \DCT. Thus, we choose $d'$ satisfying the property that for any distinct $x,y,z\in X'$, $d'(x,y)+d'(y,z)=d'(x,z)$ if and only if $d(x,y)+d(y,z)=d(x,z)$. Equivalently, we select $d'$ in such a way that (i)~$d'(x,x)=0$ for every $x\in X'$, (ii)~$d'(x,y)>0$ for all pairs of distinct $x,y\in X'$, and (iii)~for all triples $x,y,z$ of distinct elements of $X'$,
\[
\begin{cases}
d'(x,y)+d'(y,z)=d'(x,z)&\text{if }d(x,y)+d(y,z)=d(x,z),\\
d'(x,y)+d'(y,z)>d'(x,z)&\text{if }d(x,y)+d(y,z)>d(x,z).
\end{cases}
\]
By scaling, we obtain that there is $d'$ satisfying (i) such that (ii$^*$)~$d'(x,y)\geq 1$ 
for all pairs of distinct $x,y\in X'$, and (iii$^*$)~for all triples $x,y,z$ of distinct elements of $X'$,
\[
\begin{cases}
d'(x,y)+d'(y,z)=d'(x,z)&\text{if }d(x,y)+d(y,z)=d(x,z),\\
d'(x,y)+d'(y,z)-d'(x,z)\geq 1&\text{if }d(x,y)+d(y,z)>d(x,z).
\end{cases}
\]
Solving the system of linear inequalities defined by (ii$^*$) and (iii$^*$), where the values $d'(x,y)$ for pairs $\{x,y\}\in\binom{X'}{2}$ are considered to be variables, we conclude that the values of $d'(x,y)$ can be chosen to be integers upper bounded by $p^{p/2}$ for $p=\binom{|X'|}{2}$. The crucial observations here are that the coefficients of the system are from $\{-1,0,1\}$, the number of variables is $p$, and the absolute value of the determinant of a $t\times t$-matrix whose elements are from $\{-1,0,1\}$ is at most $t^{t/2}$ by Hadamard's inequality.  This way, 
we obtain an integer-valued $d'$ such that $d'(x,y)=2^{\Oh(k^4\log k)}$ for $x,y\in X'$.

\subparagraph{Kernelization lower bound: \Cref{fig:methods-kernel-lower-bound}.}
\MethodsKernelLowerBoundFigure
The restricted witness structure is not a convenience but a feature of strict
convexity that general metrics lack, and~\Cref{thm:lb-kernel} makes this precise.
We show the kernelization lower bound by presenting a polynomial parameter transformation from the \textsc{Exact Cover} problem, which asks, given a \emph{universe} set $U$ and a family $\Scal=\{S_1,\ldots,S_m\}$ of subsets of $U$, whether there is a subfamily $\Scal'\subseteq\Scal$ of size $k$ \emph{exactly covering} $U$, that is, each element of $U$ is in a unique set of $\Scal'$. We consider the special variant of the problem where all sets of $\Scal$ have the same size $t$. By the result of Dell and Marx in~\cite{DellM12}, this variant of \textsc{Exact Cover} does not admit a polynomial kernel when parameterized by both $k$ and $t$,
unless $\mathrm{coNP}\subseteq\mathrm{NP}/\mathrm{poly}$. The reduction idea is based on the following observation. Assume that $s,u_1,\ldots,u_t\in X$ for a metric space $(X,d)$, and suppose that $d(s,u_1)=\dots=d(s,u_t)=\frac{1}{2}d(u_i,u_j)$ for distinct $i,j\in\{1,\ldots,t\}$. Then $s$ is a witness for all pairs $\{u_i,u_j\}$ for distinct $i,j\in\{1,\ldots,t\}$, and the star with the center $s$ and the leaves $u_1,\ldots,u_t$ strongly preserves the pairs.  We use this to encode an instance of \textsc{Exact Cover}: the universe $U$ is encoded as a set of $n$ points at distance $4$ from each other, each set $S_i\in \Scal$ is encoded by a star connecting its center in a point $s_i$ with $d(s_i,u_j)=2$ for $u_j\in S_i$. Then by adding one additional root point $r$ and complementing the values of distances, we construct an instance of \DCT with $\ell=1$ equivalent to the considered instance of \textsc{Exact Cover}. In this reduction, a single vertex $s_i$ witnesses a whole set
of pairs $\binom{S_i\cup\{r\}}{2}$ at once. 
Notice that the reduction exploits the freedom to select a witness for the same pair of points among multiple candidate points at the same distances to the vertices of the pair, and this is precisely what the strict convexity forbids.

\subparagraph{NP-hardness in the plane: \Cref{fig:methods-nphardness}.}
\MethodsNPHardnessFigure
To show that \DCT is \textup{NP}-hard for $\ell=1$ when $X$ is a set of points in $\R^2$ with integer coordinates and $d(x,y)=\|x-y\|_2$ for $x,y\in X$ in~\Cref{thm:nphard}, we again reduce from
\textsc{Exact Cover} with the same-size sets. This problem is \textup{NP}-complete even when each set is of size three, and each element of the universe occurs in exactly three sets~\cite{Gonzalez85}. While we used stars in~\Cref{thm:lb-kernel}, here we rely on paths to encode the considered instance of \textsc{Exact Cover}—note that if $P=v_0\dots v_s$ is a shortest path in the weighted graph $G=K_X$, then $P$ strongly preserves each pair $\{v_i,v_j\}$ with $i\leq j-2$. We encode the elements $u_i\in U$ by parallel horizontal lines, and add a ``spine'' line $L$.
Then we add sufficiently big sets of points $R_i $ on each of these lines that enforce that all the points on the same line should be in a shortest path in a solution tree. The sets of $\Scal$ are encoded by parallel vertical lines—we put a point on the intersection of the vertical line encoding $S_i$ and $L$, and we put points on the intersections with horizontal lines encoding the elements of the universe if the elements are in $S_i$. Then we argue that a solution tree $T$ should consist of the inclusion-maximal shortest paths in $G$ corresponding to the horizontal lines and the inclusion-maximal shortest paths in $G$ corresponding to the vertical lines, corresponding to an exact cover of the universe.   

\subsection{Organization of the paper}
In \Cref{sec:prelim}, we introduce notation and make some basic observations. We continue in~\Cref{sec:struct} by proving structural results exploited in the paper. 
In \Cref{sec:FPT}, we show our main algorithmic result~\Cref{thm:fpt}. We discuss kernelization for \DCT and prove \Cref{thm:kernel-euclid} and \Cref{thm:lb-kernel} in~\Cref{sec:kernelization}. In~\Cref{sec:NPc} we prove~\Cref{thm:nphard}. We conclude in~\Cref{sec:conclusion} by stating some open problems. 

\section{Preliminaries}\label{sec:prelim}
In this section, we introduce the notation used throughout the paper and state some auxiliary claims. 

\subparagraph{Parameterized complexity and kernelization.} We refer to the monograph~\cite{CyganFKLMPPS15} for an introduction to the area.  Here, we just briefly recall basic notions. A \emph{parameterized problem} is a language $L\subseteq\Sigma^*\times\mathbb{N}$, where $\Sigma^*$ is the set of strings over a finite alphabet $\Sigma$, and its instance is a pair $(x,k)$, where $x\in\Sigma^*$ and $k\in\mathbb{N}$ is a \emph{parameter}. A parameterized problem $L$ is  \emph{fixed-parameter tractable} (\FPT) if there exists an algorithm solving $L$ in $f(k)\cdot|x|^{\Oh(1)}$ time for a computable function $f$. 
A \emph{kernelization} (or \emph{kernel}) for a parameterized problem $L$ is a polynomial-time algorithm that for any instance $(x,k)$, outputs an instance $(x',k')$ of $L$ such that (i)~$(x,k)\in L$ if and only if $(x',k')\in L$ and (ii)~$|x'|+k'\leq f(k)$ for a computable function $f$. A kernel is \emph{polynomial} if $f$ is polynomial. It is known that a decidable parameterized problem is in \FPT if and only if it admits a kernel, but it is unlikely that all fixed-parameter tractable parameterized problems admit polynomial kernels. It is standard to exclude the existence of a polynomial kernel under the assumption that $\mathrm{coNP}\not\subseteq\mathrm{NP}/\mathrm{poly}$ by providing a \emph{polynomial parameter transformation}, that is, a polynomial reduction from a parameterized problem   
for which a kernelization lower bound is already established such that the obtained value of the parameter is polynomial in the value of the parameter of the reduced problem.

\subparagraph{Sets.} For a finite set $X$, we use $\binom{X}{2}$ to denote the set of unordered pairs $\{x,y\}$ of distinct elements of $X$.

\subparagraph{Graphs.}
We consider only simple finite undirected graphs and use the standard graph-theoretic notation. We refer to~\cite{Dieste12} for notions not defined here. Given a graph $G$, we denote by $V(G)$ and $E(G)$ its vertex and edge sets, respectively. Throughout the paper, we write $n$ and $m$ for the number of vertices and edges of a graph if this does not create confusion. 
A graph $H$ such that $V(H)\subseteq V(G)$ and $E(H)\subseteq E(G)$ is a \emph{subgraph} of $G$. 
A subgraph is \emph{spanning} if $V(H)=V(G)$.
For a set $U\subseteq V(G)$, we denote by $G[U]$ the \emph{subgraph of $G$ induced by $U$}, that is, the subgraph with the vertex set $U$ such that any two distinct vertices $u,v\in U$ are adjacent in $G[U]$ if and only if they are adjacent in $G$. 
For an edge set $S\subseteq E(G)$, we define $G[S]$ to be the graph whose vertices are the endpoints of the edges of $S$, and the edge set is $S$.
We write $G-U$ for the subgraph of $G$ induced by $V(G)\setminus U$. Similarly, for $S\subseteq E(G)$, $G-S$ is used to denote the subgraph of $G$ obtained by deleting the edges of $S$. A \emph{forest} is a graph without cycles, and a \emph{tree} is a connected graph without cycles.  
A \emph{matching} is a set of edges with pairwise distinct endpoints. 
 We write $P=v_0\dots v_s$ for a \emph{path} with the vertices $v_0,\dots,v_s$ where $v_{i-1}v_i\in E(G)$ for $i\in\{1,\ldots,s\}$; the vertices $v_0$ and $v_s$ are the \emph{endpoints} of $P$, the other vertices are \emph{internal}, and we say that $P$ is a \emph{$v_0$--$v_s$-path}. We use $\deg_G(v)$ to denote the \emph{degree} of a vertex $v$ of $G$, that is, the number of neighbors of $v$, and we write $N_G(v)$ to denote the open neighborhood of $v$, i.e., the set  of neighbors of $v$ in $G$.

A \emph{weighted} graph is a pair $(G,w)$, where $G$ is a graph and $w\colon E(G)\rightarrow \R_{> 0}$ is a \emph{weight} function. 
Slightly abusing notation, we may write $G$ for a weighted graph assuming that a weight function is given. 
The \emph{length} of a path 
$P=v_0\dots v_s$ in a weighted graph $G$ is $\sum_{i=1}^sw(v_{i-1}v_i)$. A $v_0$--$v_s$-path $P$ of minimum length is a \emph{shortest} $v_0$--$v_s$-path. The function $d_G\colon V(G)\times V(G)\rightarrow \R_{\geq 0}$, where $d_G(u,v)$ is the length of a shortest $u$--$v$-path in $G$ if such a path exists and $d_G(u,v)=+\infty$, otherwise, defines the \emph{graph metric}. 

\subparagraph{Metric spaces.}
We recall that a \emph{metric space} is a pair $(X,d)$, where $X$ is a set of points and $d\colon X\times X\rightarrow \R_{\geq 0}$ is a \emph{metric}, that is, 
\begin{itemize}
\item $d(x,x)=0$ for all $x\in X$,
\item for every $x,y\in X$, $d(x,y)=d(y,x)$,
\item for every $x,y,z\in X$, $d(x,z)\leq d(x,y)+d(y,z)$.
\end{itemize}
Throughout the paper, we assume that $d(x,y)>0$ for distinct $x,y\in X$. 
In particular, the graph metric defines the metric space with the set of points $V(G)$ under the assumption that some distances may be $+\infty$.

We consider only finite metric spaces $(X,d)$ and use $n$ to denote $|X|$ if this does not create confusion. As it was pointed out in the introduction, we view $(X,d)$ as the complete weighted graph $G=K_X$ on the vertex set $X$, where each edge $xy$ is assigned weight $d(x,y)$.
Then a \emph{spanning tree} for $X$ is a spanning tree of $G$. 

\medskip
In the introduction, we defined preserved pairs for spanning trees. However, for technical reasons, it is convenient to extend the definition for (not necessarily spanning) forests.  

\begin{definition}[Preserved pairs]
Given a forest $F$ in $G=K_X$, a pair $\{x,y\}$ of distinct $x,y\in X$ is \emph{preserved} by $F$ if $x,y\in V(F)$ and $d_F(x,y)=d(x,y)$. We also say that $\{x,y\}$ is \emph{strongly preserved} by $F$ if $\{x,y\}$ is preserved and $xy\notin E(F)$. For a family of forests $\F=\{F_1,\ldots,F_\ell\}$, 
a pair $\{x,y\}$ is \emph{preserved} if there is $i\in\{1,\ldots,\ell\}$ such that $\{x,y\}$ is preserved by $F_i$, and $\{x,y\}$ is \emph{strongly preserved} by $\F$ if $\{x,y\}$ is preserved by $\F$, and for every $i\in\{1,\ldots,\ell\}$, $xy\notin E(F_i)$. 
\end{definition}

For a family $\F=\{F_1,\dots,F_\ell\}$ of forests, we define
\[
  \pres(\F)=
  \big\lvert\big\{
      \{x,y\}\in\binom{X}{2}\mid\text{there is }i\in\{1,\ldots,\ell\}\text{ such that }\{x,y\}\text{ is preserved by }F_i\big\}\big\rvert
\]
to be the total number of preserved pairs; each preserved pair is counted once, regardless of how many forests preserve it. We also write 
\[
\pres^*(\F)=\pres(\F)-\lvert\bigcup_{i=1}^\ell E(F_i)\rvert
\]
for the total number of strongly preserved pairs.
For a single-element family $\F=\{F\}$, we write $\pres(F)$ and $\pres^*(F)$ instead of
$\pres(\F)$ and $\pres^*(\F)$, respectively.

The following simple observation will be useful for us.

\begin{observation}\label{obs:extension}
Let $(X,d)$ be a finite metric space, and let $k\geq 0$ be an integer. Then $G=K_X$ has a spanning tree $T$ with $\pres(T)\geq k+(n-1)$ if and only if $G$ has a forest $F$ with $\pres^*(F)\geq k$.
\end{observation}

\begin{proof}
The forward implication is trivial, as any tree $T$ preserving at least $k+(n-1)$ pairs strongly preserves at least $k$ pairs. For the opposite implication, notice that a forest $F$ strongly preserving $k$ pairs can be greedily extended to a spanning tree $T$. Because for any pair $\{x,y\}$ strongly preserved by $F$, $x$ and $y$ are in the same connected component of $F$, $xy\notin E(T)$. Thus, $T$ preserves at least  $k+(n-1)$ pairs.
\end{proof}

\begin{definition}[Strong preservation forests]
Let $S\subseteq \binom{X}{2}$ for a metric space $(X,d)$. We say that a forest $F$ in the weighted graph $G=K_X$ is a \emph{strong $S$-preservation forest} if every pair of $S$ is strongly preserved by $F$. A strong $S$-preservation forest is \emph{minimal} if $F$ is inclusion-minimal, with respect to vertex and edge deletions, among strong $S$-preservation forests.
A family $\F=\{F_1,\ldots,F_\ell\}$ of forests is a \emph{strong $S$-preservation forest family} if each pair of $S$ is strongly preserved by $\F$, and $\F$ is \emph{minimal}  if $\F$ is inclusion-minimal with respect to the removal of forests from $\F$, and vertex and edge deletions from the forests in $\F$.
\end{definition}

\begin{definition}[Restrictions]
Let $S\subseteq \binom{X}{2}$ for a metric space $(X,d)$. Let also $\T=\{T_1,\ldots,T_\ell\}$ be a strong $S$-preservation family of spanning trees of $G=K_X$. Then a minimal strong $S$-preservation forest family $\F=\{F_1,\ldots,F_p\}$ such that there are distinct $h_1,\ldots,h_p\in\{1,\ldots,\ell\}$ 
with $F_i$ being a subgraph of $T_{h_i}$ for $i\in\{1,\ldots,p\}$ is called an \emph{$S$-restriction} of $\T$. 
\end{definition}

Notice that, by definition, for any strong $S$-preservation family of trees $\T$, an $S$-restriction always exists.

\begin{definition}[Extensions]
Let $(X,d)$ be a metric space. 
Given a family of forests $\F=\{F_1,\ldots,F_p\}$ in $G=K_X$, a family $\T=\{T_1,\ldots,T_\ell\}$ of spanning trees of $G$ for $\ell\geq p$ such that $E(F_i)\subseteq E(T_i)$ for all $i\in\{1,\ldots,p\}$ is said to be an \emph{extension} of $\F$. If $S\subseteq \binom{X}{2}$ and $\F$ is a strong $S$-preservation family, then an extension $\T$ is a \emph{nice $S$-extension} if $\T$ is a strong $S$-preservation family and for all distinct $i,j\in\{1,\ldots,\ell\}$, $E(T_i)\cap E(T_j)\subseteq E(F_i)\cap E(F_j)$; here we assume that $E(F_h)=\emptyset$ for $h\in\{p+1,\ldots,\ell\}$.     
\end{definition}

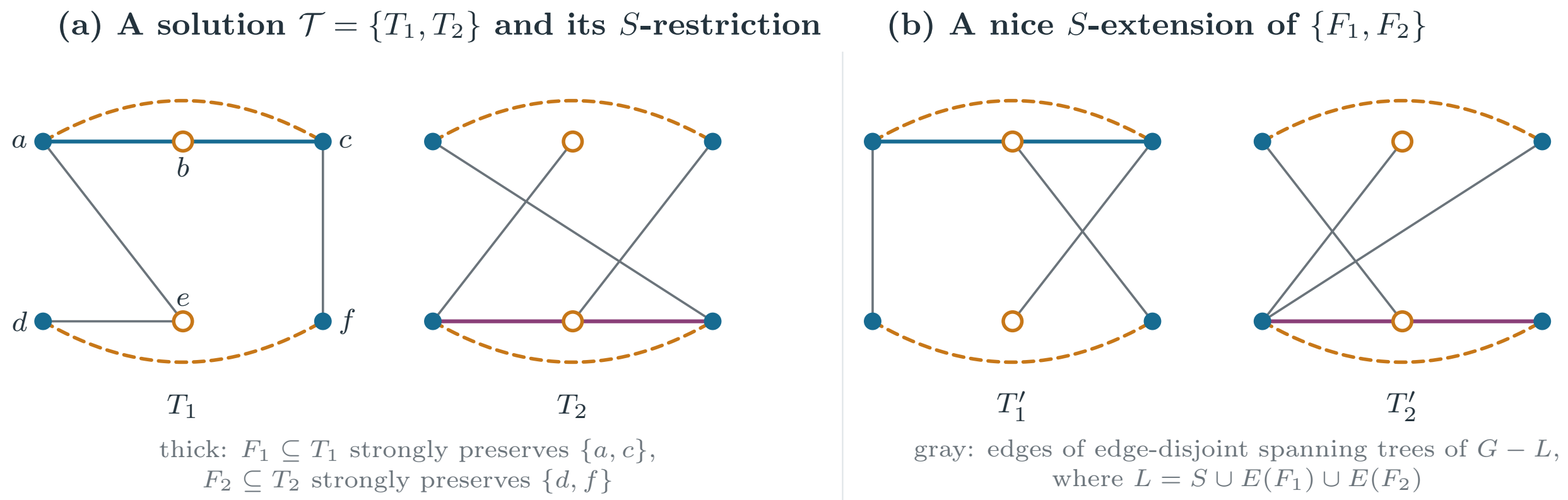
\begin{figure}[t]
  \centering
  \resizebox{\linewidth}{!}{%
\begin{tikzpicture}[x=1cm,y=1cm,font=\fontsize{9}{11}\selectfont,text=mdfInk]
  \node[mdf panel] at (0,2.35) {(a) A solution $\T=\{T_1,T_2\}$ and its $S$-restriction};
  \node[mdf panel] at (8.3,2.35) {(b) A nice $S$-extension of $\{F_1,F_2\}$};
  \draw[mdfLight] (8.0,-2.4)--(8.0,2.1);
  \foreach \sx/\lab/\tree in {0/{T_1}/1, 3.9/{T_2}/2, 8.3/{T_1'}/3, 12.2/{T_2'}/4}{%
    \begin{scope}[shift={(\sx,0)}]
      \coordinate (a) at (0,1.2); \coordinate (b) at (1.4,1.2); \coordinate (c) at (2.8,1.2);
      \coordinate (d) at (0,-.6);  \coordinate (e) at (1.4,-.6);  \coordinate (f) at (2.8,-.6);
      \ifcase\tree\or
        \draw[mdf aux] (a)--(e) (c)--(f) (e)--(d);
        \draw[mdf route] (a)--(b)--(c);
      \or
        \draw[mdf aux] (b)--(d) (a)--(f) (c)--(e);
        \draw[mdf route,draw=mdfPurple] (d)--(e)--(f);
      \or
        \draw[mdf aux] (a)--(d) (b)--(f) (c)--(e);
        \draw[mdf route] (a)--(b)--(c);
      \or
        \draw[mdf aux] (d)--(c) (e)--(a) (b)--(d);
        \draw[mdf route,draw=mdfPurple] (d)--(e)--(f);
      \fi
      \draw[mdf pair,draw=mdfOrange] (a) to[bend left=30] (c);
      \draw[mdf pair,draw=mdfOrange] (d) to[bend right=30] (f);
      \foreach \p in {a,c,d,f}\node[mdf vertex] at (\p) {};
      \node[mdf witness] at (b) {}; \node[mdf witness] at (e) {};
      \node[mdf note] at (1.4,-1.45) {$\lab$};
    \end{scope}}
  \node[mdf small] at (3.65,-2.05) {thick: $F_1\subseteq T_1$ strongly preserves $\{a,c\}$,\\$F_2\subseteq T_2$ strongly preserves $\{d,f\}$};
  \node[mdf small] at (11.95,-2.05) {gray: edges of edge-disjoint spanning trees of $G-L$,\\where $L=S\cup E(F_1)\cup E(F_2)$};
  \begin{scope}[shift={(0,0)}]
    \node[mdf note,anchor=east] at (-.08,1.2) {$a$}; \node[mdf note,anchor=north] at (1.4,1.12) {$b$};
    \node[mdf note,anchor=west] at (2.88,1.2) {$c$};
    \node[mdf note,anchor=east] at (-.08,-.6) {$d$}; \node[mdf note,anchor=south] at (1.4,-.52) {$e$};
    \node[mdf note,anchor=west] at (2.88,-.6) {$f$};
  \end{scope}
\end{tikzpicture}}
  \caption{Restrictions and nice extensions for $S=\{\{a,c\},\{d,f\}\}$. (a)~Each tree of a solution contains a minimal forest strongly preserving its share of $S$; the forests $F_1,F_2$ form an $S$-restriction of $\T$. (b)~By \Cref{lem:nice}, the forests are completed with edges of edge-disjoint spanning trees of $G-L$, where $L=S\cup E(F_1)\cup E(F_2)$. Hence two trees of the extension can share only edges of $E(F_1)\cap E(F_2)$, and no pair of $S$ becomes a tree edge.}
  \label{fig:restriction-extension}
\end{figure}

\Cref{fig:restriction-extension} illustrates restrictions and nice extensions.

Next, we make the following observation about pairs that can be preserved.

\begin{observation}\label{obs:preservance}
Let $(X,d)$ be a finite metric space. For any pair $\{x,y\}\in\binom{X}{2}$, $G=K_X$ has a forest $F$ strongly preserving $\{x,y\}$ if and only if there is $z\in X\setminus\{x,y\}$ such that 
$d(x,y)=d(x,z)+d(z,y)$.
\end{observation}

\begin{proof}
Suppose that $F$ is a forest strongly preserving $\{x,y\}$. Then $xy\notin E(F)$ and $F$ has an $x$--$y$-path $P=xv_1\dots v_sy$ whose length is $d_F(x,y)=d(x,y)$. Because $P$ is a shortest $x$--$y$-path in $G=K_X$, 
we have that $d_F(v_1,y)=d(v_1,y)$. Then $d(x,y)=d(x,z)+d(z,y)$ for $z=v_1$. For the opposite implication, observe that if $d(x,y)=d(x,z)+d(z,y)$ for some $z\in X\setminus\{x,y\}$, then the forest
$F=(\{x,y,z\},\{xz,yz\})$ in $G$ strongly preserves $\{x,y\}$.
\end{proof}

Using \Cref{obs:preservance}, we define the special graph $H_X$.

\begin{definition}[Preservation graphs]
Given a metric space $(X,d)$, the     
\emph{preservation graph} $H_X$ is the graph with $V(H_X)=X$ whose edge set consists of pairs $\{x,y\}\in \binom{X}{2}$ such that there is $z\in X\setminus\{x,y\}$, called a \emph{witness}, with
$d(x,y)=d(x,z)+d(z,y)$. 
Thus, the edges of $H_X$ correspond to the pairs that may be strongly preserved by the trees in a solution to an instance of \DCT.
\end{definition}

The next two  observations concern minimal strong $S$-preservation forests and preservation graphs.

\begin{observation}\label{obs:longpath}
Let $(X,d)$ be a finite metric space, and let $G=K_X$. Then for any shortest path $P$ in $G$ with at least four vertices, its vertices are non-isolated vertices of $H_X$.
\end{observation}

\begin{proof}
It is sufficient to observe that if $P=v_0\dots v_s$ for $s\geq 3$, then for each $i\in\{2,\ldots,s\}$, $v_{i-2}v_i\in E(H_X)$.     
\end{proof}

\begin{observation}\label{obs:pres_forest}
Let $(X,d)$ be a finite metric space, and let $S\subseteq\binom{X}{2}$ such that for every $\{x,y\}\in S$, $xy\in E(H_X)$. Then for every minimal strong $S$-preservation forest $F$ and every $v\in V(F)$, either $v$ is a non-isolated vertex of $H_X$, or 
there is $\{x,y\}\in S$ such that $v$ is a witness for $\{x,y\}$ and $xv,vy\in E(F)$. Furthermore, for any $uv\in E(F)$, $u$ or $v$ is a non-isolated vertex of $H_X$, and if $v$ is an isolated vertex of $H_X$, then there is a pair $\{u,x\}\in S$ such that $v$ is a witness for the pair.  
\end{observation}

\begin{proof}
Suppose that $v\in V(F)$ is an isolated vertex of $H_X$. 
Since $v\in V(F)$, by minimality, there is $\{x,y\}\in S$ such that $F$ contains a shortest $x$--$y$-path $P=xv_1\dots v_sy$ of $G$ with at least two edges, such that $v$ is an internal vertex of $P$. If $s\geq 2$, then by~\Cref{obs:longpath}, $v_1,\ldots,v_s$ are non-isolated vertices of $H_X$. This would contradict that $v$ is an isolated  vertex of $H_X$. Thus, $s=1$ and $v=v_1$. This means that $v$ is a witness for $\{x,y\}$ and $xv,vy\in E(F)$. 

Similarly, for the second claim, we have that $uv$ is an edge of a shortest $x$--$y$-path $P=xv_1\dots v_sy$ of $G$, which is the unique $x$--$y$-path in $F$, for some $\{x,y\}\in S$.
If $s\geq 2$, both $u$ and $v$ are non-isolated vertices of $H_X$ by~\Cref{obs:longpath}. If $s=1$, then $u\in\{x,y\}$ or $v\in\{x,y\}$, that is, at least one of the endpoints of $uv$ is a non-isolated vertex of $H_X$. By symmetry, we can assume that if $v$ is an isolated vertex of $H_X$, then $u=y$. Then we have that $v$ is a witness for  $\{u,x\}\in S$. This concludes the proof.
\end{proof}

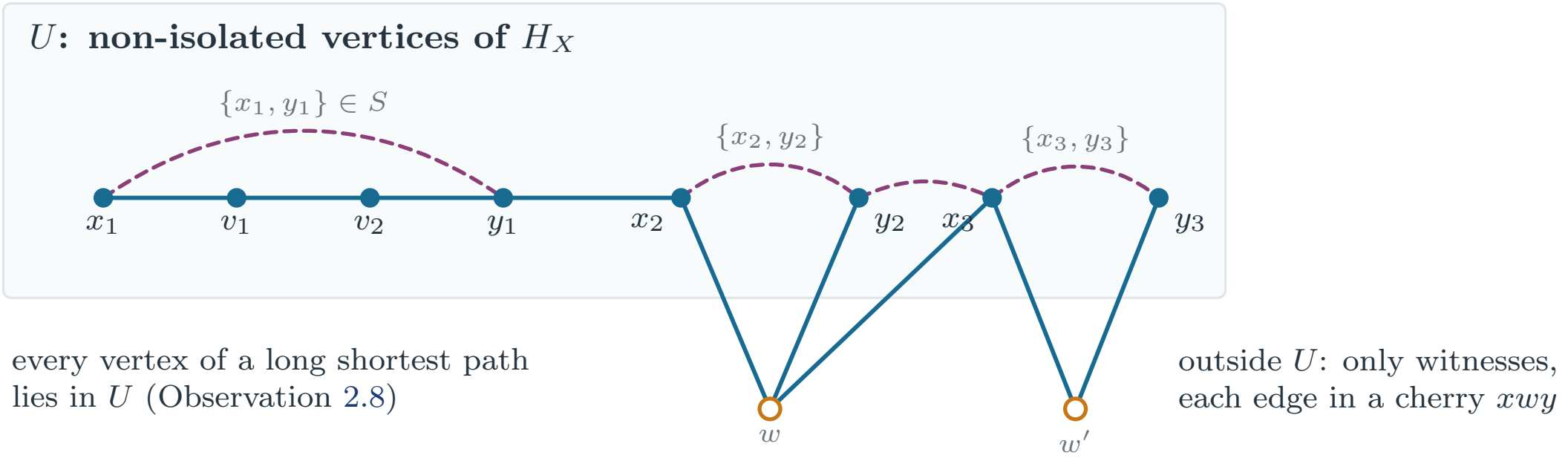
\begin{figure}[t]
  \centering
  \resizebox{.85\linewidth}{!}{%
\begin{tikzpicture}[x=1cm,y=1cm,font=\fontsize{9}{11}\selectfont,text=mdfInk]
  \draw[mdf box] (-.5,-.6) rectangle (10.5,2.05);
  \node[mdf panel,anchor=north west] at (-.4,2.0) {$U$: non-isolated vertices of $H_X$};
  \coordinate (x1) at (0.4,.3); \coordinate (v1) at (1.6,.3); \coordinate (v2) at (2.8,.3); \coordinate (y1) at (4.0,.3);
  \coordinate (x2) at (5.6,.3); \coordinate (y2) at (7.2,.3);
  \coordinate (x3) at (8.4,.3); \coordinate (y3) at (9.9,.3);
  \draw[mdf route] (x1)--(v1)--(v2)--(y1);
  \draw[mdf route] (y1)--(x2);
  \draw[mdf pair] (x1) to[bend left=35] node[above=1pt,mdf small] {$\{x_1,y_1\}\in S$} (y1);
  \coordinate (w) at (6.4,-1.6); \coordinate (w2) at (9.15,-1.6);
  \draw[mdf route] (x2)--(w)--(y2);
  \draw[mdf route] (w)--(x3);
  \draw[mdf route] (x3)--(w2)--(y3);
  \draw[mdf pair] (x2) to[bend left=40] node[above=1pt,mdf small] {$\{x_2,y_2\}$} (y2);
  \draw[mdf pair] (x3) to[bend left=40] node[above=1pt,mdf small] {$\{x_3,y_3\}$} (y3);
  \draw[mdf pair] (y2) to[bend left=25] (x3);
  \foreach \p/\l in {x1/x_1,v1/v_1,v2/v_2,y1/y_1}
     \node[mdf vertex,label={[mdf note]below:$\l$}] at (\p) {};
  \foreach \p/\l in {x2/x_2,x3/x_3}
     \node[mdf vertex,label={[mdf note]below left:$\l$}] at (\p) {};
  \foreach \p/\l in {y2/y_2,y3/y_3}
     \node[mdf vertex,label={[mdf note]below right:$\l$}] at (\p) {};
  \node[mdf witness,label={[mdf small]below:$w$}] at (w) {};
  \node[mdf witness,label={[mdf small]below:$w'$}] at (w2) {};
  \node[mdf note,anchor=west,align=left] at (-.5,-1.35) {every vertex of a long shortest path\\lies in $U$ (\Cref{obs:longpath})};
  \node[mdf note,anchor=west,align=left] at (10.0,-1.35) {outside $U$: only witnesses,\\each edge in a cherry $xwy$};
\end{tikzpicture}}
  \caption{The shape of a minimal strong $S$-preservation forest (\Cref{obs:pres_forest}). Inside $U$, the forest may use arbitrary edges and longer shortest paths. A vertex outside $U$ is a witness $w$ for some pairs of $S$, and all its edges lie on two-edge paths $xwy$ for such pairs; here $w$ witnesses $\{x_2,y_2\}$ and $\{y_2,x_3\}$, and $w'$ witnesses $\{x_3,y_3\}$. Dashed curves are pairs of $S$.}
  \label{fig:minimal-forest}
\end{figure}

The structure of minimal strong $S$-preservation forests described in \Cref{obs:pres_forest} is shown in \Cref{fig:minimal-forest}.

\section{Structural results}\label{sec:struct}
In this section, we consider structural properties of preservation graphs.
We use the Nash-Williams--Tutte theorem~\cite{NashWilliams61,Tutte61} to give a  sufficient condition for the existence of a nice extension. 
In particular, we use this condition 
to identify certain yes-instances of \DCT. Informally, we show that if $H_X$ has sufficiently many edges with respect to $k$, then $(X,d,\ell,k)$ is a yes-instance. 

We start by stating  the Nash-Williams--Tutte theorem~\cite{NashWilliams61,Tutte61}. 
Given a partition 
$\Pcal=\{V_1,V_2,\dots,V_p\}$ of the vertex set of a graph $G$, it is said that an edge $xy\in E(G)$ is \emph{crossing} if $x$ and $y$ are in distinct sets of $\Pcal$.

\begin{theorem}[\cite{NashWilliams61,Tutte61}]\label{thm:NWT}
A graph $G$ has $\ell$ edge-disjoint spanning trees if and only if for every partition
$\Pcal=\{V_1,V_2,\dots,V_p\}$
of $V(G)$ into $p\ge 2$ nonempty parts, the number of crossing edges of $G$ is at least
$\ell(p-1)$.
\end{theorem}

As a corollary of~\Cref{thm:NWT}, we obtain the following lemma.

\begin{lemma}\label{lem:kn-minus-s}
Let $G=K_n$ be the complete graph on $n\geq 2$ vertices, and let $S\subseteq E(G)$ satisfy
$|S|<\frac n2$.
Then for every positive integer $\ell \le \left\lceil\frac n2\right\rceil-1$,
$G-S$ contains $\ell$ edge-disjoint spanning trees.
\end{lemma}

\begin{proof}
Notice that it is sufficient to show the claim for 
$\ell=\left\lceil\frac n2\right\rceil-1\leq \frac{1}{2}(n-1)$. Similarly, because $|S|<\frac{n}{2} $ and $|S|$ is an integer,  
we can assume that $|S|=\left\lceil\frac n2\right\rceil-1\leq \frac{1}{2}(n-1)$.
Then by~\Cref{thm:NWT}, it is sufficient to show that for 
every partition $\Pcal=\{V_1,V_2,\dots,V_p\}$
of $V(G)$ for $p\ge 2$, the number of crossing edges is at least 
\begin{equation}\label{eq:one_S}
\frac{1}{2}(n-1)p=\frac{1}{2}(n-1)(p-1)+\frac{1}{2}(n-1)\geq \ell(p-1)+|S|.
\end{equation}

Consider a partition $\Pcal=\{V_1,V_2,\dots,V_p\}$ of $V(G)$ such that the number of crossing edges is minimum. Let $n_i=|V_i|$. Then the number of crossing edges is 
\[f(n_1,\ldots,n_p)=\binom{n}{2}-\sum_{i=1}^p\binom{n_i}{2}.\]
By the concavity of $f(n_1,\ldots,n_p)$ and the assumption that the number of crossing edges is minimum, we obtain that  $\Pcal$ contains at most one set of size at least two. 
Thus, we can assume without loss of generality that 
$n_1=\dots=n_{p-1}=1$ and $n_p=n-(p-1)$. Then the number of crossing edges is
\begin{equation}\label{eq:two_S}
\binom{n}{2}-\binom{n-p+1}{2}=(p-1)(n-\frac{p}{2}).
\end{equation}

Combining~\Cref{eq:one_S} and \Cref{eq:two_S}, we have that it is sufficient to show that 
\begin{equation}\label{eq:three_S}
g(p)=(p-1)(n-\frac{p}{2})-\frac{1}{2}(n-1)p\geq 0.
\end{equation}
The function $g(p)$ is concave and $2\leq p\leq n$. Thus, the minimum value of $g(p)$ is achieved when either $p=2$ or $p=n$.
Because $g(2)=g(n)=0$,
we conclude that \Cref{eq:three_S} holds. This
means that $G-S$ contains $\ell$ edge-disjoint spanning trees and completes the proof.
\end{proof}

Using~\Cref{lem:kn-minus-s}, we obtain the following lemma.

\begin{lemma}\label{lem:pres_one}
Let $(X,d)$ be a finite metric space, and let $k\geq 0$ and $1\leq \ell<\lceil\frac{n}{2}\rceil$ be integers. Suppose that $G=K_X$ has a forest $F$ such that (i)~$|E(F)|<\frac{n}{2}-k$ and (ii)~$\pres^*(F)\geq k$. Then $G$ has a family $\T$ of $\ell$ spanning trees 
such that $\pres(\T)\geq k+\ell(n-1)$.
\end{lemma}

\begin{proof}
Let $R$ be a set of $k$ edges $xy\in E(G)$  such that the pair $\{x,y\}$ is strongly preserved by $F$. 
Consider $S=R\cup E(F)$. By condition~(i), $|S|<\frac{n}{2}$. Then by~\Cref{lem:kn-minus-s}, $G-S$ has $\ell$ edge-disjoint spanning trees $T_1,\ldots,T_\ell$. Because $F$ is a forest, $G$ has a spanning tree $T_1'$ with $E(F)\subseteq E(T_1')\subseteq E(F)\cup E(T_1)$. Then the trees $T_1',T_2,\dots,T_\ell$ are edge-disjoint. Notice that each pair $\{x,y\}$ for $xy\in R$ is strongly preserved by $T_1'$ because this pair is strongly preserved by $F$ and $xy\notin E(T_1')$ as $x$ and $y$ are in the same connected component of $F$. Furthermore, since $R\subseteq S$, $xy\notin E(T_i)$ for each $xy\in R$ and $i\in\{2,\ldots,\ell\}$. Because $\T=\{T_1',T_2,\ldots,T_\ell\}$ preserves the edges of $R\cup E(T_1')\cup\bigcup_{i=2}^\ell E(T_i)$, we have that $\pres(\T)\geq k+\ell(n-1)$. This concludes the proof.
\end{proof}

Also, by applying~\Cref{lem:kn-minus-s}, we prove the following sufficient condition for the existence of a nice $S$-extension.

\begin{lemma}\label{lem:nice}
Let $S\subseteq \binom{X}{2}$ for a metric space $(X,d)$. Let also $\F=\{F_1,\ldots,F_p\}$ be a strong $S$-preservation forest family in $G=K_X$ such that $|S\cup\bigcup_{i=1}^p E(F_i)|<\frac{n}{2}$. Then for any $\ell$ such that $p\leq\ell\leq\lceil\frac{n}{2}\rceil-1$,
$\F$ admits a nice $S$-extension $\T=\{T_1,\ldots,T_\ell\}$.
\end{lemma}

\begin{proof}
Let $L=S\cup \bigcup_{i=1}^p E(F_i)$. As $|L|<\frac{n}{2}$, $G-L$ has $\ell$ edge-disjoint spanning trees $T_1,\ldots,T_\ell$. 
We construct $T_1',\ldots,T_\ell'$ as follows. For $i\in\{1,\ldots,p\}$, we complement $F_i$ to a spanning tree $T_i'$ by using the edges of $T_i$. Thus, $T_i'$ is a spanning tree of $G$ with $E(F_i)\subseteq E(T_i')\subseteq E(F_i)\cup E(T_i)$. For $i\in\{p+1,\ldots,\ell\}$, we set $T_i'=T_i$. By the construction, $xy\notin E(T_i')$ for $\{x,y\}\in S$ and $i\in\{1,\ldots,\ell\}$. Thus, $\T'$ is a strong $S$-preservation family of spanning trees. Also, by the construction,
we have that $E(T_i')\cap E(T_j')\subseteq E(F_i)\cap E(F_j)$ for distinct $i,j\in\{1,\ldots,\ell\}$ (assuming  that $E(F_h)=\emptyset$ for $h\in\{p+1,\ldots,\ell\}$).  This means that $\T'$ is a nice  $S$-extension and concludes the proof.
\end{proof}

Now we show our sufficient conditions for yes-instances.

\begin{lemma}\label{lem:bigH}
Let $(X,d)$ be a finite metric space, and let $k\geq 0$ and $1\leq \ell<\lceil\frac{n}{2}\rceil$ be integers. Suppose that $n>6k$ and the preservation graph $H_X$ has at least $(2k-1)k+1$ non-isolated vertices or at least $(2k-1)k+1$ edges. Then $G=K_X$ has a family $\T$ of $\ell$ spanning trees 
such that $\pres(\T)\geq k+\ell(n-1)$.
\end{lemma}

\begin{proof}
We argue that if $H_X$ contains a matching $M$ of size $2k$ or a vertex $x$ with $\deg_{H_X}(x)\geq k$, then $G=K_X$ has a family $\T$ of $\ell$ spanning trees 
such that $\pres(\T)\geq k+\ell(n-1)$.   

\begin{claim}\label{cl:matching}
If $H_X$ has a matching $M$ of size at least $2k$, then $G$ has a family $\T$ of $\ell$ spanning trees such that $\pres(\T)\geq k+\ell(n-1)$.     
\end{claim}

\begin{proof}[Proof of~\Cref{cl:matching}]
For each edge $xy\in M$, denote by $z_{xy}\in X$ a witness for the pair $\{x,y\}\in \binom{X}{2}$, that is, a point with the property that $d(x,z_{xy})+d(z_{xy},y)=d(x,y)$. We claim that there is a submatching $M'\subseteq M$ of size $k$ such that 
the subgraph $G[R_{M'}]$ with $R_{M'}=\bigcup_{xy\in M'}\{xz_{xy},z_{xy}y\}$ is a forest. 

For this, we use induction and show that for every positive $t\leq k$, $M$ has a submatching $M'$ of size $t$ satisfying our condition. This is trivial for $t=1$ as we can pick an arbitrary $xy\in M$. Suppose that $M'$
is a submatching of size $t$ such that $G[R_{M'}]$ is a forest.
We show that we can add at least one edge of $M$ to $M'$ without violating the condition. Consider the set $Z$ of witnesses for the pairs $\{x,y\}$ for $xy\in M'$.  We have that $|Z|\leq t$  (note that distinct pairs may have the same witnesses). Because the endpoints of the edges of $M$ are distinct, at most $|Z|\leq t$ edges of $M$ have their endpoints in $Z$. Since 
$|M|\geq 2k$ and $|M'|=t<k$, $|M\setminus M'|=2k-t>t$. Then at least $(2k-t)-t=2(k-t)\geq 1$ edges of $M\setminus M'$ have no endpoint in $Z$. Choose one such edge $uv$ and set $M''=M'\cup\{uv\}$.
Since $u,v\notin Z$ and $u,v$ are distinct from the endpoints of the edges of $M'$, 
the vertices $u$ and $v$ have degree one in the spanning subgraph of $G$ with the edge set $R_{M''}$. Thus, $R_{M''}$ forms a forest. This proves that there is $M'$ of size $k$ such that 
 $G[R_{M'}]$ for $R_{M'}=\bigcup_{xy\in M'}\{xz_{xy},z_{xy}y\}$ is a forest. 

Consider the forest $F=G[R_{M'}]$. Notice that for any $xy\in M'$, $xy\notin E(F)$ as $xz_{xy},z_{xy}y\in E(F)$ and $F$ is a forest. Then, because for every $xy\in M'$, $xz_{xy},z_{xy}y\in E(F)$, $F$ strongly preserves every pair $\{x,y\}$ with $xy\in M'$. Thus, $\pres^*(F)\geq k$.
Because $|E(F)|=2k$ and $n>6k$, $|E(F)|<\frac{n}{2}-k$. Then 
$G$ has a family $\T$ of $\ell$ spanning trees such that $\pres(\T)\geq k+\ell(n-1)$ by \Cref{lem:pres_one}. This proves the claim.
\end{proof}

\begin{claim}\label{cl:bigdeg}
If $H_X$ has a vertex $x$ with $\deg_{H_X}(x)\geq k$, then $G$ has a family $\T$ of $\ell$ spanning trees such that $\pres(\T)\geq k+\ell(n-1)$.     
\end{claim}

\begin{figure}[t]
  \centering
  \resizebox{.8\linewidth}{!}{%
\begin{tikzpicture}[x=1cm,y=1cm,font=\fontsize{9}{11}\selectfont,text=mdfInk]
  \coordinate (x) at (0,0);
  \coordinate (z1) at (1.4,0); \coordinate (y1) at (2.8,0); \coordinate (y3) at (4.8,0);
  \coordinate (z2) at (1.3,1.25); \coordinate (y2) at (2.6,2.5);
  \draw[mdf route] (x)--(z1)--(y1)--(y3);
  \draw[mdf route] (x)--(z2)--(y2);
  \draw[mdf pair] (x) to[bend right=28] (y1);
  \draw[mdf pair] (x) to[bend right=38] (y3);
  \draw[mdf pair] (x) to[bend left=30] (y2);
  \node[mdf vertex,label=left:$x$] at (x) {};
  \node[mdf witness,label=above:$z_1$] at (z1) {};
  \node[mdf witness,label=below right:$z_2$] at (z2) {};
  \node[mdf vertex,label={above:{$y_1=z_3$}}] at (y1) {};
  \node[mdf vertex,label=above:$y_2$] at (y2) {};
  \node[mdf vertex,label=right:$y_3$] at (y3) {};
  \node[mdf note,anchor=west,align=left] at (5.6,1.4) {$Q$ has the edges $xz_i,z_iy_i$, except the edges $xy_1,\dots,xy_k$};
  \node[mdf note,anchor=west,align=left] at (5.6,.65) {the witness $z_3$ of $\{x,y_3\}$ is $y_1$,\\so $xz_3=xy_1\notin E(Q)$};
  \node[mdf note,anchor=west,align=left] at (5.6,-.4) {yet $d_Q(x,y_3)\le d_Q(x,y_1)+d(y_1,y_3)$\\$\phantom{d_Q(x,y_3)}=d(x,y_3)$, as $d(x,y_1)<d(x,y_3)$};
\end{tikzpicture}}
  \caption{The proof of \Cref{cl:bigdeg} for $k=3$. Dashed curves are the pairs $\{x,y_i\}$ to be strongly preserved. A witness of one pair may be another neighbor $y_j$ of $x$ in $H_X$; then the edge $xy_j$ is missing from $Q$, but the distance to $y_i$ is still realized through $y_j$. Induction on $d(x,y_i)$ makes this precise.}
  \label{fig:bigdeg}
\end{figure}
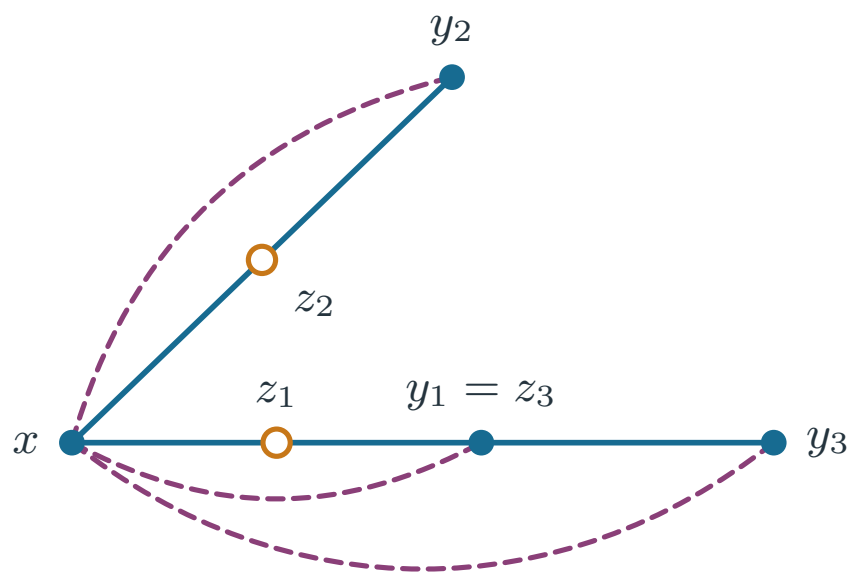

\begin{proof}[Proof of~\Cref{cl:bigdeg}]
See \Cref{fig:bigdeg} for an illustration.
Suppose that $\deg_{H_X}(x)\geq k$. Denote by $y_1,\ldots,y_k$ any $k$ distinct neighbors of $x$ in $H_X$.
Denote by $z_i$ a witness for the pair $\{x,y_i\}$ for $i\in\{1,\ldots,k\}$. Consider the subgraph $Q$ of $G$ with the edge set $\bigcup_{i=1}^k\{xz_i,z_iy_i\}\setminus \{xy_1,\ldots,xy_k\}$. We claim that the distances from $x$ to $y_1,\ldots,y_k$ in $Q$ are the same as in $G$. 

The proof is by contradiction. Assume that there is $i\in\{1,\ldots,k\}$ such that $d_Q(x,y_i)>d(x,y_i)$. We select such an index $i\in\{1,\ldots,k\}$ for which $d(x,y_i)$ is minimum. Because $z_i$ is a witness for $\{x,y_i\}$, $d(x,y_i)=d(x,z_i)+d(z_i,y_i)$. Since 
$z_iy_i\notin\{xy_1,\ldots,xy_k\}$, we obtain that $z_iy_i\in E(Q)$. Then because $d_Q(x,y_i)>d(x,y_i)$, $xz_i\notin E(Q)$, that is, $z_i=y_j$ for some $j\in \{1,\ldots,k\}$. However, $d(x,y_j)<d(x,y_i)$. Therefore, by the choice of $i$, $d_Q(x,y_j)=d(x,y_j)$. This means that 
$d_Q(x,y_i)\leq d_Q(x,y_j)+d_Q(z_i,y_i)=d(x,z_i)+d(z_i,y_i)=d(x,y_i)$, contradicting the assumption that $d_Q(x,y_i)>d(x,y_i)$. This proves that the distances in $Q$ from $x$ to $y_1,\ldots,y_k$ are the same as the distances in $G$.

Consider an arbitrary tree $F$ of shortest paths in $Q$ from $x$ to $y_1,\ldots,y_k$. We have that $|E(F)|\leq |E(Q)|\leq 2k$. Furthermore, because $xy_1,\ldots,xy_k\notin E(Q)$ and the 
distances in $Q$ from $x$ to $y_1,\ldots,y_k$ are the same as the distances in $G$, 
the pairs $\{x,y_i\}$ for $i\in\{1,\ldots,k\}$ are strongly preserved by $F$.
This means that $\pres^*(F)\geq k$.
Since $|E(F)|\leq 2k$ and $n>6k$, $|E(F)|<\frac{n}{2}-k$. Then 
$G$ has a family $\T$ of $\ell$ spanning trees such that $\pres(\T)\geq k+\ell(n-1)$ by \Cref{lem:pres_one}. This completes the proof.
\end{proof}

Combining \Cref{cl:matching} and \Cref{cl:bigdeg}, we obtain that if $H_X$ contains a matching $M$ of size $2k$ or a vertex $x$ with $\deg_{H_X}(x)\geq k$, then $G=K_X$ has a family $\T$ of $\ell$ spanning trees such that $\pres(\T)\geq k+\ell(n-1)$. Assume that these are not the cases.  We claim that $H_X$  has at most $(2k-1)k$ vertices of positive degree and at most $(2k-1)k$ edges. 

Because the maximum degree of $H_X$ is at most $k-1$, $H_X$ admits a proper edge-coloring with at most $k$ colors by Vizing's theorem (see, e.g.,~\cite{Dieste12}). This implies that $H_X$ has a color class, that is, a matching, of size at least $\frac{|E(H_X)|}{k}\leq 2k-1$. Thus, $|E(H_X)|\leq (2k-1)k$.  
To upper bound the number of non-isolated vertices, consider a maximum-size matching $M$ in $H_X$. By our assumption, $|M|\leq 2k-1$. Notice that for each $uv\in M$, $u$ and $v$ have no distinct neighbors 
that are not endpoints of the edges of $M$  because $M$ is maximum. Thus, if both $u$ and $v$ have neighbors that are not endpoints of the edges of $M$, then 
any such a neighbor of $u$ or $v$ is a common neighbor of $u$ and $v$. 
Therefore, both $u$ and $v$
have at most $k-2$ neighbors in $H_X$ distinct from the endpoints of the edges of $M$. Since every non-isolated vertex of $H_X$ is either endpoint of an edge of $M$ or adjacent to such an endpoint, 
the total number of the non-isolated vertices is at most $(2k-1)k$.
This completes the proof.
\end{proof}

\section{An \FPT algorithm for \DCT}\label{sec:FPT}
In this section, we prove \Cref{thm:fpt}, which we restate here.

\FPTalg*

\begin{proof}
Let $(X,d,\ell,k)$ be an instance of \DCT. We let $G=K_X$ be the weighted complete graph encoding the metric space $(X,d)$.

If $k=0$, then by~\Cref{thm:NWT}, $G$ has $\ell$ edge-disjoint spanning trees. Thus, $(X,d,\ell,k)$ is a yes-instance, and we report this and stop.

Assume first that $n\leq 14k^4$. Then we solve the problem by brute force. The number of spanning trees of $G$ is $n^{n-2}$ by Cayley's formula~\cite{Cayley09}. Furthermore, it is well-known that spanning trees (of any graph) can be enumerated with polynomial delay (see, e.g.,~\cite{Winter86}). Therefore, we can list all spanning trees and guess $\ell\leq \frac{n}{2}\leq 7k^4$ of them by trying all $k^{\Oh(k^8)}+\Oh(1)$ possible choices. 
For every choice of a set of $\ell$ trees $\T=\{T_1,\ldots,T_\ell\}$, we verify in polynomial time whether $\pres(\T)\geq k+\ell(n-1)$ by computing $\min\{d_{T_i}(x,y)\mid 1\leq i\leq \ell\}$ for all pairs $\{x,y\}\in\binom{X}{2}$. 
The overall running time of this procedure is $k^{\Oh(k^8)}+\Oh(1)$. 

From now on, we assume that $n>14k^4$. 

We start by constructing the preservation graph $H_X$. Note that this can be done in $\Oh(n^3)$ time.
If $k=1$, then $n>14$. If $E(H_X)\neq\emptyset$, then $H_X$ has at least two non-isolated vertices, and $(X,d,\ell,k)$ is a yes-instance by~\Cref{lem:bigH}. Otherwise, no pair can be strongly preserved by~\Cref{obs:preservance}, so any family of $\ell$ spanning trees preserves at most $\ell(n-1)$ pairs, and $(X,d,\ell,k)$ is a no-instance. In either case, we report the answer and stop.

If $H_X$ has at least $(2k-1)k+1$ non-isolated vertices or
$(2k-1)k+1$ edges, then $G$ has a family $\T$ of $\ell$ spanning trees 
such that $\pres(\T)\geq k+\ell(n-1)$ by~\Cref{lem:bigH}, that is, 
$(X,d,\ell,k)$ is a yes-instance. Accordingly, we report this and stop. 
In the following, we assume that the set $U$ of non-isolated vertices of $H_X$ is of size at most $(2k-1)k\leq 2k^2$, and $H_X$ has at most $(2k-1)k\leq 2 k^2$ edges. Furthermore, $k\geq 2$.  

Recall that by the definition of the preservation graph and \Cref{obs:preservance}, a pair $\{x,y\}\in\binom{X}{2}$ can be strongly preserved by some forest if and only if $xy\in E(H_X)$. Therefore, to preserve pairs that are not edges of spanning trees, we have to preserve pairs corresponding to the edges of $H_X$. Thus, we guess the pairs that will be strongly preserved by the family of spanning trees forming a potential solution $\T$. First, we guess $S_0\subseteq R$ for $R=\{\{x,y\}\in\binom{X}{2}\mid xy\in E(H_X)\}$, which is the subset of pairs that will not be strongly preserved. Then the pairs of $R\setminus S_0$ should be strongly preserved by $\T$. Therefore, they should be preserved by the minimal strong $(R\setminus S_0)$-preservation forest family $\F=\{F_1,\ldots,F_p\}$ for some $p\leq \ell$ forming a $(R\setminus S_0)$-restriction of $\T$. Assign each pair of $R\setminus S_0$ to one forest of $\F$ that strongly preserves it, and let $S_i$ consist of the pairs assigned to $F_i$ for $i\in\{1,\ldots,p\}$. By the minimality of $\F$, each $S_i$ is non-empty and $F_i$ is a minimal strong $S_i$-preservation forest. Indeed, if $S_i$ were empty, removing $F_i$ would leave every pair of $R\setminus S_0$ strongly preserved by its assigned forest. Similarly, any vertex or edge deletion from $F_i$ that preserves all pairs of $S_i$ would leave every pair of $R\setminus S_0$ strongly preserved by the family, contradicting its minimality. Thus, we consider all possible partitions $\Scal=\{S_0,\ldots,S_p\}$
of 
$R=\{\{x,y\}\in\binom{X}{2}\mid xy\in E(H_X)\}$ into at most $\ell+1$ sets, where only  $S_0 $ may be empty. 
The total number of partitions of an $m$-element set into non-empty sets is the Bell number $B_m=m^{\Oh(m)}$ (see, e.g., \cite{BerendT10}). This implies that we have to consider $k^{\Oh(k^2)}$ partitions. 
Note that $p\leq |E(H_X)|\leq 2k^2$.
Our assumption for such a choice of a partition is that 
\begin{itemize}
\item[(i)] each pair of $S_0$ (which may be empty) is not strongly preserved or the corresponding edge of $H_X$ is in one of the trees in a potential solution $\T$. 
\item[(ii)] the pairs of $R\setminus S_0$ are strongly preserved by $\T$, and for each $i\in\{1,\ldots,p\}$, $F_i$ is a minimal strong $S_i$-preservation forest in a $(R\setminus S_0)$-restriction $\F=\{F_1,\ldots,F_p\}$ of $\T$.
\end{itemize}
As we consider all possible partitions $\Scal$, a solution $\T$ satisfying (i) and (ii) should exist for at least one choice, unless 
$(X,d,\ell,k)$ is a no-instance.
From now on, we assume that $\Scal$ is given. Then our task is to try to extend this choice to a solution.

The general idea is to guess the forests $F_1,\ldots,F_p$ and then extend them to a solution by complementing them into spanning trees using~\Cref{lem:kn-minus-s}. 
However,
we will argue that it is sufficient to guess a family $\F'=\{F_1',\ldots,F_p'\}$ of forests such that  
\begin{itemize}
\item[(a)] $\F'$ is a strong $(R\setminus S_0)$-preservation forest family such that $F_i'$ is a minimal strong $S_i$-preservation forest for $i\in\{1,\ldots,p\}$,
\item[(b)] for each $i\in\{1,\ldots,p\}$, $|E(F_i')|=|E(F_i)|$,
\item[(c)] $|\bigcup_{i=1}^p E(F_i)|\leq |\bigcup_{i=1}^p E(F_i')|$.
\end{itemize}

The crucial fact that allows us to find $\F'$ is that, by~\Cref{obs:pres_forest}, each vertex $v$ of $F_i'$ outside $U$ is a witness for some pair $\{x,y\}\in S_i$ and $xv,vy\in E(F_i')$, and each edge of $F_i'$ incident to $v$ belongs to $\{xv,vy\}$ for some pair $\{x,y\}\in S_i$. 
We use this to restrict the choice of the vertices of $F_i'$ outside $U$ and incident edges.
Consider the auxiliary bipartite \emph{witness} graph $W$. One side of the bipartition is the set $R$ of pairs $\{x,y\}$ for $xy\in E(H_X)$. The other side consists of the vertices $v\in V(G)\setminus U$ such that $v$ is a witness of some pair $\{x,y\}\in R$, and we make $v$ adjacent to each pair $\{x,y\}$ which is witnessed by $v$. Denote by $A$ the set of vertices of $V(G)\setminus U$ in $W$. We say that $u,v\in A$ are \emph{equivalent} if these vertices have the same neighbors in $R$. Denote by $A_1,\ldots,A_r$ the equivalence classes of $A$. Since $|R|\leq 2k^2$, $r\leq 2^{2k^2}$. For each $i\in\{1,\ldots,r\}$, we define 
$A_i^*=A_i$ if $|A_i|\leq 2k^2$, and we set $A_i^*$ to be a set of arbitrary $2k^2$ elements of $A_i$, otherwise. We set $A^*=\bigcup_{i=1}^rA_i^*$. Observe that $|A^*|\leq 2k^2\cdot2^{2k^2}$. Note also that $W$ (and, therefore, $A^*$) can be constructed in $2^{\Oh(k^2)}\cdot n^{\Oh(1)}$ time. 
We claim that the choice of witnesses outside $U$ can be restricted by $A^*$.

\begin{claim}\label{cl:resrict_witness}
Suppose that $\F=\{F_1,\ldots,F_p\}$ is an $(R\setminus S_0)$-restriction of a solution $\T$ satisfying conditions (i) and (ii). Then there is a family $\F'=\{F_1',\ldots,F_p'\}$ of forests satisfying (a)--(c) such that for each $i\in\{1,\ldots,p\}$ and each $v\in V(F_i')\setminus U$,
$v\in A^*$.
\end{claim}

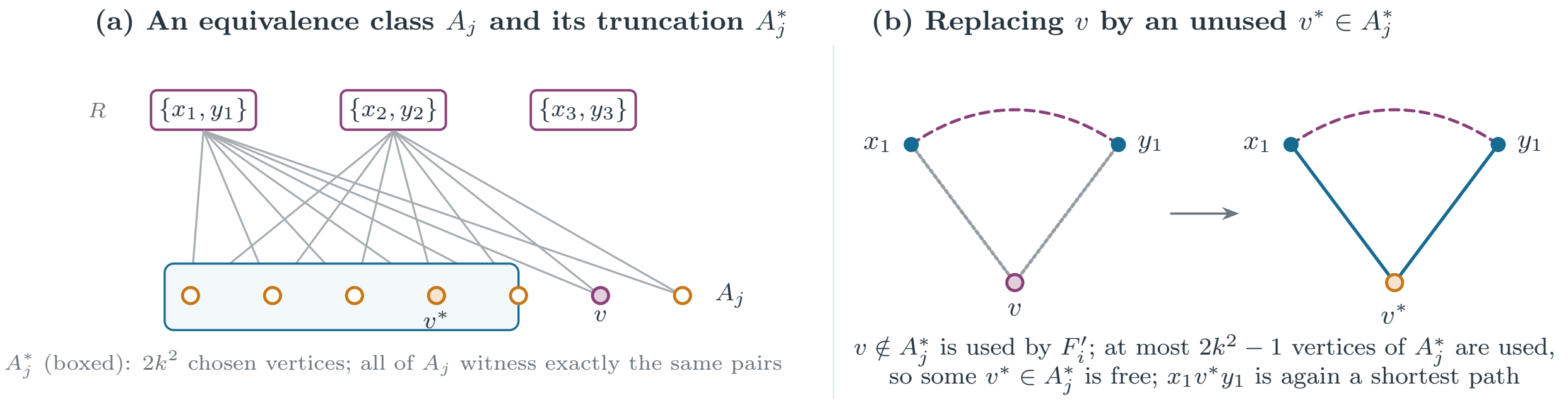
\begin{figure}[t]
  \centering
  \resizebox{\linewidth}{!}{%
\begin{tikzpicture}[x=1cm,y=1cm,font=\fontsize{9}{11}\selectfont,text=mdfInk]
  \node[mdf panel] at (0,2.6) {(a) An equivalence class $A_j$ and its truncation $A_j^*$};
  \node[mdf panel] at (9.0,2.6) {(b) Replacing $v$ by an unused $v^*\in A_j^*$};
  \draw[mdfLight] (8.7,-1.75)--(8.7,2.35);
  \foreach \i/\xx in {1/1.4,2/3.6,3/5.8}
    \node[draw=mdfPurple,rounded corners=2pt,fill=white,line width=.8pt,inner sep=2.5pt,font=\fontsize{8}{9}\selectfont] (r\i) at (\xx,1.6) {$\{x_\i,y_\i\}$};
  \node[mdf small,anchor=east] at (.35,1.6) {$R$};
  \foreach \t in {1,...,7}{
    \coordinate (a\t) at ({.3+\t*.95},-.55);
    \draw[mdf aux,draw=mdfGray!60] (a\t)--(r1.south);
    \draw[mdf aux,draw=mdfGray!60] (a\t)--(r2.south);}
  \draw[draw=mdfBlue,rounded corners=3pt,line width=.7pt,fill=mdfBlue!5] (.95,-.95) rectangle (5.05,-.18);
  \foreach \t in {1,...,7}{
    \ifnum\t=4 \node[mdf witness,fill=mdfOrange!20] at (a\t) {};
    \else\ifnum\t=6 \node[mdf witness,draw=mdfPurple,fill=mdfPurple!25] at (a\t) {};
    \else \node[mdf witness] at (a\t) {};\fi\fi}
  \node[mdf note,anchor=north] at ($(a4)+(0,-.1)$) {$v^*$};
  \node[mdf note,anchor=north] at ($(a6)+(0,-.1)$) {$v$};
  \node[mdf note,anchor=west] at (7.25,-.55) {$A_j$};
  \node[mdf small] at (3.6,-1.35) {$A_j^*$ (boxed): $2k^2$ chosen vertices; all of $A_j$ witness exactly the same pairs};
  \begin{scope}[shift={(9.6,0)}]
    \coordinate (x) at (0,1.2); \coordinate (y) at (2.4,1.2);
    \coordinate (v) at (1.2,-.4);
    \draw[mdf route,draw=mdfGray!70,densely dotted] (x)--(v)--(y);
    \draw[mdf pair] (x) to[bend left=35] (y);
    \node[mdf vertex,label=left:$x_1$] at (x) {}; \node[mdf vertex,label=right:$y_1$] at (y) {};
    \node[mdf witness,draw=mdfPurple,fill=mdfPurple!25,label=below:$v$] at (v) {};
    \draw[mdf arrow] (3.0,.4)--(3.8,.4);
    \begin{scope}[shift={(4.4,0)}]
      \coordinate (x) at (0,1.2); \coordinate (y) at (2.4,1.2); \coordinate (v) at (1.2,-.4);
      \draw[mdf route] (x)--(v)--(y);
      \draw[mdf pair] (x) to[bend left=35] (y);
      \node[mdf vertex,label=left:$x_1$] at (x) {}; \node[mdf vertex,label=right:$y_1$] at (y) {};
      \node[mdf witness,fill=mdfOrange!20,label=below:$v^*$] at (v) {};
    \end{scope}
    \node[mdf note,align=center] at (3.4,-1.35) {$v\notin A_j^*$ is used by $F_i'$; at most $2k^2-1$ vertices of $A_j^*$ are used,\\so some $v^*\in A_j^*$ is free; $x_1v^*y_1$ is again a shortest path};
  \end{scope}
\end{tikzpicture}}
  \caption{The exchange argument in \Cref{cl:resrict_witness}. (a)~In the witness graph $W$, equivalent vertices outside $U$ have the same neighbors in $R$; the algorithm keeps only $2k^2$ of them. (b)~If a forest uses a vertex $v\in A_j\setminus A_j^*$, then $v$ is replaced by an unused $v^*\in A_j^*$ with the same edges. The forest stays minimal, and the union of edges does not shrink.}
  \label{fig:witness-swap}
\end{figure}

\begin{proof}[Proof of \Cref{cl:resrict_witness}]
The exchange argument is illustrated in \Cref{fig:witness-swap}.
Notice that a family of forests $\F'$ satisfying (a)--(c) exists, as we can take $\F'=\F$.
Choose such a family $\F'$ minimizing
$\sum_{h=1}^p |V(F_h')\setminus (U\cup A^*)|$. We claim that this sum is zero. For the sake of contradiction, assume that there is $i\in\{1,\ldots,p\}$ and $v\in V(F_i')\setminus U$ such that $v\notin A^*$. By~\Cref{obs:pres_forest}, $v\in A$, and hence $v\in A_j$ for some $j\in\{1,\ldots,r\}$ with $|A_j|\geq 2k^2+1$.

For each $h\in\{1,\ldots,p\}$, every vertex $u\in V(F_h')\setminus U$ is the unique internal vertex of a two-edge shortest path $xuy$ in $F_h'$ for some pair $\{x,y\}\in S_h$, by~\Cref{obs:pres_forest}. Two distinct such vertices cannot serve the same pair in $F_h'$, since the corresponding two-edge paths would form a cycle. Therefore, $|V(F_h')\setminus U|\leq |S_h|$. Since $S_1,\ldots,S_p$ partition $R\setminus S_0$, it follows that
\[
\big\lvert\big(\bigcup_{h=1}^p V(F_h')\big)\setminus U\big\rvert
\leq \sum_{h=1}^p |V(F_h')\setminus U|
\leq \sum_{h=1}^p |S_h|
=|R\setminus S_0|\leq |R|\leq 2k^2.
\]
As $v$ belongs to this union but not to $A_j^*$, at most $2k^2-1$ vertices of $A_j^*$ belong to any forest of $\F'$. Since $|A_j^*|=2k^2$, there is a vertex $v^*\in A_j^*$ that does not belong to any forest of $\F'$.

Modify $F_i'$ by replacing $v$ by $v^*$ and making $v^*$ adjacent to the same vertices as $v$, leaving the other forests unchanged. This replacement is an isomorphism of the underlying unweighted graphs, so the modified graph is a forest with the same number of edges, and (b) holds. To verify (a), consider the unique path in $F_i'$ for any pair $\{x,y\}\in S_i$. If it avoids $v$, it is unchanged. Otherwise, since $x,y\in U$ and $v\notin U$, \Cref{obs:longpath} implies that this path is $xvy$. Because $v$ and $v^*$ have the same neighbors in $W$, the replacement path $xv^*y$ is also shortest.
By the minimality of the original $F_i'$, every vertex and edge of $F_i'$ belongs to the unique path for some pair of $S_i$. This remains true after the replacement, so deleting any vertex or edge destroys the preservation of some pair of $S_i$. Thus, the modified forest is still minimal. Moreover, all new edges are incident to $v^*\notin U$, so none belongs to $E(H_X)$ or corresponds to a pair in $R\setminus S_0$. Consequently, the modified family still strongly preserves $R\setminus S_0$, proving (a).

Finally, all new edges incident to $v^*$ are absent from the other forests, since $v^*$ does not belong to any of them. The replacement adds as many new edges to the union as there were edges incident to $v$ in $F_i'$, and removes at most that many edges from the union. Hence the size of the edge union does not decrease, preserving (c). However, replacing $v\notin U\cup A^*$ by $v^*\in A^*$ in $F_i'$ decreases the minimized sum by exactly one, even if $v$ occurs in other forests. This contradicts the choice of $\F'$ and proves the claim.
\end{proof}

Next, we show that $\F'$ admits a nice $(R\setminus S_0)$-extension to a family $\T'=\{T_1',\ldots,T_\ell'\}$ of spanning trees.

\begin{claim}\label{cl:extend_forest}
Let $\F'=\{F_1',\ldots,F_p'\}$ be a strong $(R\setminus S_0)$-preservation forest family such that $F_i'$ is a minimal strong $S_i$-preservation forest for $i\in\{1,\ldots,p\}$. Then $\F'$ has a nice $(R\setminus S_0)$-extension to a family $\T'=\{T_1',\ldots,T_\ell'\}$ of spanning trees for $p\leq\ell<\lceil\frac{n}{2}\rceil$.
\end{claim}

\begin{proof}
Consider the set 
$L=\bigcup_{i=1}^pE(F_i')\cup\bigcup_{i=1}^p\{xy\mid\{x,y\}\in S_i\}$. By~\Cref{obs:pres_forest}, each vertex $v\notin U$ of $F_i'$ for $i\in\{1,\ldots,p\}$
is a witness for some pair $\{x,y\}\in S_i$, and for each edge $vx$ incident to $v$, there is $y\in U$ such that $\{x,y\}\in  S_i$. Since $|R|\leq 2k^2$ and $\{S_1,\ldots,S_p\}$ is a partition of a subset of $R$, we have that $L$ contains at most $4k^2$ edges incident to vertices $v\notin U$. Because $|U|\leq 2k^2$ and for each pair $\{x,y\}\in R$, $x,y\in U$, we have that 
$|L|\leq \binom{2k^2}{2}+4k^2\leq 7k^4<\frac{n}{2}$. Then the claim follows from~\Cref{lem:nice}.
\end{proof}

Now we are ready to explain the steps of our algorithm for the considered partition $\Scal=\{S_0,\ldots,S_p\}$ of $R$. For every $i\in\{1,\ldots,p\}$, we guess $F_i'$ by considering all possible choices in the following steps:
\begin{itemize}
\item Initialize $E(F_i')$ by selecting a subset $E_i\subseteq E(G[U])\setminus\{xy\mid \{x,y\}\in\bigcup_{j=1}^pS_j\}$ such that  $G[E_i]$ is a forest. 
\item For each $\{x,y\}\in S_i$ such that $\{x,y\}$ is not strongly preserved by the forest selected in the first step, select $v\in A^*$ such that $v$ is a witness for $\{x,y\}$ and set 
$E_i:=E_i\cup\{xv,vy\}$ (note that the same witness may be chosen for distinct pairs).
\item Set $F_i'=G[E_i]$. Discard the choice if $F_i'$ contains a cycle or is not a minimal strong $S_i$-preservation forest.  
\end{itemize}
The total number of possible choices of $E_i\subseteq E(G[U])\setminus\{xy\mid \{x,y\}\in\bigcup_{j=1}^pS_j\}$ is at most \linebreak
$(|U|+1)^{|U|-1}\leq(2k^2+1)^{2k^2-1}$ by Cayley's formula~\cite{Cayley09}. Indeed, any forest on $U$ can be extended to a tree on $U$ and one additional vertex by joining the new vertex to one vertex of each component, and distinct forests give distinct trees. 
Since $p\leq |R|\leq 2k^2$, the total number of all initial choices of the sets $E_i$ is at most 
$(2k^2+1)^{4k^4}$. 
Because the total number of witnesses $v\notin U$ selected in the second step is at most $|R|\leq 2k^2$ and $|A^*|\leq 2k^2\cdot2^{2k^2}$, there are at most 
$(2k^2\cdot2^{2k^2})^{2k^2}$ choices of these witnesses. Thus, the overall number of possible choices of $F_i'$ for $i\in\{1,\ldots,p\}$ is $k^{\Oh(k^4)}$.

Notice that we can verify whether $F_i'=G[E_i]$ is a minimal strong $S_i$-preservation forest in polynomial time by first checking that $F_i'$ is acyclic, and then checking whether it strongly preserves each pair of $S_i$ and the deletion of any vertex or edge violates this property. 

Assume that $\F'=\{F_1',\ldots,F_p'\}$ is given. 
We check whether 
\begin{equation}\label{eq:yes}
 \sum_{i=1}^p|S_i|-(\sum_{i=1}^p|E(F_i')|-|\bigcup_{i=1}^p E(F_i')|)\geq k,  
\end{equation}
and return that $(X,d,\ell,k)$ is a yes-instance of \DCT and stop. Otherwise, we discard the current choice of $\F'$. If we exhaust the list of possible choices of $\F'$ for the considered $\Scal$, we  discard the choice of $\Scal$.

Finally, recall that we consider all possible choices of $\Scal$. If we find that there is no solution for every selection of $\Scal$, we report that $(X,d,\ell,k)$ is a no-instance of \DCT. This completes the description of the algorithm.

To argue correctness, assume that there is a family of forests $\F'=\{F_1',\ldots,F_p'\}$ constructed for a partition $\Scal=\{S_0,\ldots,S_p\}$ of $R=\{\{x,y\}\in\binom{X}{2}\mid xy\in E(H_X)\}$ for $p\leq\ell$ such that \Cref{eq:yes} holds. 
 By the construction,  $\F'$ is a strong $(R\setminus S_0)$-preservation forest family such that $F_i'$ is a minimal strong $S_i$-preservation forest for $i\in\{1,\ldots,p\}$. Then
 by \Cref{cl:extend_forest}, $\F'$ has a nice $(R\setminus S_0)$-extension to a family $\T'=\{T_1',\ldots,T_\ell'\}$ of spanning trees for $p\leq\ell<\lceil\frac{n}{2}\rceil$. 
Because $\T'$ is a nice $(R\setminus S_0)$-extension,
\begin{align*}
|\bigcup_{i=1}^\ell E(T_i')|=&|\bigcup_{i=1}^p E(F_i')|+\sum_{i=1}^\ell|E(T_i')\setminus E(F_i')|\\
=&|\bigcup_{i=1}^p E(F_i')|+\sum_{i=1}^\ell|E(T_i')|-\sum_{i=1}^p|E(F_i')|\\
=&|\bigcup_{i=1}^p E(F_i')|+\ell(n-1)-\sum_{i=1}^p|E(F_i')|
\end{align*}
(we assume that $E(F_i')=\emptyset$ if $i\geq p+1$).
Then
\begin{align*}
\pres(\T')\geq& \sum_{i=1}^p|S_i|+|\bigcup_{i=1}^\ell E(T_i')|\\
=&\sum_{i=1}^p|S_i|+(|\bigcup_{i=1}^p E(F_i')|+\ell(n-1)-\sum_{i=1}^p|E(F_i')|)\\
=&[\sum_{i=1}^p|S_i|-(\sum_{i=1}^p|E(F_i')|-|\bigcup_{i=1}^pE(F_i')|)]+\ell(n-1)\\
\geq&k+\ell(n-1).
\end{align*}
This means that $(X,d,\ell,k)$ is a yes-instance of \DCT. Thus, if our algorithm reports a yes-instance, then this is a correct answer.

For the opposite implication, suppose that $(X,d,\ell,k)$ is a yes-instance of \DCT.  Then there is a family  $\T$ of spanning trees such that $\pres(\T)\geq k+\ell(n-1)$.
By~\Cref{obs:preservance} and the definition of the preservation graph $H_X$, a pair $\{x,y\}$ can be strongly preserved by a forest if and only if $xy\in E(H_X)$. 
Then every pair $\{x,y\}\in \binom{X}{2}$ such that $\{x,y\}$ is strongly preserved by 
$\T$
should be in 
$R=\{\{x,y\}\in\binom{X}{2}\mid xy\in E(H_X)\}$.  
Set $S_0$ to be the set of pairs of $R$ that are not strongly preserved by $\T$, and take a $(R\setminus S_0)$-restriction $\F=\{F_1,\ldots,F_p\}$ of $\T$. As above, assign each pair of $R\setminus S_0$ to one forest of $\F$ that strongly preserves it, and let $S_i$ be the set of pairs assigned to $F_i$. By the minimality of $\F$, the sets $S_1,\ldots,S_p$ are non-empty and each $F_i$ is a minimal strong $S_i$-preservation forest. Thus, $\Scal=\{S_0,\ldots,S_p\}$ is a partition of $R$ into $p+1\leq \ell+1$ sets satisfying (i) and (ii).
Furthermore, it holds that $\pres(\T)=\sum_{i=1}^p|S_i|+|\bigcup_{i=1}^\ell E(T_i)|$.
By symmetry, we assume without loss of generality that $F_i$ is a subgraph of $T_i$ for $i\in\{1,\ldots,p\}$.

By~\Cref{cl:resrict_witness}, there is a family of forests $\F'=\{F_1',\ldots,F_p'\}$ such that (a)~$\F'$ is a strong $(R\setminus S_0)$-preservation forest family such that $F_i'$ is a minimal strong $S_i$-preservation forest for $i\in\{1,\ldots,p\}$,
(b)~for each $i\in\{1,\ldots,p\}$, $|E(F_i')|=|E(F_i)|$, (c)~$|\bigcup_{i=1}^p E(F_i)|\leq |\bigcup_{i=1}^p E(F_i')|$, and  furthermore, 
for each $i\in\{1,\ldots,p\}$ and each $v\in V(F_i')\setminus U$,
$v\in A^*$. We show that our algorithm considers this family $\F'$. For each $i\in\{1,\ldots,p\}$, the set $E(F_i'[U])$ is a valid choice of $E_i$ in the first step: it is the edge set of a forest, and none of its edges corresponds to a pair of $R\setminus S_0$, since $\F'$ strongly preserves all these pairs. Consider any pair $\{x,y\}\in S_i$ that is not strongly preserved by $F_i'[U]$. Its unique path in $F_i'$ is shortest in $G$ and contains a vertex outside $U$. By~\Cref{obs:longpath}, this path has only two edges and is of the form $xvy$ for some $v\in V(F_i')\setminus U\subseteq A^*$. Therefore, the second step includes the choice of this witness $v$. All edges added for these choices belong to $F_i'$. Moreover, by the minimality of $F_i'$, every edge outside $E(F_i'[U])$ belongs to the unique path for some pair of $S_i$, and this pair cannot also be strongly preserved by $F_i'[U]$, as $F_i'$ is a forest. Hence these choices generate exactly $F_i'$, which passes the checks in the third step. It follows that the algorithm considers $\F'$. Then, by~\Cref{cl:extend_forest}, $\F'$ admits a nice $(R\setminus S_0)$-extension $\T'=\{T_1',\ldots,T_\ell'\}$. 
Notice that because $|E(F_i)|=|E(F_i')|$ for all $i\in\{1,\ldots,p\}$,
$|E(T_i)\setminus E(F_i)|=|E(T_i')\setminus E(F_i')|=(n-1)-|E(F_i')|$.
Recall also that every spanning tree has $n-1$ edges.
Then, because of (c), 
\begin{align*}
k+\ell(n-1)\leq&\pres(\T)=\sum_{i=1}^p|S_i|+|\bigcup_{i=1}^\ell E(T_i)|\\
\leq &\sum_{i=1}^p|S_i|+|\bigcup_{i=1}^p E(F_i)|+\sum_{i=1}^\ell|E(T_i)\setminus E(F_i)|\\%
\leq & \sum_{i=1}^p|S_i|+|\bigcup_{i=1}^p E(F_i')|+\sum_{i=1}^\ell|E(T_i')\setminus E(F_i')|\\%
= & \sum_{i=1}^p|S_i|+|\bigcup_{i=1}^p E(F_i')|+\ell(n-1)-\sum_{i=1}^p|E(F_i')|
\end{align*}
(we assume that $E(F_i)=E(F_i')=\emptyset$ if $i\geq p+1$).
Therefore,
\begin{equation*}
\sum_{i=1}^p|S_i|-(\sum_{i=1}^p|E(F_i')|-|\bigcup_{i=1}^p E(F_i')|)\geq k.    
\end{equation*}
We obtain that \Cref{eq:yes} holds for $\Scal$ and $\F'$. Then, the algorithm should conclude that $(X,d,\ell,k)$ is a yes-instance of \DCT. This completes the correctness proof.

To evaluate the running time, recall that in the case $n
\leq 14k^4$, we solve the problem by brute force in $k^{\Oh(k^8)}+\Oh(1)$ time. When $n>14k^4$, we construct $H_X$ in polynomial time. Then we use~\Cref{lem:bigH} to rule out an easy yes-instance. Further, we 
construct the witness graph $W$ and find the restricted set of witnesses $A^*$ in  
 $2^{\Oh(k^2)}\cdot n^{\Oh(1)}$ time. Next, we
consider $k^{\Oh(k^2)}$ partitions 
$\Scal=\{S_0,\ldots,S_p\}$ of
$R=\{\{x,y\}\in\binom{X}{2}\mid xy\in E(H_X)\}$. For each partition, we try $k^{\Oh(k^4)}$
choices of $\F'$, for which we verify \Cref{eq:yes}. This implies that the running time for the case $n>14k^4$ is $k^{\Oh(k^4)}\cdot n^{\Oh(1)}$. Thus, the overall running time is $k^{\Oh(k^8)}\cdot n^{\Oh(1)}$. This completes the proof.
\end{proof}

\section{Kernelization for \DCT}\label{sec:kernelization}
In this section, we discuss kernelization for \DCT parameterized by $k$. In~\Cref{sec:kern_conv}, we show compression for the case of a strictly convex metric space. In~\Cref{sec:kern_lb}, we complement this result by demonstrating a kernelization lower bound for the general case.

\subsection{Kernelization for a strictly convex metric}\label{sec:kern_conv}
In this subsection, we prove \Cref{thm:kernel-euclid}. 
Our kernelization algorithm uses the properties of \emph{enhanced preservation graphs}.

\begin{definition}[Enhanced preservation graphs]
Given a metric space $(X,d)$, we say that $z_{xy}\in X\setminus\{x,y\}$ is a \emph{special witness} for a pair $\{x,y\}\in \binom{X}{2}$ if $z_{xy}$ is a \emph{unique} witness for $\{x,y\}$.
Then the      
\emph{enhanced preservation graph} is the graph $H_X^*$ with $V(H_X^*)=X$ whose edge set is
$E(H_X)\cup\bigcup_{z\in X}\{xz,zy\mid z\text{ is a special witness for }\{x,y\}\in\binom{X}{2}\}$
\end{definition}

\begin{lemma}\label{lem:enhanced}
Let $(X,d)$ be a finite strictly convex  metric space, and let $S\subseteq\binom{X}{2}$ such that for every $\{x,y\}\in S$, $xy\in E(H_X)$. Then for every minimal strong $S$-preservation forest $F$, $F$ is a subgraph of $H_X^*$.    
\end{lemma}

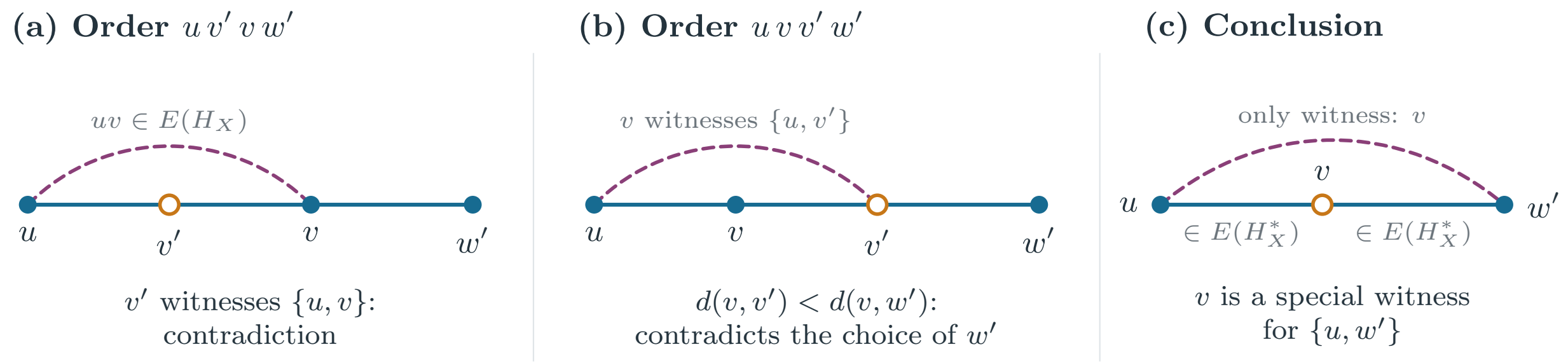
\begin{figure}[t]
  \centering
  \resizebox{\linewidth}{!}{%
\begin{tikzpicture}[x=1cm,y=1cm,font=\fontsize{9}{11}\selectfont,text=mdfInk]
  \node[mdf panel] at (0,1.75) {(a) Order $u\,v'\,v\,w'$};
  \node[mdf panel] at (5.6,1.75) {(b) Order $u\,v\,v'\,w'$};
  \node[mdf panel] at (11.2,1.75) {(c) Conclusion};
  \draw[mdfLight] (5.3,-1.55)--(5.3,1.5);
  \draw[mdfLight] (10.9,-1.55)--(10.9,1.5);
  \begin{scope}[shift={(.3,0)}]
    \coordinate (u) at (0,0); \coordinate (vp) at (1.4,0); \coordinate (v) at (2.8,0); \coordinate (w) at (4.4,0);
    \draw[mdf route] (u)--(vp)--(v)--(w);
    \draw[mdf pair] (u) to[bend left=45] node[above=1pt,mdf small] {$uv\in E(H_X)$} (v);
    \node[mdf vertex,label=below:$u$] at (u) {};
    \node[mdf witness,label=below:$v'$] at (vp) {};
    \node[mdf vertex,label=below:$v$] at (v) {};
    \node[mdf vertex,label=below:$w'$] at (w) {};
    \node[mdf note] at (2.2,-1.1) {$v'$ witnesses $\{u,v\}$:\\contradiction};
  \end{scope}
  \begin{scope}[shift={(5.9,0)}]
    \coordinate (u) at (0,0); \coordinate (v) at (1.4,0); \coordinate (vp) at (2.8,0); \coordinate (w) at (4.4,0);
    \draw[mdf route] (u)--(v)--(vp)--(w);
    \draw[mdf pair] (u) to[bend left=45] node[above=1pt,mdf small] {$v$ witnesses $\{u,v'\}$} (vp);
    \node[mdf vertex,label=below:$u$] at (u) {};
    \node[mdf vertex,label=below:$v$] at (v) {};
    \node[mdf witness,label=below:$v'$] at (vp) {};
    \node[mdf vertex,label=below:$w'$] at (w) {};
    \node[mdf note] at (2.2,-1.1) {$d(v,v')<d(v,w')$:\\contradicts the choice of $w'$};
  \end{scope}
  \begin{scope}[shift={(11.5,0)}]
    \coordinate (u) at (0,0); \coordinate (v) at (1.6,0); \coordinate (w) at (3.4,0);
    \draw[mdf route] (u)--node[below=2pt,mdf small] {$\in E(H_X^*)$}(v)--node[below=2pt,mdf small] {$\in E(H_X^*)$}(w);
    \draw[mdf pair] (u) to[bend left=40] node[above=1pt,mdf small] {only witness: $v$} (w);
    \node[mdf vertex,label=left:$u$] at (u) {};
    \node[mdf witness,label=above:$v$] at (v) {};
    \node[mdf vertex,label=right:$w'$] at (w) {};
    \node[mdf note] at (1.7,-1.1) {$v$ is a special witness\\for $\{u,w'\}$};
  \end{scope}
\end{tikzpicture}}
  \caption{The proof of \Cref{lem:enhanced}. Here $uv\in E(F)\setminus E(H_X)$, $v$ witnesses $\{u,w'\}$, and $w'$ is chosen closest to $v$. If $v'\neq v$ were another witness for $\{u,w'\}$, strict convexity would put $u,v,v',w'$ on a shortest path in one of the two orders (a) and (b), and both are impossible. Hence (c) $v$ is the unique witness, and $uv\in E(H_X^*)$.}
  \label{fig:enhanced}
\end{figure}

\begin{proof}
See \Cref{fig:enhanced} for an illustration of the argument.
Consider an arbitrary edge $uv$ of $F$. We claim that $uv\in E(H_X^*)$. This holds if $uv\in E(H_X)$. Assume that this is not the case and $uv\notin E(H_X)$. 
Because $F$ is a minimal strong $S$-preservation forest, there is a pair $\{x,y\}\in S$ such that 
$uv$ is an edge of a shortest $x$--$y$-path $P$ with at least two edges. As $P$ contains at least two edges, by symmetry, we can assume that $P$ contains a subpath $uvw$. Thus, $v$ is a witness for the pair $\{u,w\}$ for some $w\in X\setminus\{u,v\}$. For all pairs $\{u,w'\}\in\binom{X}{2}$ witnessed by $v$, we choose $w'\in X\setminus\{u,v\}$ at minimum distance from $v$. We show that $v$ is a unique witness for $\{u,w'\}$. 

For the sake of contradiction, assume that $v'\neq v$ is a witness for $\{u,w'\}$. Then,
$d(u,w')=d(u,v)+d(v,w')=d(u,v')+d(v',w')$. Because $(X,d)$ is a strictly convex metric space, either $uv'vw'$ or $uvv'w'$ is a shortest $u$--$w'$-path in $G=K_X$. 
However, in the first case, we have that $v'$ is a witness for $\{u,v\}$ implying that $uv\in E(H_X)$, but this contradicts the assumption that $uv\notin E(H_X)$. Thus, $uvv'w'$ is a shortest $u$--$w'$-path. Then $d(v,v')<d(v,w')$, but this contradicts the choice of $w'$. This proves that 
$v$ is a unique witness for $\{u,w'\}$. 

Because $v$ is a unique witness for $\{u,w'\}$, we have that $uv,vw'\in E(H_X^*)$ by the definition of this graph. In particular, $uv\in E(H_X^*)$, and this completes the proof.
\end{proof}

To compress the distances, we use the betweenness property by applying the following observation that immediately follows from~\Cref{obs:preservance} and the definition of the strict convexity.

\begin{observation}\label{obs:betweenness}
Let $(X,d)$ and $(X,d')$ be finite metric spaces with the property that for any distinct $x,y,z\in X$, $d(x,y)+d(y,z)=d(x,z)$ if and only if 
 $d'(x,y)+d'(y,z)=d'(x,z)$. Then $(X,d,\ell,k)$ and $(X,d',\ell,k)$ are equivalent instances of \DCT for all $k\geq 1$ and $1\leq\ell<\lceil\frac{n}{2}\rceil$. Moreover,
 $(X,d)$ is a strictly convex metric space if and only if $(X,d')$ is strictly convex.  
\end{observation}

Now we restate \Cref{thm:kernel-euclid}. 

\Compression*

\begin{proof}
We start by showing that there is a polynomial-time algorithm that, given an instance
\((X,d,\ell,k)\) of \DCT where $(X,d)$ is a finite strictly convex metric space, 
either solves the problem, or outputs an equivalent instance \((X',d,\ell',k)\), such that $X'\subseteq X$ is of size $\Oh(k^2)$. The proof is based on the same structural results as the proof of \Cref{thm:fpt}.

Let $(X,d,\ell,k)$ be an instance of \DCT. We let $G=K_X$ be the weighted complete graph encoding the metric space $(X,d)$. If $k=0$, then by~\Cref{thm:NWT}, $(X,d,\ell,k)$ is a yes-instance, and we report this and stop. Hence, we assume that $k\geq 1$. 
If $n< 12k^2$, the algorithm simply returns the input instance, which has bounded size. From now on, we assume that $n\geq 12k^2$.

We construct the preservation graph $H_X$ and the enhanced preservation graph $H_X^*$. This can be done in $\Oh(n^3)$ time by going over $\binom{n}{2}$ pairs of elements of $X$ and checking whether a pair has a witness. Simultaneously, we find the set $U$ of non-isolated vertices of $H_X$ and the set $W$ of special witnesses. 
If $|E(H_X)|<k$, then by~\Cref{obs:preservance}, any family of $\ell$ spanning trees preserves at most $|E(H_X)|+\ell(n-1)<k+\ell(n-1)$ pairs. In this case, we report a no-instance and stop.
If $H_X$ has at least $(2k-1)k+1$ non-isolated vertices or
$(2k-1)k+1$ edges, then $G$ has a family $\T$ of $\ell$ spanning trees 
such that $\pres(\T)\geq k+\ell(n-1)$ by~\Cref{lem:bigH}. Then we report a yes-instance and stop. Otherwise, we have that the set $U$ of non-isolated vertices of $H_X$ is of size at most $(2k-1)k< 2k^2$, and $H_X$ has at most $(2k-1)k< 2k^2$ edges. 
Since $|W|\leq |E(H_X)|<2k^2$, we have that $|U\cup W|<4k^2$. Furthermore, 
$|E(H_X^*)|\leq |E(H_X)|+2|E(H_X)|<6k^2$. 

Now we construct the new instance $(X',d,\ell',k)$:
\begin{itemize}
\item set $X'\subseteq X$ to be an arbitrary superset of $U\cup W$ of size $12k^2$,
\item set $\ell'=\min\{\ell, |E(H_X)|\}$.
\end{itemize}
Clearly, $(X',d,\ell',k)$ can be constructed in polynomial time. Note that $|X'|= 12k^2$ and $1\leq\ell'< 2k^2\leq \lceil\frac{|X'|}{2}\rceil-1$.
To complete the proof of the first part of~\Cref{thm:kernel-euclid}, we show the following claim.

\begin{claim}\label{cl:equiv_inst}
$(X,d,\ell,k)$ is a yes-instance of \DCT if and only if $(X',d,\ell',k)$ is a yes-instance.
\end{claim}

\begin{proof}[Proof of~\Cref{cl:equiv_inst}]
Suppose that $(X,d,\ell,k)$ is a yes-instance. Then $G$ has a family $\T=\{T_1,\ldots,T_\ell\}$ of spanning trees such that $\pres(\T)\geq k+\ell(n-1)$. Denote by $S\subseteq\binom{X}{2}$ the set of pairs strongly preserved by $\T$. Recall that if $\{x,y\}\in S$ then $xy\in E(H_X)$ by~\Cref{obs:preservance}. Let a family of forests $\F=\{F_1,\ldots,F_p\}$ in $G$ be an $S$-restriction of $\T$. Notice that $p\leq\min\{\ell,|E(H_X)|\}$. We assume without loss of generality that $F_i$ is a subgraph of $T_i$ for $i\in\{1,\ldots,p\}$.
By \Cref{lem:enhanced}, we have that for every $i\in\{1,\ldots,p\}$, $F_i$ is a forest in 
$H_X^*$. Moreover, as $F_i$ has no isolated vertices, $V(F_i)\subseteq U\cup W\subseteq X'$.
This implies that $L=S\cup\bigcup_{i=1}^p E(F_i)\subseteq E(H_X^*)$. Hence,
$|L|\leq |E(H_X^*)|<6k^2=\frac{|X'|}{2}$.
Since $U\cup W\subseteq X'$, we also have that $\F$ is a family of forests in $G'=K_{X'}$.

Assuming that $E(F_i)=\emptyset$ for $i\geq p+1$,
we have that
\begin{align*}
k+\ell(|X|-1)\leq&\pres(\T)=|S|+|\bigcup_{i=1}^\ell E(T_i)|\\
\leq&|S|+|\bigcup_{i=1}^pE(F_i)|+\sum_{i=1}^\ell|E(T_i)\setminus E(F_i)|\\
=&|S|+|\bigcup_{i=1}^pE(F_i)|+\sum_{i=1}^\ell|E(T_i)|-\sum_{i=1}^p |E(F_i)|\\
=&|S|+|\bigcup_{i=1}^pE(F_i)|+\ell(|X|-1)-\sum_{i=1}^p |E(F_i)|,
\end{align*}
and $k\leq |S|+|\bigcup_{i=1}^pE(F_i)|-\sum_{i=1}^p |E(F_i)|$.
Since $\F$ is a strong $S$-preservation forest family in $G'$, $|L|<\frac{|X'|}{2}$, and $p\leq\ell'<\lceil\frac{|X'|}{2}\rceil$, $\F$ admits a nice $S$-extension $\T'=\{T_1',\ldots,T_{\ell'}'\}$ in $G'$ by~\Cref{lem:nice}. Then
\begin{align*}
\pres(\T')\geq& |S|+|\bigcup_{i=1}^{\ell'}E(T_i')|\\
=&|S|+|\bigcup_{i=1}^{p}E(F_i)|+\sum_{i=1}^{\ell'}|E(T_i')\setminus E(F_i)|\\
=&|S|+|\bigcup_{i=1}^{p}E(F_i)|+\sum_{i=1}^{\ell'}|E(T_i')|-\sum_{i=1}^p|E(F_i)|\\
=&[|S|+|\bigcup_{i=1}^pE(F_i)|-\sum_{i=1}^p |E(F_i)|]+\ell'(|X'|-1)\\
\geq k+\ell'(|X'|-1).
\end{align*}
Therefore, $(X',d,\ell',k)$ is a yes-instance.

For the opposite implication, suppose that $(X',d,\ell',k)$ is a yes-instance, and let $\T'=\{T_1',\ldots,T_{\ell'}'\}$ be a solution. Denote by $S'$ the set of pairs strongly preserved by $\T'$, and let $\F'=\{F_1',\ldots,F_p'\}$ be an $S'$-restriction of $\T'$. 
Since $X'\subseteq X$ and the distances are unchanged, $E(H_{X'})\subseteq E(H_X)$.
We claim that the same holds for the enhanced preservation graphs, that is, $E(H_{X'}^*)\subseteq E(H_X^*)$. Consider an arbitrary pair $\{x,y\}\in \binom{X}{2}$ such that $xy\in E(H_X)$. Suppose that $\{x,y\}$ has two distinct witnesses $z_1,z_2\in X\setminus\{x,y\}$, that is,
$d(x,y)=d(x,z_1)+d(z_1,y)=d(x,z_2)+d(z_2,y)$. Because $(X,d)$ is strictly convex and $x,z_1,z_2,y$ are pairwise distinct, either $xz_1z_2y$ or $xz_2z_1y$ is a shortest $x$--$y$-path in $G$.
By~\Cref{obs:longpath}, $z_1$ and $z_2$ are non-isolated vertices of $H_X$, that is, $z_1,z_2\in U$. Since $U\subseteq X'$, we obtain that $\{x,y\}$ has at least two witnesses in $X'$. 
This implies that $E(H_{X'}^*)\subseteq E(H_X^*)$.
Then, by~\Cref{lem:enhanced}, $E(F_i')\subseteq E(H_X^*)$. 
Therefore, for $L'=S'\cup\bigcup_{i=1}^p E(F_i')$, we have that $L'\subseteq E(H_X^*)$ and
$|L'|\leq |E(H_X^*)|<6k^2=\frac{|X'|}{2}$.

As $p\leq\ell'\leq\ell<\lceil\frac{|X|}{2}\rceil$, \Cref{lem:nice} gives a nice $S'$-extension $\T$ of $\F'$ to $\ell$ spanning trees of $G$. The same counting argument as above, applied to the restriction $\F'$ of $\T'$, yields
\[
k\leq |S'|+|\bigcup_{i=1}^p E(F_i')|-\sum_{i=1}^p|E(F_i')|.
\]
Because the extension is nice, it follows that
\[
\pres(\T)\geq |S'|+|\bigcup_{i=1}^p E(F_i')|
+\ell(|X|-1)-\sum_{i=1}^p|E(F_i')|
\geq k+\ell(|X|-1).
\]
Thus, $(X,d,\ell,k)$ is a yes-instance. This completes the proof.
\end{proof} 

Next, we compress the distances. Using \Cref{obs:betweenness}, we define a metric $d'\colon X'\times X'\to \mathbb{Z}_{\geq 0}$ such that $d(x,y)+d(y,z)=d(x,z)$ if and only if $d'(x,y)+d'(y,z)=d'(x,z)$ for all distinct $x,y,z\in X'$. Consider the system 
of linear inequalities for triples $(x,y,z)$ of distinct $x,y,z\in X'$
over 
$p=\binom{|X'|}{2}$ variables $\chi_{x,y}$ for pairs $\{x,y\}\in \binom{X'}{2}$:
\begin{equation}\label{eq:system}
\begin{cases}
\chi_{x,y}+\chi_{y,z}=\chi_{x,z}&\text{for }(x,y,z)\text{ s.t. }d(x,y)+d(y,z)=d(x,z),\\
\chi_{x,y}+\chi_{y,z}>\chi_{x,z}&\text{for }(x,y,z)\text{ s.t. }d(x,y)+d(y,z)>d(x,z),\\
\chi_{x,y}>0&\text{for }\{x,y\}\in\binom{X'}{2}.
\end{cases}
\end{equation}
By positive scaling, System~\ref{eq:system} is feasible if and only if the following system is feasible: 
\begin{equation}\label{eq:system_two}
\begin{cases}
\chi_{x,y}+\chi_{y,z}=\chi_{x,z}&\text{for }(x,y,z)\text{ s.t. }d(x,y)+d(y,z)=d(x,z),\\
\chi_{x,y}+\chi_{y,z}-\chi_{x,z}\geq 1&\text{for }(x,y,z)\text{ s.t. }d(x,y)+d(y,z)>d(x,z),\\
\chi_{x,y}\geq 1&\text{for }\{x,y\}\in\binom{X'}{2}.
\end{cases}
\end{equation}
Clearly, the assignment $\chi_{x,y}:=d(x,y)$ satisfies System~\ref{eq:system}, that is, the system is feasible. Therefore, System~\ref{eq:system_two} is feasible as well. Notice that the coefficients of System~\ref{eq:system_two} are integers from $\{-1,0,1\}$. Then the system admits a rational-valued solution $\chi_{x,y}^*$ for $\{x,y\}\in\binom{X'}{2}$, where the values $\chi_{x,y}^*$ have a common denominator $\delta>0$, and such a solution can be found in polynomial time (see, e.g., the monograph of Schrijver~\cite{Schrijver99}). Moreover, given that the coefficients are from $\{-1,0,1\}$, the fact that the absolute value of the determinant of a $t\times t$-matrix with such coefficients is at most $t^{t/2}$ by Hadamard's inequality (see, e.g., \cite{Lange14}), a solution can be chosen in such a way that the absolute values of the numerators and the denominator of the solution elements are at most $p^{p/2}$. Observe that $\chi_{x,y}^{**}=\delta \chi_{x,y}^*$ for $\{x,y\}\in \binom{X'}{2}$ is also a solution for System~\ref{eq:system_two}. We define $d'(x,y)$ by setting
$d'(x,x)=0$ for $x\in X'$ and defining $d'(x,y)=\chi_{x,y}^{**}$ for distinct $x,y\in X'$. By the definition of $d'$, we immediately obtain that $d'$ is an integer-valued metric. Then, by~\Cref{obs:betweenness}, the instances $(X',d,\ell',k)$ and $(X',d',\ell',k)$ are equivalent, and $d'$ is strictly convex.
Since $p=\binom{|X'|}{2}$ and $|X'|=\Oh(k^2)$, we have that $d'(x,y)=2^{\Oh(k^4\log k)}$. Finally, because each value $d'(x,y)$ can be encoded by a string of length $\Oh(k^4\log k)$ and $|X'|=\Oh(k^2)$, the instance $(X',d',\ell',k)$ can be encoded by $\Oh(k^8\log k)$ bits. This concludes the proof.
\end{proof}

\subsection{Kernelization lower bound}\label{sec:kern_lb}
In this subsection, we show that it is unlikely that \DCT admits a polynomial kernel when parameterized by $k$. 

\NoKernel*

\begin{proof}
We show the theorem by constructing a polynomial parameter transformation from the \textsc{Exact Cover} problem. This problem asks, given a \emph{universe} set $U$ and a family $\Scal=\{S_1,\ldots,S_m\}$ of subsets of $U$, whether there is a subfamily $\Scal'\subseteq\Scal$ of size $k$ \emph{exactly covering} $U$, that is, each element of $U$ is in a unique set of $\Scal'$. We consider the special variant of the problem where all sets of $\Scal$ have the same size $t$. It was shown by Dell and Marx in~\cite{DellM12} that this variant of \textsc{Exact Cover} does not admit a polynomial kernel when parameterized by both $k$ and $t$,
unless $\mathrm{coNP}\subseteq\mathrm{NP}/\mathrm{poly}$.

\begin{figure}[th]
\centering
\scalebox{0.8}{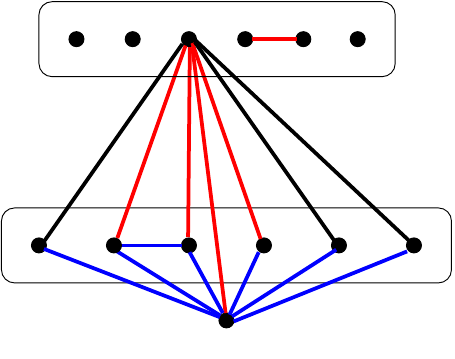}
\caption{Construction of $(X,d)$ for $S_i=\{u_p,u_q,u_s\}$ shown as the weighted graph $G=K_X$; the edges of weight $2$ are red, the edges of weight $4$ are blue, and the edges of weight $3$ are drawn in black.}
\label{fig:kernel_lb_a}
\end{figure}

Let $(U,\Scal,k)$ be an instance of \textsc{Exact Cover} with $U=\{u_1,\ldots,u_n\}$ and $\Scal=\{S_1,\ldots,S_m\}$ where $|S_1|=\dots=|S_m|=t$. We assume that $U=\bigcup_{i=1}^mS_i$ and  
$n=kt$ as $(U,\Scal,k)$ is a trivial no-instance otherwise. We construct the following metric space $(X,d)$ (see~\Cref{fig:kernel_lb_a}).
\begin{itemize}
\item Construction of $X$:
\begin{itemize}
\item construct $n$ points $u_1,\ldots,u_n$ and $m$ points $s_1,\ldots,s_m$,
\item construct a \emph{root} point $r$.
\end{itemize}
\item Construction of $d$:
\begin{itemize}
\item for all $x\in X$, set $d(x,x)=0$,
\item for all pairs $\{u_i,u_j\}$ with $1\leq i<j\leq n$, set $d(u_i,u_j)=4$,
\item for every $i\in\{1,\ldots,n\}$, set $d(r,u_i)=4$,
\item for all pairs $\{s_i,s_j\}$ with $1\leq i<j\leq m$, set $d(s_i,s_j)=2$,
\item for every $i\in\{1,\ldots,m\}$, set $d(r,s_i)=2$,
\item for every $i\in\{1,\ldots,m\}$ and each $j\in\{1,\ldots,n\}$,
set $d(s_i,u_j)=2$ if $u_j\in S_i$, and set $d(s_i,u_j)=3$ otherwise.
\end{itemize}
\end{itemize}
In the above construction, $u_1,\ldots,u_n$ correspond to the elements of the universe $U$, and we do not distinguish these elements of $X$ and $U$. In particular, we may write that $u_i\in S_j$ for $u_i\in X$.
It is straightforward to verify that $(X,d)$ is a metric space as the distances between distinct points are $2$, $3$, or $4$. We define $k'=k\binom{t+1}{2}$. We claim that $(U,\Scal,k)$ is a yes-instance of \textsc{Exact Cover} if and only if $G=K_X$ has a spanning tree $T$ with $\pres(T)\geq k'+|X|-1$. 

Suppose that $(U,\Scal,k)$ is a yes-instance of \textsc{Exact Cover}. Let $i_1,\ldots,i_k$ be distinct indices from $\{1,\ldots,m\}$ such that $\{S_{i_1},\ldots,S_{i_k}\}$ is a solution, that is, a partition of $U$. For $j\in\{1,\ldots,k\}$, consider 
$E_j=\{s_{i_j}u_h\mid u_h\in S_{i_j}\}$. Then define \[L=(\bigcup_{j=1}^kE_j)\cup\{rs_{i_j}\mid 1\leq j\leq k\}.\] 
Because $\{S_{i_1},\ldots,S_{i_k}\}$ is a partition of $U$, $F=G[L]$ is a tree. By the definition of $d(x,y)$,  
\begin{itemize}
\item for every $j\in\{1,\ldots,k\}$, $F$ strongly preserves every pair $\{u_p,u_q\}$ with distinct $u_p,u_q\in S_{i_j}$, and
\item $F$ strongly preserves every pair $\{r,u_i\}$ for $i\in \{1,\ldots,n\}$.
\end{itemize}
Thus, $F$ strongly preserves at least $k\binom{t+1}{2}=k'$ pairs of $\binom{X}{2}$.  By~\Cref{obs:extension}, we have that $G$ has a spanning tree $T$ with $\pres(T)\geq k'+|X|-1$.

For the opposite implication, assume that $G$ has a spanning tree $T$ with $\pres(T)\geq k'+|X|-1$. 
Let $R$ be the set of pairs $\{x,y\}\in \binom{X}{2}$ that are strongly preserved by $T$. In particular, $\pres(T)=|R|+|X|-1$ and $|R|\geq k'$. Consider a forest $F$ that is an $R$-restriction of $T$, that is, $F$ is a minimal strong $R$-preservation forest. We argue that $F$ has a special structure.

Consider the preservation graph $H_X$. By the construction of $(X,d)$,
\[E(H_X)=\{ru_i\mid 1\leq i\leq n\}\cup\bigcup_{i=1}^m\{u_pu_q\mid \{u_p,u_q\}\in \binom{S_i}{2}\}.\]
Then $R\subseteq\{\{x,y\}\mid xy\in E(H_X)\}$. Observe that for each $\{x,y\}\in \binom{X}{2}$ with $xy\in E(H_X)$, a witness $z\in\{s_1,\ldots,s_m\}$. Furthermore, 
\begin{itemize}
\item for $i\in \{1,\ldots,n\}$, $s_j$ is a witness for $\{r,u_i\}$ if and only if $u_i\in S_j$, and
\item for every $\{u_p,u_q\}$ for distinct $p,q\in \{1,\ldots,n\}$, $s_j$ is a witness for $\{u_p,u_q\}$ if and only if $u_p,u_q\in S_j$.
\end{itemize}
Also, the construction of $(X,d)$ implies that for any $\{x,y\}\in R$, a shortest $x$--$y$-path in $G$ distinct from $xy$ contains exactly two edges, and the middle vertex is a witness for the pair. Let $W=V(F)\cap\{s_1,\ldots,s_m\}=\{s_{i_1},\ldots,s_{i_p}\}$.
By the minimality of $F$, every vertex and edge of $F$ belongs to a two-edge shortest path for some pair of $R$. Thus, $F$ is bipartite with parts $W$ and $V(F)\setminus W\subseteq U\cup\{r\}$, and every vertex of $W$ has degree at least two. Furthermore, $N_F(s_{i_j})\subseteq S_{i_j}\cup\{r\}$ for $j\in\{1,\ldots,p\}$, 
and 
$R=\bigcup_{j=1}^p\binom{N_F(s_{i_j})}{2}$.

Since $N_F(s_{i_j})\subseteq S_{i_j}\cup\{r\}$, $\binom{|N_F(s_{i_j})|}{2}\leq\binom{t+1}{2}$ for $j\in\{1,\ldots,p\}$. Then because $|R|\geq k'=k\binom{t+1}{2}$, $p\geq k$.
Since $F$ is a bipartite forest and $|V(F)\setminus W|\leq n+1$, we have that
\[
\sum_{j=1}^p |N_F(s_{i_j})|=|E(F)|\leq |V(F)|-1\leq p+n.
\]
Thus, $\sum_{j=1}^p(|N_F(s_{i_j})|-1)\leq n=kt$. Consider the function
\[f(x_1,\ldots,x_p)=\sum_{i=1}^p \frac{x_i(x_i-1)}{2}.\]
For $1\leq x_i\leq t+1$, we have that
$\frac{x_i(x_i-1)}{2}\leq\frac{t+1}{2}(x_i-1)$,
with equality if and only if $x_i=1$ or $x_i=t+1$. Hence, if $\sum_{i=1}^p x_i\leq kt+p$, then
\[
f(x_1,\ldots,x_p)\leq \frac{t+1}{2}\sum_{i=1}^p(x_i-1)
\leq k\binom{t+1}{2}.
\]
Equality holds if and only if $\sum_{i=1}^p(x_i-1)=kt$ and every $x_i$ is either $1$ or $t+1$. Apply this with $x_j=|N_F(s_{i_j})|$ for $j\in\{1,\ldots,p\}$. Because $|R|\leq f(x_1,\ldots,x_p)$ and $|R|\geq k\binom{t+1}{2}$, equality must hold. Since every vertex of $W$ has degree at least two, no $x_j$ is $1$. Consequently, $p=k$ and $|N_F(s_{i_j})|=t+1$ for $j\in\{1,\ldots,k\}$. 
Since $N_F(s_{i_j})\subseteq S_{i_j}\cup\{r\}$ and $|S_{i_j}|=t$, we obtain that $r\in N_F(s_{i_j})$ and  
$N_F(s_{i_j})\setminus\{r\}=S_{i_j}$ for $j\in\{1,\ldots,k\}$. Because $F$ is a forest and $r$ is adjacent to $s_{i_1},\ldots,s_{i_k}$ in $F$, the sets $S_{i_1},\ldots,S_{i_k}$ are disjoint. This means that $\{S_{i_1},\ldots,S_{i_k}\}$ is a partition of $U$. Therefore, 
$(U,\Scal,k)$ is a yes-instance of \textsc{Exact Cover}. This completes the proof.
\end{proof}

\section{NP-hardness of \DCT for the Euclidean plane}\label{sec:NPc}
In this section, we prove \Cref{thm:nphard}. Throughout the section, we consider metric spaces $(X,d)$, where $X\subseteq \R^2$ and $d(x,y)=\|x-y\|_2$. The weighted graph $G=K_X$ is always defined according to this metric.

We need some additional notation and auxiliary claims.

\begin{definition}[Grids]\label{def:grid}
For a positive integer $N$, 
we say that a set of $N^2$ distinct points $p_{i,j}=(x_i,y_j)$ with integer coordinates
for $i,j\in\{0,\ldots,N-1\}$ is a \emph{$N\times N$-grid}. An $N\times N$-grid is a \emph{general-position} grid if for any three distinct points $p_{i_1,j_1}$, $p_{i_2,j_2}$, and $p_{i_3,j_3}$, the points are on the same line in $\R^2$ if and only if either $i_1=i_2=i_3$, or $j_1=j_2=j_3$, that is, any line which is not parallel to the coordinate axes contains at most two points.     
\end{definition}

\begin{lemma}\label{lem:grid}
For any $N\geq 1$, there is a general-position grid $\Gamma_N=\{p_{i,j}=(x_i,y_j)\mid 0\leq i,j\leq N-1 \}$ such that $0\leq x_i,y_j\leq N^4$ for all $i,j\in\{0,\ldots,N-1\}$.    
\end{lemma}

\begin{proof}
The lemma is trivial for $N=1$. Assume that $N\geq 2$.
Consider $\Gamma_N=\{p_{i,j}=(x_i,y_j)\mid 0\leq i,j\leq N-1\}$ with $x_i=i$ and $y_j=N^2j^2+j$ for $i,j\in\{0,\ldots,N-1\}$. Clearly, $0\leq x_i,y_j\leq N^4$ for all $i,j\in\{0,\ldots,N-1\}$. Consider any three points $(p,N^2h^2+h),(q,N^2i^2+i),(r,N^2j^2+j)\in\Gamma_N$ with $p<q<r$.
If these points are on the same line, then
\[\frac{N^2i^2+i-N^2h^2-h}{q-p}=\frac{N^2j^2+j-N^2i^2-i}{r-q},\]
and
\begin{equation}\label{eq:grid}
(N^2(i^2-h^2)+(i-h))(r-q)=(N^2(j^2-i^2)+(j-i))(q-p).
\end{equation}
If $i=h$, the equality of slopes implies that $j=i$, and the three points are on the same horizontal line. Assume that $i\neq h$. Since $y_0<\dots<y_{N-1}$ and $p<q<r$, the equality of slopes implies that $i-h$ and $j-i$ have the same sign. Hence $(i-h)(r-q)$ and $(j-i)(q-p)$ have the same sign and both have absolute value less than $N^2$, so their difference has absolute value less than $N^2$. Taking \Cref{eq:grid} modulo $N^2$, we obtain
$(i-h)(r-q)=(j-i)(q-p)$.
It follows that $N^2(i^2-h^2)(r-q)=N^2(j^2-i^2)(q-p)$. Since $i^2-h^2=(i+h)(i-h)$ and $j^2-i^2=(j+i)(j-i)$, we can divide by the common nonzero factor $(i-h)(r-q)=(j-i)(q-p)$ and obtain $i+h=j+i$, that is, $h=j$. This contradicts that $i-h$ and $j-i$ have the same sign. Therefore, $h=i=j$, and the three points are on the same horizontal line. This completes the proof.
\end{proof}

We use a general-position grid to choose points of $X\subseteq\R^2$ in our reduction. 
We can make the next observation, which follows from \Cref{obs:preservance} and the properties of the Euclidean metric.

\begin{observation}\label{obs:grid_choice}
Let  $X\subseteq \Gamma_N$ for a general position grid $\Gamma_N=\{p_{i,j}=(x_i,y_j)\mid 0\leq i,j\leq N-1\}$. Then for any pair $\{v_1=(x_1,y_1),v_2=(x_2,y_2)\}\in\binom{X}{2}$, $G=K_X$ has a forest $F$ strongly preserving $\{v_1,v_2\}$ if and only if either $x_1=x_2$ or $y_1=y_2,$ and there is $z=(x,y)\in X$ that is an internal point of the segment of either vertical or horizontal line with the endpoints $v_1$ and $v_2$.  
\end{observation}

For $X\subseteq \Gamma_N=\{p_{i,j}=(x_i,y_j)\mid 0\leq i,j\leq N-1\}$, we define 
$X_i=X\cap\{p_{i,0},\ldots,p_{i,N-1}\}$ and 
$X^j=X\cap \{p_{0,j},\ldots,p_{N-1,j}\}$ for $i,j\in\{0,\ldots,N-1\}$. 
Let $Y=\{v_1=(x_i,y_{j_1}),\ldots,v_\ell=(x_i,y_{j_\ell})\}\subseteq X_i$ 
with $y_{j_1}<\dots<y_{j_\ell}$
($Y=\{v_1=(x_{i_1},y_j),\ldots,v_\ell=(x_{i_\ell},y_j)\}\subseteq X^j$ with
$x_{i_1}<\dots<x_{i_\ell}$, respectively). Then we say that $P=v_1\dots v_\ell$ is a \emph{straight $Y$-path}.  We use the following properties.

\begin{observation}\label{obs:max_preserve}
Let $X\subseteq \Gamma_N$ for a general position grid $\Gamma_N$, and let $F$ be a forest in $G=K_X$. Then $F$ strongly preserves at most $\binom{|V(F)|}{2}-|V(F)|+1$ pairs of $\binom{X}{2}$ if $F$ is non-empty, and the maximum $\binom{|V(F)|}{2}-|V(F)|+1$ is achieved when $F$ is a straight $V(F)$-path. Furthermore, if $F$ has at least two connected components, it strongly preserves at most $\binom{|V(F)|-1}{2}-|V(F)|+2$ pairs.   
\end{observation}

\begin{proof}
The first bound is immediate if $F$ is empty. Otherwise, suppose that $F$ has $r$ connected components with $s_1,\ldots,s_r$ vertices, respectively. A strongly preserved pair has its endpoints in the same component and does not correspond to an edge of $F$. Therefore, the component with $s_i$ vertices strongly preserves at most $\binom{s_i}{2}-s_i+1$ pairs, and $F$ strongly preserves at most
\[
f(s_1,\ldots,s_r)=\sum_{i=1}^r\left(\binom{s_i}{2}-s_i+1\right)
\]
pairs. By the convexity of $f$ and the symmetry between variables, for fixed $r$ the maximum is achieved when $s_1=|V(F)|-r+1$ and $s_2=\dots=s_r=1$. This gives at most $\binom{|V(F)|}{2}-|V(F)|+1$ pairs. A straight $V(F)$-path attains this bound, since it preserves all pairs of its vertices. If $r\geq 2$, the maximum over the possible values of $r$ is achieved when $r=2$, giving the bound $\binom{|V(F)|-1}{2}-|V(F)|+2$.     
\end{proof}

\NPhard*

\begin{proof}
We prove the theorem by reducing from \textsc{Exact Cover}. Recall that the task of the problem is, given a universe set $U$ and a family $\Scal=\{S_1,\ldots,S_m\}$ of subsets of $U$, to decide whether there is a subfamily $\Scal'\subseteq\Scal$ of size $k$ exactly covering $U$, that is, each element of $U$ is in a unique set of $\Scal'$. The problem is known to be \textup{NP}-complete when each set is of size three, and each element of the universe occurs in exactly three sets~\cite{Gonzalez85}. 

\begin{figure}[th]
\centering
\scalebox{0.75}{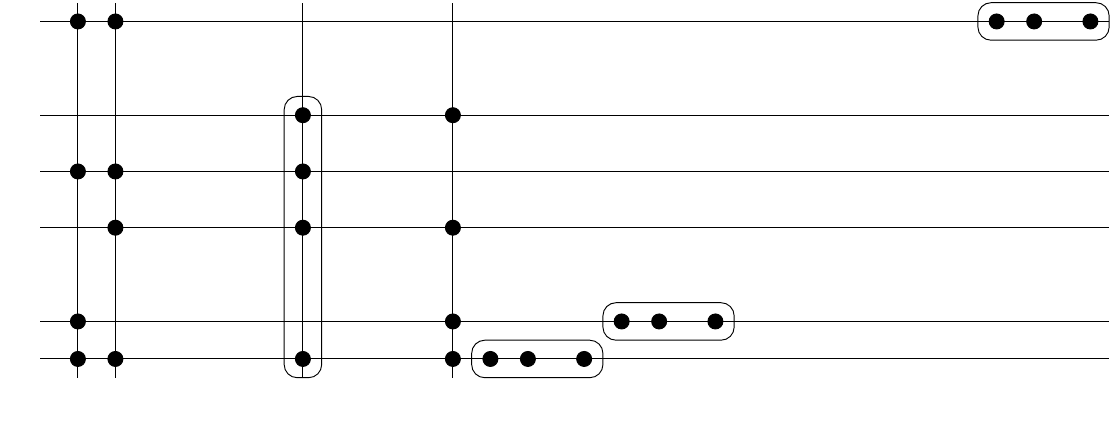}
\caption{Construction of $X\subseteq \Gamma_N$.}
\label{fig:npc_a}
\end{figure}

Let $(U,\mathcal{S},k)$ with $U=\{u_1,\ldots,u_n\}$ and $\Scal=\{S_1,\ldots,S_m\}$ be such an instance of \textsc{Exact Cover} with $n=3k$.
We set $N=3m(n+1)+m+1$ and construct 
a general position grid $\Gamma_N=\{p_{i,j}=(x_i,y_j)\mid 0\leq i,j\leq N-1\}$ using~\Cref{lem:grid}. In the construction, we select a set of points $X\subseteq \Gamma_N$. Notice that, by~\Cref{lem:grid}, the absolute values of the coordinates of the points are upper bounded by a polynomial in $n$ and $m$. Then $X$ is constructed as follows (see~\Cref{fig:npc_a}).
\begin{itemize}
\item For each $j\in\{1,\ldots,m\}$, assume that $S_j=\{u_{i_1},u_{i_2},u_{i_3}\}$  and 
construct a set
$\hat{S}_j=\{p_{j,0},p_{j,i_1},p_{j,i_2},p_{j,i_3}\}$.
\item For each $i\in\{0,\ldots,n\}$, construct a \emph{forcing} set $R_i=\{p_{m+3mi+1,i},\ldots,p_{m+3m(i+1),i}\}$ of size $3m$.
\item Set $X=\big(\bigcup_{j=1}^m\hat{S}_j\big)\cup\big(\bigcup_{i=0}^n R_i\big)$
\end{itemize}
We set 
\begin{equation}\label{eq:kprime}
k'=3k+(\binom{4m}{2}-4m+1)+\sum_{i=1}^n(\binom{3m+3}{2}-3m-2).   
\end{equation}
We show that $(U,\Scal,k)$ is a yes-instance of \textsc{Exact Cover} if and only if $G=K_X$ has a spanning tree $T$ with $\pres(T)\geq k'+|X|-1$. 

Suppose that $(U,\Scal,k)$ is a yes-instance, and let $\mathcal{S}'=\{S_{j_1},\ldots,S_{j_k}\}$ be an exact cover of $U$. Recall that $X_i=X\cap\{p_{i,0},\ldots,p_{i,N-1}\}$ and 
$X^i=X\cap \{p_{0,i},\ldots,p_{N-1,i}\}$ for $i\in\{0,\ldots,N-1\}$.
For $i\in\{0,\ldots,n\}$, let $P_i$ be the straight $X^i$-path in $G$, that is, $P_i$ is the shortest path in $G$ with $V(P_i)=X^i$. Also, for $i\in\{1,\ldots,k\}$, let $P_i'$ be the straight $X_{j_i}$-path. Consider the graph $T$ obtained by taking the union of $P_0,\ldots,P_n$ and $P_1',\ldots,P_k'$. Observe that $T$ is a spanning tree. To see this, note that $P_0,\ldots,P_n$ are disjoint paths spanning $V(G)$. Then for each $i\in\{1,\ldots,k\}$, $P_i'$ connects the path $P_0$ with the paths $P_h$ containing the vertices of $\hat{S}_{j_i}$, that is, $P_0$ is joined with $P_h$ if and only if $u_h\in S_{j_i}$. Because $\Scal'$ is an exact cover, we obtain that $T$ is a spanning tree.

The path $P_0$ in $T$ strongly preserves 
$\binom{|X^0|}{2}-|X^0|+1=\binom{4m}{2}-4m+1$ pairs $\{u,v\}\in\binom{X^0}{2}$ by~\Cref{obs:max_preserve}.
For $i\in\{1,\ldots,n\}$, $P_i$ in $T$ strongly preserves $\binom{3m+3}{2}-3m-2$ pairs, as  $u_i$ occurs in exactly three sets of $\Scal$. For each $i\in\{1,\ldots,k\}$, $P_i'$ strongly preserves $3$ pairs, because $X_{j_i}=\hat{S}_{j_i}$ is of size $4$. Thus, 
$\pres^*(T)=3k+(\binom{4m}{2}-4m+1)+\sum_{i=1}^n(\binom{3m+3}{2}-3m-2)$, and 
$\pres(T)\geq k'+|X|-1$. 

For the opposite implication, suppose that $G=K_X$ has a spanning tree $T$ with $\pres(T)\geq k'+|X|-1$. Let $L\subseteq \binom{X}{2}$ be the set of pairs strongly preserved by $T$. Then $|L|\geq k'$, and $T$ contains a minimal strong $L$-preservation forest $F$. 
For $i\in\{0,\ldots,n\}$, denote by $P_i$ the straight $X^i$-path in $G$, and for $j\in\{1,\ldots,m\}$, let $P_j'$ be the straight $X_j$-path in $G$.
Observe that because $\Gamma_N$ is a general-position grid, by the construction of $X$, $P_0,\ldots,P_n$ and $P_1',\ldots,P_m'$ are the inclusion-maximal shortest paths in $G$ with at least three vertices. In particular, any pair $\{u,v\}\in L$ is a pair of distinct vertices of one of these paths. Moreover, for every $Y\in\{X^0,\ldots,X^n,X_1,\ldots,X_m\}$, any shortest path in $G$ between two vertices of $Y$ uses only vertices of $Y$, since all its vertices lie on the segment joining its endpoints. Thus, a pair of $\binom{Y}{2}$ is strongly preserved by $F$ if and only if it is strongly preserved by $F[Y]$. We use this to argue that for each $i\in\{0,\ldots,n\}$, $X^i\subseteq V(F)$ and $F[X^i]$ is connected. 
 
The proof is by contradiction. We use the fact that for each $i\in\{0,\ldots,n\}$, $X^i$ contains the forcing set $R_i$.
Assume that there is $i\in\{0,\ldots,n\}$ such that either 
$X^i\setminus V(F)\neq\emptyset$ or $F[X^i]$ is disconnected. Consider first the case $i=0$. Recall that $|X^0|=4m$. If $X^0\setminus V(F)\neq\emptyset$, then $F$ strongly preserves at most $\binom{4m-1}{2}-4m+2$ pairs $\{u,v\}\in\binom{X^0}{2}$ by~\Cref{obs:max_preserve}.
For each $i\in\{1,\ldots,n\}$, $F$ can strongly preserve at most $\binom{3m+3}{2}-3m-2$ pairs of $\binom{X^i}{2}$ since $|X^i|=3m+3$, and for each $j\in\{1,\ldots,m\}$, $F$ can strongly preserve at most three pairs of $\binom{X_j}{2}$ as $|X_j|=4$. Thus, $F$ can strongly preserve at most
\begin{align*}
 &(\binom{4m-1}{2}-4m+2)+\sum_{i=1}^n(\binom{3m+3}{2}-3m-2)+3m\\
 =&(\binom{4m}{2}-4m+1)+\sum_{i=1}^n(\binom{3m+3}{2}-3m-2)\\
 +&3m-((\binom{4m}{2}-4m+1)-(\binom{4m-1}{2}-4m+2))\\
 =&(\binom{4m}{2}-4m+1)+\sum_{i=1}^n(\binom{3m+3}{2}-3m-2)-m+2<k',
\end{align*}
contradicting that $F$ strongly preserves at least $k'$ pairs.
If $X^0\subseteq V(F)$ but $F[X^0]$ is disconnected, then the arguments are exactly the same as $F$ strongly preserves at most $\binom{4m-1}{2}-4m+2$ pairs of $\binom{X^0}{2}$ by~\Cref{obs:max_preserve}.
Assume now that $i\in\{1,\ldots,n\}$. The arguments are almost the same. We note that 
$F$ strongly preserves at most $\binom{3m+2}{2}-3m-1$ pairs in $\binom{X^i}{2}$ by~\Cref{obs:max_preserve}.
Therefore, the total number of strongly preserved pairs is at most 
\begin{align*}
 &(\binom{4m}{2}-4m+1)+\sum_{i=1}^n(\binom{3m+3}{2}-3m-2)\\+&3m-((\binom{3m+3}{2}-3m-2)-(\binom{3m+2}{2}-3m-1))\\
 =&(\binom{4m}{2}-4m+1)+\sum_{i=1}^n(\binom{3m+3}{2}-3m-2)-1<k'.
\end{align*}
Again, we obtain a contradiction.
Thus,  $X^i\subseteq V(F)$ and $F[X^i]$ is connected for $i\in\{0,\ldots,n\}$.

Note that $F$ strongly preserves at most 
\[(\binom{4m}{2}-4m+1)+\sum_{i=1}^n(\binom{3m+3}{2}-3m-2)\]
pairs from the sets $\binom{X^i}{2}$ for $i\in\{0,\ldots,n\}$.
Then, besides these pairs,
$F$ strongly preserves at least $3k$ pairs $\{u,v\}\in \binom{X_j}{2}$ for $j\in\{1,\ldots,m\}$.
Let $j_1,\ldots,j_r\in\{1,\ldots,m\}$ be the indices such that for every $i\in\{1,\ldots,r\}$,
$F$ strongly preserves at least one pair of points of $X_{j_i}$. Since $|X_j|=4$ for $j\in\{1,\ldots,m\}$, $F$ strongly preserves at most three pairs in each $\binom{X_j}{2}$. Then, by~\Cref{eq:kprime}, $r\geq k$. 

Because $|X_{j_i}|\leq 4$ for $i\in\{1,\ldots,r\}$ and $F$ strongly preserves a pair $\{u,v\}\in \binom{X_{j_i}}{2}$, $F[X_{j_i}]$ contains one connected component $F_i$ without isolated vertices by~\Cref{obs:preservance}, and $F_i$ has at least three vertices. 
Because $F[X^h]$ is connected for each $h\in\{0,\ldots,n\}$, we can contract these $n+1$ connected subforests to single vertices. The edges of each $F_j$ have their endpoints in distinct sets $X^h$, and the edge sets $E(F_1),\ldots,E(F_r)$ are disjoint. Since $F$ is a forest, after the contractions these edges form an acyclic edge set on $n+1$ vertices. Therefore, $\sum_{j=1}^r|E(F_j)|\leq n$. By~\Cref{obs:max_preserve}, we obtain that 
\[\sum_{i=1}^r(\binom{|V(F_i)|}{2}-|V(F_i)|+1)\geq 3k.\]
Consider 
\[f(x_1,\ldots,x_r)=\sum_{i=1}^r(\frac{x_i(x_i-1)}{2}-x_i+1)\]
and observe that if $1\leq x_i\leq 4$ for $i\in\{1,\ldots,r\}$ and  
$\sum_{i=1}^rx_i\leq n+r=3k+r$, then $f$ achieves its maximum value if exactly $k$ variables are equal to $4$, and other variables are ones by the convexity and the symmetry with respect to the variables. Furthermore, the maximum value is $3k$. Thus, 
\[\sum_{i=1}^r(\binom{|V(F_i)|}{2}-|V(F_i)|+1)\geq 3k\]
if and only if $r=k$ and $|V(F_i)|=4$. This means that $r=k$ and $F[X_{j_i}]$ is connected for each $i\in\{1,\ldots,k\}$.  Then because $F$ is a forest, we obtain that there is no $i\in\{1,\ldots,n\}$ such that $X^i\cap X_{j_h}\neq\emptyset$ and $X^i\cap X_{j_{h'}}\neq\emptyset$
for two distinct $h,h'\in\{1,\ldots,k\}$. Then the construction of $X$ implies that the sets $S_{j_1},\ldots,S_{j_k}\subseteq U$ are disjoint. As $n=3k$, we have that 
$\Scal'=\{S_{j_1},\ldots,S_{j_k}\}$ is an exact cover of $U$. Thus, $(U,\Scal,k)$ is a yes-instance. This completes the proof.
\end{proof}

\section{Conclusion}\label{sec:conclusion}

We studied distance coverage by a bounded number of spanning trees,
viewing it as a  maximum-coverage counterpart of exact tree
covers. Our results establish fixed-parameter tractability above the
guarantee $\ell(n-1)$ and identify a distinction between general metrics
and strictly convex metrics concerning polynomial compression.

A natural extension of \DCT allows the trees to contain additional vertices,
called \emph{Steiner points}. For geometric inputs, this leads to the
following model. Let $X$ be a finite subset of a normed vector space
$(V,\|\cdot\|)$. Each solution tree has vertex set $X\cup S$, where
$S\subseteq V\setminus X$ is a finite set of Steiner points, and each edge
$uv$ has length $\|u-v\|$. A pair $\{x,y\}\in\binom{X}{2}$ is preserved
if its distance in at least one solution tree equals $\|x-y\|$.
Only pairs of input points contribute to the objective.

A norm is \emph{strictly convex} if
 $ \left\|\frac{1}{2}(a+b)\right\|<1$
  {whenever } $\|a\|=\|b\|=1$ { and } $a\neq b$.
Consequently, equality
$
  \|x-y\|=\|x-z\|+\|z-y\|
$
holds precisely when $z$ belongs to the line segment $[x,y]$.
More generally, every path whose length equals the distance between its
endpoints follows their line segment in order. This also implies the
finite strict convexity property used in this paper: if $z$ and $w$ both
lie on shortest paths between $x$ and $y$, then there is a shortest
$x$--$y$-path containing both $z$ and $w$.

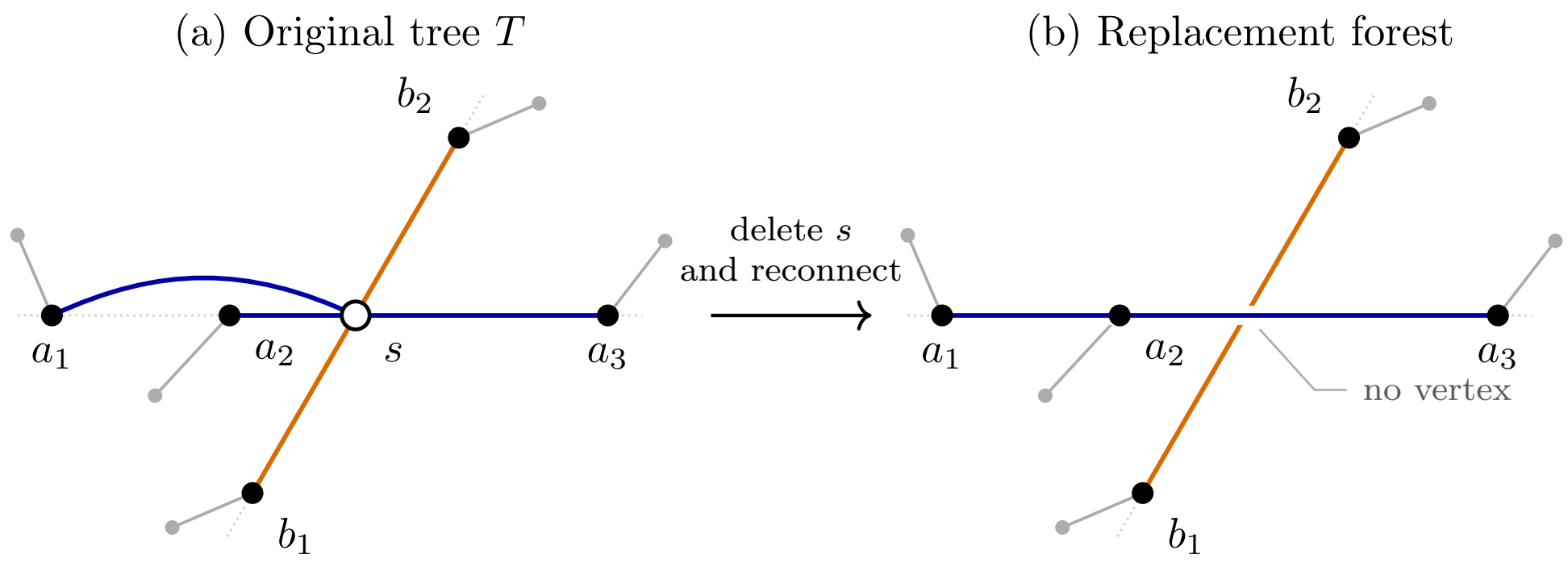
\begin{figure}[t]
\centering
\begin{tikzpicture}[
    x=0.94cm,y=0.94cm,font=\small,
    terminal/.style={circle,fill=black,inner sep=1.8pt},
    auxiliary/.style={circle,fill=gray!65,inner sep=1.2pt},
    retained/.style={draw=gray!65,line width=0.7pt},
    groupa/.style={draw=blue!65!black,line width=1.1pt},
    groupb/.style={draw=orange!85!black,line width=1.1pt},
    guide/.style={draw=gray!40,densely dotted,line width=0.5pt}
]
\foreach \offset/\stage in {0/0,7.3/1} {
  \begin{scope}[xshift=\offset cm]
    \coordinate (s)  at (0,0);
    \coordinate (a1) at (-2.65,0);
    \coordinate (a2) at (-1.10,0);
    \coordinate (a3) at (2.20,0);
    \coordinate (b1) at (-0.90,-1.55);
    \coordinate (b2) at (0.90,1.55);

    \draw[guide] (-2.95,0) -- (2.50,0);
    \draw[guide] (-1.12,-1.9289) -- (1.12,1.9289);

    \draw[retained] (a1) -- (-2.95,0.70)
      node[auxiliary] {};
    \draw[retained] (a2) -- (-1.75,-0.70)
      node[auxiliary] {};
    \draw[retained] (a3) -- (2.70,0.65)
      node[auxiliary] {};
    \draw[retained] (b1) -- (-1.60,-1.85)
      node[auxiliary] {};
    \draw[retained] (b2) -- (1.60,1.85)
      node[auxiliary] {};

    \ifnum\stage=0
      \draw[groupa] (s) to[out=155,in=25] (a1);
      \draw[groupa] (a2) -- (s) -- (a3);
      \draw[groupb] (b1) -- (s) -- (b2);
      \node[circle,draw=black,fill=white,line width=0.9pt,
            inner sep=2.3pt,label=below right:{$s$}] at (s) {};
      \node at (-0.05,2.45) {(a) Original tree $T$};
    \else
      \draw[groupb] (b1) -- (b2);
      \draw[groupa,preaction={draw=white,line width=4.5pt}]
        (a1) -- (a2) -- (a3);
      \draw[draw=gray!65,line width=0.5pt]
        (0.12,-0.12) -- (0.60,-0.65) -- (0.88,-0.65);
      \node[anchor=west,font=\scriptsize,text=gray!70!black]
        at (0.88,-0.65) {no vertex};
      \node at (-0.05,2.45) {(b) Replacement forest};
    \fi

    \node[terminal,label=below:{$a_1$}] at (a1) {};
    \node[terminal,label=below right:{$a_2$}] at (a2) {};
    \node[terminal,label=below:{$a_3$}] at (a3) {};
    \node[terminal,label=below right:{$b_1$}] at (b1) {};
    \node[terminal,label=above left:{$b_2$}] at (b2) {};
  \end{scope}
}
\draw[->,thick] (3.10,0) -- (4.50,0)
  node[midway,above=4pt,align=center,font=\scriptsize]
  {delete $s$\\and reconnect};
\end{tikzpicture}
\caption{Eliminating a geometric Steiner vertex $s$. The neighbors
  $a_1,a_2,a_3$ lie on one line through $s$, and $b_1,b_2$ lie on another.
  After deleting $s$, each group is connected in its order along the
  corresponding line, producing the paths $a_1a_2a_3$ and $b_1b_2$.
  Gray edges represent unchanged parts of the distinct components of
  $T-s$.}
\label{fig:steiner-elimination}
\end{figure}

In this geometric Steiner model, strict convexity ensures that Steiner
points cannot improve the optimum. To see this, consider a solution
tree $T$ and a Steiner vertex $s$. Delete $s$ and its incident edges.
Group its former neighbors according to the unoriented lines through
$s$. Within each group, connect the neighbors by a path in their order
along the corresponding line, see \Cref{fig:steiner-elimination}. The resulting graph is a forest:
distinct neighbors of $s$ belonged to distinct components after its
deletion, and the added edges connect these components without creating
a cycle.

Every previously preserved input pair remains connected by a path of
the same length.  Repeating this operation removes all
Steiner vertices while retaining every preserved input pair. Finally,
the resulting forest on $X$ can be extended to a spanning tree without
changing any of its existing paths.

Applying this argument separately to each solution tree shows that,
for every budget $\ell$, the geometric Steiner model and the spanning
tree model have the same optimum on finite subsets of strictly convex
normed spaces. Thus our FPT algorithm, the compression,
 and the NP-hardness result for the Euclidean plane extend
to this model without changing the parameter $k$ or the target
$\ell(n-1)+k$. For rational distances, the polynomial bit compression
also remains available with the same output-class qualification as
in~\Cref{thm:kernel-euclid}: the compressed instance need not remain
Euclidean.

A broader research direction is to allow \emph{abstract Steiner tree
metrics}. Here the additional vertices need not have positions in an
ambient space, and edge lengths may be chosen freely, subject to the 
non-contraction condition 
$
  d_{T_i}(x,y)\geq d(x,y)
$
  {for all } $x,y\in X $  { and } $i\in\{1,\ldots,\ell\}$.
 
The objective is again to maximize the number of input pairs for which
equality holds in at least one tree. This freedom can substantially
increase coverage. For example, the metric consisting of $n$ points at
unit pairwise distance is represented exactly by a single star with a
Steiner center and edges of length $1/2$, whereas a spanning tree on
the input points preserves only its $n-1$ edges.

Determining the parameterized complexity and kernelization landscape
of this abstract Steiner variant is an interesting open direction.
In particular, which universal coverage guarantees provide useful
baselines, and when does parameterizing the excess above such a
guarantee yield FPT algorithms or polynomial kernels? Our equivalence
for the geometric model does not apply to this setting, and the
general-metric kernel lower bound for spanning trees does not
automatically extend to it.

\subparagraph*{AI Statement.} 
The initial results presented in this work were obtained and verified by the authors without the assistance of any AI tools. ChatGPT Astra was used in the final stage of manuscript preparation for improvement suggestions and proofreading. The suggestions were used in \Cref{thm:kernel-euclid} to improve the number of points in the compression from $\Oh(k^4)$ to $\Oh(k^2)$ and reduce the bit-size of the kernel. Finally, ChatGPT Astra was used to draw some figures.

\bibliographystyle{siam}
\bibliography{mdp-blb}

\end{document}